\documentclass[twoside]{article}
\usepackage{epsfig}

\def\beq{\begin{eqnarray}} 
\def\eeq{\end{eqnarray}}
\def\Slash#1{#1\hskip-0.65em/}

\catcode`\@=11
\long\def\@makefntext#1{
\protect\noindent \hbox to 3.2pt {\hskip-.9pt  
$^{{\eightrm\@thefnmark}}$\hfil}#1\hfill}               

\def\thefootnote{\fnsymbol{footnote}}
\def\@makefnmark{\hbox to 0pt{$^{\@thefnmark}$\hss}}    
\def\slash#1{#1\hskip-0.5em/}
        
\def\ps@myheadings{\let\@mkboth\@gobbletwo
\def\@oddhead{\hbox{}
\rightmark\hfil\eightrm\thepage}   
\def\@oddfoot{}\def\@evenhead{\eightrm\thepage\hfil
\leftmark\hbox{}}\def\@evenfoot{}
\def\sectionmark##1{}\def\subsectionmark##1{}}

\oddsidemargin=\evensidemargin
\renewcommand{\thefootnote}{\fnsymbol{footnote}}

\newcounter{sectionc}\newcounter{subsectionc}\newcounter{subsubsectionc}
\renewcommand{\section}[1] {\vspace{12pt}\addtocounter{sectionc}{1} 
\setcounter{subsectionc}{0}\setcounter{subsubsectionc}{0}\noindent 
        {\tenbf\thesectionc. #1}\par\vspace{5pt}}
\renewcommand{\subsection}[1] {\vspace{12pt}\addtocounter{subsectionc}{1} 
        \setcounter{subsubsectionc}{0}\noindent 
        {\bf\thesectionc.\thesubsectionc. {\kern1pt \bfit #1}}\par\vspace{5pt}}
\renewcommand{\subsubsection}[1] {\vspace{12pt}\addtocounter{subsubsectionc}{1}
        \noindent{\tenrm\thesectionc.\thesubsectionc.\thesubsubsectionc.
        {\kern1pt \tenit #1}}\par\vspace{5pt}}
\newcommand{\nonumsection}[1] {\vspace{12pt}\noindent{\tenbf #1}
        \par\vspace{5pt}}

\newcounter{appendixc}
\newcounter{subappendixc}[appendixc]
\newcounter{subsubappendixc}[subappendixc]
\renewcommand{\thesubappendixc}{\Alph{appendixc}.\arabic{subappendixc}}
\renewcommand{\thesubsubappendixc}
        {\Alph{appendixc}.\arabic{subappendixc}.\arabic{subsubappendixc}}

\renewcommand{\appendix}[1] {\vspace{12pt}
        \refstepcounter{appendixc}
        \setcounter{figure}{0}
        \setcounter{table}{0}
        \setcounter{lemma}{0}
        \setcounter{theorem}{0}
        \setcounter{corollary}{0}
        \setcounter{definition}{0}
        \setcounter{equation}{0}
        \renewcommand{\thefigure}{\Alph{appendixc}.\arabic{figure}}
        \renewcommand{\thetable}{\Alph{appendixc}.\arabic{table}}
        \renewcommand{\theappendixc}{\Alph{appendixc}}
        \renewcommand{\thelemma}{\Alph{appendixc}.\arabic{lemma}}
        \renewcommand{\thetheorem}{\Alph{appendixc}.\arabic{theorem}}
        \renewcommand{\thedefinition}{\Alph{appendixc}.\arabic{definition}}
        \renewcommand{\thecorollary}{\Alph{appendixc}.\arabic{corollary}}
        \renewcommand{\theequation}{\Alph{appendixc}.\arabic{equation}}
        \noindent{\tenbf Appendix \theappendixc #1}\par\vspace{5pt}}
\newcommand{\subappendix}[1] {\vspace{12pt}
        \refstepcounter{subappendixc}
        \noindent{\bf Appendix \thesubappendixc. {\kern1pt \bfit #1}}
        \par\vspace{5pt}}
\newcommand{\subsubappendix}[1] {\vspace{12pt}
        \refstepcounter{subsubappendixc}
        \noindent{\rm Appendix \thesubsubappendixc. {\kern1pt \tenit #1}}
        \par\vspace{5pt}}

\newcommand{\textlineskip}{\baselineskip=13pt}
\newcommand{\smalllineskip}{\baselineskip=10pt}

\def\eightcirc{
\begin{picture}(0,0)
\put(4.4,1.8){\circle{6.5}}
\end{picture}}
\def\eightcopyright{\eightcirc\kern2.7pt\hbox{\eightrm c}} 

\newcommand{\copyrightheading}[1]
        {\vspace*{-2.5cm}\smalllineskip{\flushleft
        {\footnotesize International Journal of Modern Physics A, #1}\\
        {\footnotesize $\eightcopyright$\, World Scientific Publishing
         Company}\\
         }}

\def\abstracts#1#2#3{{
        \centering{\begin{minipage}{4.5in}\baselineskip=10pt\footnotesize
        \parindent=0pt #1\par 
        \parindent=15pt #2\par
        \parindent=15pt #3
        \end{minipage}}\par}} 

\renewenvironment{thebibliography}[1]
        {\frenchspacing
         \ninerm\baselineskip=11pt
         \begin{list}{\arabic{enumi}.}
        {\usecounter{enumi}\setlength{\parsep}{0pt}
         \setlength{\leftmargin 12.7pt}{\rightmargin 0pt} 
         \setlength{\itemsep}{0pt} \settowidth
        {\labelwidth}{#1.}\sloppy}}{\end{list}}

\newcounter{itemlistc}
\newcounter{romanlistc}
\newcounter{alphlistc}
\newcounter{arabiclistc}

\newcommand{\fcaption}[1]{
        \refstepcounter{figure}
        \setbox\@tempboxa = \hbox{\footnotesize Fig.~\thefigure. #1}
        \ifdim \wd\@tempboxa > 5in
           {\begin{center}
        \parbox{5in}{\footnotesize\smalllineskip Fig.~\thefigure. #1}
            \end{center}}
        \else
             {\begin{center}
             {\footnotesize Fig.~\thefigure. #1}
              \end{center}}
        \fi}

\newcommand{\tcaption}[1]{
        \refstepcounter{table}
        \setbox\@tempboxa = \hbox{\footnotesize Table~\thetable. #1}
        \ifdim \wd\@tempboxa > 5in
           {\begin{center}
        \parbox{5in}{\footnotesize\smalllineskip Table~\thetable. #1}
            \end{center}}
        \else
             {\begin{center}
             {\footnotesize Table~\thetable. #1}
              \end{center}}
        \fi}

\def\@citex[#1]#2{\if@filesw\immediate\write\@auxout
        {\string\citation{#2}}\fi
\def\@citea{}\@cite{\@for\@citeb:=#2\do
        {\@citea\def\@citea{,}\@ifundefined
        {b@\@citeb}{{\bf ?}\@warning
        {Citation `\@citeb' on page \thepage \space undefined}}
        {\csname b@\@citeb\endcsname}}}{#1}}

\newif\if@cghi
\def\cite{\@cghitrue\@ifnextchar [{\@tempswatrue
        \@citex}{\@tempswafalse\@citex[]}}
\def\citelow{\@cghifalse\@ifnextchar [{\@tempswatrue
        \@citex}{\@tempswafalse\@citex[]}}
\def\@cite#1#2{{$\null^{#1}$\if@tempswa\typeout
        {IJCGA warning: optional citation argument 
        ignored: `#2'} \fi}}

\def\pmb#1{\setbox0=\hbox{#1}
        \kern-.025em\copy0\kern-\wd0
        \kern.05em\copy0\kern-\wd0
        \kern-.025em\raise.0433em\box0}

\def\fnt#1#2{\footnotetext{\kern-.3em
        {$^{\mbox{\scriptsize #1}}$}{#2}}}

\def\fpage#1{\begingroup
\voffset=.3in
\thispagestyle{empty}\begin{table}[b]\centerline{\footnotesize #1}
        \end{table}\endgroup}

\def\runninghead#1#2{\pagestyle{myheadings}
\markboth{{\protect\footnotesize\it{\quad #1}}\hfill}
{\hfill{\protect\footnotesize\it{#2\quad}}}}
\font\tenrm=cmr10
\font\tenit=cmti10 
\font\tenbf=cmbx10
\font\bfit=cmbxti10 at 10pt
\font\ninerm=cmr9

\font\eightrm=cmr8

\def\qed{\hbox{${\vcenter{\vbox{                        
   \hrule height 0.4pt\hbox{\vrule width 0.4pt height 6pt
   \kern5pt\vrule width 0.4pt}\hrule height 0.4pt}}}$}}

\renewcommand{\thefootnote}{\fnsymbol{footnote}}        

\begin{document}

\runninghead{Quark Structure of Pseudoscalar Mesons}{Quark Structure of Pseudoscalar Mesons}

\normalsize\textlineskip
\thispagestyle{empty}
\setcounter{page}{1}

\copyrightheading{}                     

\vspace*{0.88truein}

\fpage{1}
\centerline{\bf QUARK STRUCTURE OF PSEUDOSCALAR MESONS}
\vspace*{0.37truein}
\centerline{\footnotesize THORSTEN FELDMANN\footnote{
Address after September 1999:\\ {\it Institut f\"ur Theoretische
Physik E, RWTH Aachen, 52056 Aachen (Germany)}\/.}  }
\vspace*{0.015truein}
\centerline{\footnotesize\it Fachbereich Physik, Universit\"at
  Wuppertal, Gau\ss{}stra\ss{}e 20}
\baselineskip=10pt
\centerline{\footnotesize\it D-42097 Wuppertal, Germany}
\vspace*{0.225truein}



\vspace*{0.21truein} \abstracts{ 
I review to which extent the 
properties of pseudoscalar mesons can be understood in terms of the
underlying quark (and eventually gluon) structure. Special emphasis is put on
the progress in our understanding of $\eta$-$\eta'$ mixing.
Process-independent mixing parameters are
defined, and relations between different bases and conventions are
studied.  Both, the low-energy description in the
framework of chiral perturbation theory  and the high-energy
application in terms of light-cone wave functions for partonic Fock
states, are considered. A thorough discussion of theoretical and
phenomenological consequences of the mixing approach will be given.
Finally, I will discuss 
mixing with other states ($\pi^0$, $\eta_c$, \ldots).  }{}{}


\vspace*{1pt}\textlineskip      


\section{Introduction}  
\vspace*{-0.5pt}
\noindent
\textheight=7.8truein
\setcounter{footnote}{0}
\renewcommand{\thefootnote}{\alph{footnote}}
The fundamental degrees of
freedom in strong interactions of hadronic matter
are quarks and gluons, and
their behavior is controlled by  Quantum Chromodynamics (QCD). 
However, due to the confinement mechanism in QCD, in experiments
the only observables are hadrons which appear
as complex bound systems of quarks and gluons.
A rigorous analytical solution of how to relate 
quarks and gluons in QCD to the hadronic world
is still missing.
We have therefore
developed effective descriptions 
that allow us to derive non-trivial statements about
hadronic processes from QCD and vice versa.
It should be obvious that the notion of quark or gluon
structure may depend on the physical context. Therefore one aim is to
find process-independent concepts which allow a
comparison of different approaches.

The simplest example, which essentially reflects
our intuitive picture of hadrons, is to
assign a particular quark content to
each hadron (say, proton$\sim uud$).
It enables us to
classify the hadrons in the particle data book\cite{PDG98}
according to their flavor quantum numbers.
This concept is the basis of the numerous versions of 
constituent quark models which are used 
as an effective low-energy approximation to QCD.
These models are sufficient to obtain a reasonable
explanation of global features like the mass hierarchy in the
hadronic spectrum, relations among scattering amplitudes
and decay widths etc.

A more elaborated approach 
to the low-energy sector of hadronic physics is
based on the symmetry properties of QCD.
The (approximate)
chiral symmetry of light quarks plays a special role.
It appears to be spontaneously broken, and the
light pseudoscalar mesons can be identified as (almost) Goldstone bosons. 
These properties can be used for a systematic
construction of an effective
theory, {\em chiral perturbation theory}\/ ($\chi$PT),\cite{GaLe85}
where the non-perturbative information is encoded in the
coefficients of operators in an effective Lagrangian. By comparison
with experimental data for hadronic low-energy reactions one
extracts the values of these coefficients which provide a 
well-defined measure of the hadronic structure.

At higher energies one resolves the partonic degrees
of freedom inside the hadron.
The structure functions, measured in deep inelastic scattering 
experiments, can be successfully
described in terms of momentum distributions of
valence-, sea-quarks and gluons. These parton distributions
are process-independent. They describe the probability 
to find a certain parton with a specific momentum fraction
inside the hadron.
They depend on the resolution-scale,
but this dependence is controlled
solely in terms of perturbative QCD.
The enormous amount of data for nucleon structure functions
over a wide range of energy and momentum transfer
simultaneously allows us to extract the parton 
distributions and to test QCD 
to a rather high accuracy.\cite{GRX,MRST,CTEQ} 
From inclusive pion-nucleon scattering, taking the nucleon structure
functions as input, one may also extract the distributions
of quarks and gluons in the pion, but with much poorer 
accuracy.\cite{GRS}  
For all other hadrons experimental results
of similar quality are not available.

Furthermore, light hadrons provide a
rich phenomenology of
exclusive reactions with large momentum transfer, e.g.\
in decays of heavy particles or electroweak form factors.
These reactions are expected to be dominated by
a finite number of particular parton Fock states
of the hadrons being involved. 
The momentum distribution of the partons in
each Fock state defines the so-called distribution amplitudes
which behave in a similar way as
the parton distributions in inclusive reactions, i.e.\
they evolve with the resolution 
scale\cite{Efremov:1980qk,BrLe80,Chernyak:1984ej} 
but are otherwise process-independent.
By comparing different exclusive reactions,
one is again in the position
to obtain important
information on the hadron structure and to test our
understanding of QCD at the same time. 
In principle, the parton distributions
measured in inclusive
reactions can be reconstructed from
the light-cone wave functions
(from which the distribution amplitudes are obtained by
integration over transverse momenta)
of each individual Fock state.\cite{BrLe80}  
In practice, however, only a few Fock states are under control,
and the connection to the parton distributions can only be
exploited at large momentum fraction $x$ 
(see e.g.\ Refs.\cite{Bolz:1996sw,JaKrRa96,Diehl:1998kh}).

The main part of this review deals with a subject where already
the simple question concerning the quark flavor decomposition turns
out to be rather non-trivial, namely for pseudoscalar mesons
$\pi^0,\eta,\eta'\ldots$ with
vanishing isospin, strange\-ness etc. 
In these cases the strong interaction induces transitions
between quarks of
different flavors ($u\bar u$, $d\bar d$, $s\bar s$)
or gluons ($gg$  \ldots).
Both, in the constituent quark model at low energies
and in the parton picture at high energies,
the physical mesons  
appear as complicated mixtures of 
different quark-antiquark combinations.
The mixing phenomenon is strongly connected with the $U(1)_A$ 
anomaly\cite{ABJ} of QCD. I will show below how 
this fact can be used to define
and to quantify reasonable and process-independent
mixing parameters.
The investigation of mixing phenomena in the pseudoscalar meson sector
has a long tradition. Of particular interest is the $\eta$-$\eta'$ 
system: Since the light pseudoscalar mesons approximately fall into 
multiplets of $SU(3)_F$ flavor symmetry, mainly disturbed 
by the mass of the strange quark, one conventionally regards
$\eta$ and $\eta'$ as linear combinations of octet and singlet
basis states, parametrized by a mixing angle $\theta_P$. A
determination of its value should be achieved -- in principle --
 from the diagonalization
of suitably chosen mass matrices (motivated by e.g.\ $\chi$PT)
or from phenomenology. 

Already in 1964 Schwinger\cite{Schwinger:1964} used mass formulas
to estimate the mass of the $\eta'$ (actually the original prediction,
$1600$~MeV, is 50\% too high due to the neglect of some 
$SU(3)_F$ corrections).
An early estimate of the mixing angle has been given, for instance,
by Isgur\cite{Isgur}
 ($\theta_P\simeq-10^\circ$). Kramer et al.\cite{Kazi:1976tu}
as well as  Fritzsch and Jackson\cite{Fritzsch} have emphasized
the importance of $SU(3)_F$--breaking effects. Their ansatz for the
mass matrix is formulated in the quark-flavor basis ($u\bar u$, $d\bar d$,
etc.) and takes into account mixing with the $\eta_c$ state.
Their mixing angle (corresponding to $\theta_P\simeq-11^\circ$) has also
been tested against experimental data.
In 1981 Diakonov and Eides\cite{Diakonov:1981nv} presented a 
quantitative analysis of anomalous Ward identities and quoted a
value of $\theta_P\simeq-9^\circ$.
In subsequent publications\cite{phen1,Gilman:1987ax}, however, 
very different values for the
mixing angles have been found: The incorporation of loop corrections
in $\chi$PT to meson masses and decay constants and a comparison
with phenomenological data for various decay modes seem to favor
values of $\theta_P$ around $-20^\circ$. 
Ball, Fr{\`e}re and Tytgat\cite{BaFrTy95} concentrated on processes
where the anomalous gluonic content
of $\eta$ and $\eta'$, which can be related to the decay constants,
is probed. They also prefer values for $\theta_P$ near to $-20^\circ$.
On the other hand a quark model calculation of 
Schechter et al.\cite{Schechter:1993iz} more or less recovers the
mixing scenario of Diakonov and Eides with $\theta_P\simeq -13^\circ$.
Finally, Bramon et al.\cite{Bramon97} considered only such processes
where the light ($u\bar u$ or $d\bar d$) or strange ($s\bar s$)
component is probed and found a mixing angle $\theta_P\simeq -15^\circ$.

The situation concerning the actual value of $\theta_P$ thus seemed
to be rather dissatisfying with results ranging from $-10^\circ$ 
to $-20^\circ$. 
However, it turned out during the last two years, that a big part
of the discrepancies between different analyses can be solved by
relaxing and correcting some of the implicit assumptions
made therein. Considerations of Leutwyler and Kaiser\cite{Leutwyler97}
as well as  Kroll, Stech and 
myself\cite{FeKr97b,Feldmann:1998vh,Feldmann:1998sh} have shown
that the definition of mixing parameters requires some care.
In low-energy effective theories the decay constants
relate the physical states with the (bare) octet/singlet fields.
It turns out that this connection 
cannot be a simple rotation. Therefore, $\eta$-$\eta'$ mixing cannot
be adequately described by a single 
mixing angle $\theta_P$.
A new scheme, which can be strictly related to
the effective Lagrangian of $\chi$PT, is formulated 
on the basis of a general parametrization of 
the octet and singlet decay constants of $\eta$ and $\eta'$
mesons, respectively. 
This scheme can also be successfully applied
to hard exclusive reactions with $\eta$ or
$\eta'$ mesons where one is
sensitive to the light-cone wave functions \lq{}at the origin\rq{},
which are fixed by the decay constants, too.
The main achievement of the new mixing scheme
is that the properties of $\eta$ and $\eta'$ mesons in 
different phenomenological situations and at different energy scales
can be described in a consistent way.

The organization of this article is as follows:
In the following section I will shortly recall some
important properties of QCD under chiral symmetry transformations
and the consequences for the pseudoscalar meson spectrum, which
includes a summary of the $U(1)_A$ problem and its possible
solutions.
Section~3, which is the main part of this review, includes
a detailed discussion of $\eta$-$\eta'$ mixing: 
The chiral effective Lagrangian for
the pseudoscalar nonet, including the $\eta'$ meson, is presented,
and its parameters are related to
the octet/singlet decay constants of $\eta$ and $\eta'$ mesons.
In the quark-flavor basis, the consequent application of the
Okubo-Zweig-Iizuka--rule (OZI-rule) is shown to lead to a scheme with
a single mixing angle $\phi$.
Phenomenological estimates of mixing parameters from the literature
are compared in both, the octet-singlet and quark-flavor basis.
This is followed by a discussion of 
the two-photon decay widths
and the $VP\gamma$ coupling constants.
The $\eta\gamma$ and $\eta'\gamma$
transition form factors are analyzed within the
hard-scattering approach. For this
purpose the light-cone wave functions of 
$\eta$ and $\eta'$ mesons are introduced and compared to the pion one.
For the latter also the connection to the parton distribution functions
will be illuminated.
Interesting relations among the mixing parameters are obtained
by considering the matrix elements of pseudoscalar quark currents and
of the topological charge density.
I will further present improved versions of various
mass formulas for the $\eta$-$\eta'$ system.
Finally, the consequences of the new mixing scheme for
the pseudoscalar coupling constants of the nucleon are investigated.
Section~4 is devoted to mixing with $\pi^0$ or
$\eta_c$ mesons and includes a comment on mixing with
glueballs or excited quarkonium states.
A summary is presented in Section~5.

\section{Chiral symmetry of light quarks}
As a starting point let me recall some
important global symmetries of the strong interactions and consider
the QCD Lagrangian
\beq
  {\mathcal L}_{\rm QCD} &=&  
  \sum_{i=u,d,s,\ldots} \bar q_i \, (i\Slash D - m_i) \, q_i
 - \frac12 \, {\rm tr}[ G^{\mu\nu} G_{\mu\nu}] 
  -  \theta \, \omega \nonumber\\[0.2em]
&& \qquad \quad
  + \mbox{FP-ghost} + \mbox{gauge-fixing} \ .
\label{eq:QCD}
\eeq
Here $G^{\mu\nu}=\partial^\mu A^\nu - \partial^\nu A^\mu 
- i \, g \, [A^\mu,A^\nu]$ 
is the gluonic field strength tensor with 
$A_\mu = A_\mu^a \lambda^a/2$ denoting the color gauge fields.
Furthermore, $q_i$ denotes
quark fields of a specific flavor with mass $m_i$, and
$i \Slash D = i \slash\partial - g \, \slash A$ is the covariant
in QCD.
Faddeev-Popov ghosts and gauge-fixing terms will be unimportant
for the further considerations. A summation over color indices
is to be understood.
I have also included the $\theta$-term which reflects the non-trivial
features connected with the axial $U(1)_A$ anomaly to be discussed
below. Here $\omega(x)$ is the topological charge
density which can be written as the divergence  of 
a gauge-variant current $K^\mu$ where
\beq
  K_\mu &=& \frac{\alpha_s}{4\pi} \,
	\epsilon_{\mu\nu\rho\sigma} \,
\sum_{a,b,c} A_a^\nu \left( \partial^\rho A_a^\sigma +
	\frac{g}{3} \, f^{abc} \, A_b^\rho \, A_c^\sigma \right) \ ,
\nonumber
\\[0.2em]
\omega &=& \partial_\mu K^\mu = \frac{\alpha_s}{8\pi} G \tilde G \ .
\label{eq:Kmu}
\eeq
The term $G \tilde G$ denotes the product of the gluon field strength
and its dual, and $f^{abc}$ are the anti-symmetric
structure constants of 
$SU(3)$, $[\lambda^a,\lambda^b]=2 i\, f^{abc} \,\lambda^c$. 
The different topological sectors of QCD are classified 
by the Pontryagin-Index which is given by the topological charge     
\beq
\int d^4x \, \omega(x) = \int d\sigma_\mu \, K^\mu = n \in Z \ .
\label{Pontryagin}
\eeq

It is well known that the pseudoscalar mesons $\pi,K,\eta$
built from the light flavors $u,d,s$ play a special role in
strong interaction physics. This is connected with the
behavior of the light quark fields under global chiral transformations
\beq
 q_L = \frac{1-\gamma_5}{2} \, q &\to & L \, q_L  
\ , \cr
 q_R = \frac{1+\gamma_5}{2} \, q & \to & R \, q_R
\ .
\eeq
Here $q=(u,d,s)^T$, and $L \in SU(3)_L$ and $R \in SU(3)_R$
denote unitary $3\times 3$ matrices acting on left- and
right-handed projections of the light quark fields,
$L,R = \exp[i \, \epsilon^a_{L,R} \, \lambda^a]$,
where $\lambda^a$ are the usual Gell-Mann matrices ($a=1..8$).
Obviously, in the limit $m_{u,d,s} \to 0$ the Lagrangian~(\ref{eq:QCD})
is invariant under chiral $SU(3)_L \times SU(3)_R$ transformations.
The corresponding Noether currents,
\beq
  J_{\mu \, L}^a &=& \bar q \, \frac{\lambda^a}{\sqrt2} \, \gamma_\mu
	\, \frac{1-\gamma_5}{2} \, q \ , \cr
  J_{\mu \, R}^a &=& \bar q \, \frac{\lambda^a}{\sqrt2} \, \gamma_\mu
	\, \frac{1+\gamma_5}{2} \, q \ ,
\label{eq:currents}
\eeq
are approximately conserved ($a=1\ldots 8$)
\beq
\partial^\mu (J_{\mu \, R}^a \pm J_{\mu \, L}^a) &=&
\bar q \, \left[ \frac{\lambda^a}{\sqrt2} , 
	\hat m \right]_{\mp}
		i \gamma_5 \, q \ . \qquad
(\hat m = {\rm diag}[m_u,m_d,m_s])
\label{eq:pcac8}
\eeq
The octet of vector currents
$J_\mu^a = J_{\mu\, R}^a + J_{\mu \, L}^a$ in Eq.~(\ref{eq:pcac8}) 
reflects the (approximate) flavor symmetry, which is observed
in the physical spectrum.
On the other hand,
the conservation laws related to the octet of axial-vector
currents $J_{\mu  5}^a =
J_{\mu\, R}^a - J_{\mu \, L}^a$ in Eq.~(\ref{eq:pcac8})
cannot be observed in the hadronic world. The chiral
symmetry of the Lagrangian appears to be spontaneously broken,
$SU(3)_L\times SU(3)_R \to SU(3)_V$, 
and $\pi,K,\eta$ appear in a natural way
as an octet of 
(almost) Goldstone particles which become massless in the 
limit $m_{u,d,s} \to 0$. This is the starting point
for the systematic construction of an effective theory for the low-energy
hadronic physics, see Section~3.1.

\subsection{$U(1)_A$ anomaly}
The QCD Lagrangian (\ref{eq:QCD}) has two additional
global symmetries: The $U(1)_V$ symmetry,
$q \to \exp[i \, \epsilon_V ] q$, corresponds to the
conservation of baryon number.
The $U(1)_A$ symmetry,
$q \to \exp[i \, \epsilon_A \gamma_5 ] q$ gives rise neither to
a conserved quantum number nor to a ninth Goldstone boson in
the mesonic spectrum: the $\eta'$ meson is too heavy
to be identified as the Goldstone particle of a spontaneously
broken $U(1)_A$ symmetry.
The solution of this $U(1)_A$ puzzle is connected with
the emergence of an anomaly contributing to
the divergence of the singlet axial
vector current, and Eq.~(\ref{eq:pcac8}) is to be extended,
\beq
  \partial^\mu J_{\mu5}^{a} &=& \bar q \,
  \left\{\frac{\lambda^a}{\sqrt2}, \hat m \right\} \, i\gamma_5 \, q
 + \delta^{a0} \,2 \sqrt 3 \,  \omega
\label{eq:anomaly} \ . \qquad (a = 0 \ldots 8)
\eeq
My normalization convention is ${\rm tr}[\lambda^a \lambda^b]
= 2 \, \delta^{ab}$ for $a=0\ldots 8$, i.e.\ $\lambda^0=\sqrt{2/3}
\, {\mathbf 1}$.
In Eq.~(\ref{eq:anomaly}) 
the topological charge density $\omega$
defined in Eq.~(\ref{eq:Kmu})
leads to a non-conservation of the axial-vector current
in the flavor-singlet sector even in the limit $\hat m \to 0$.

Anomalies in quantum theories have been studied
in several ways, following the pioneering work of Adler, Bell and
Jackiw,\cite{ABJ}
In perturbation theory the $U(1)_A$ anomaly
arises when calculating the quark triangle diagram with the
axial-vector current and two gauge fields since it turns out 
to be impossible to
define a regularization prescription which preserves both, gauge
invariance and chiral invariance. (In QCD 
there are no gauge fields coupled to the axial-vector current;
thus the anomaly causes no theoretical inconsistencies here.)
An elegant derivation of Eq.~(\ref{eq:anomaly}) 
can be found in the article of Fujikawa.\cite{Fujikawa:1979ay}
The anomalous term stems from a non-trivial Jacobian which is
related to the transformation of the path integral measure of
the fermion fields
\beq
&&
\Psi(x) \to \exp[i\epsilon_A(x)\gamma_5] \, \Psi(x) \ ,\nonumber\\[0.1em]
&&
\bar\Psi(x) \to \bar\Psi(x) \, \exp[i\epsilon_A(x)\gamma_5] \nonumber\\[0.2em]
&\Rightarrow&
{\mathcal D}\bar\Psi \, 
{\mathcal D}\Psi \,  \to 
{\mathcal D}\bar\Psi \, 
{\mathcal D}\Psi \,
\exp\left[ 2 \, i \int dx \, \epsilon_A(x) \, \omega(x)\right] \ .
\label{eq:path}
\eeq

It is also interesting to see how the $U(1)_A$ anomaly arises in
lattice-QCD. Here, the discretization of space-time on a
four-dimensional lattice
can be considered as a particular regularization scheme.
The lattice action is constructed in such a way that it reproduces
the Lagrangian~(\ref{eq:QCD}) with the lattice spacing approaching
zero (continuum limit).
It has been shown that the 
anomalous contribution to the Ward identity (\ref{eq:anomaly})
is related to an {\em irrelevant}\/
operator in the lattice action.\cite{Karsten:1981wd} 
The operator itself naively vanishes in
the continuum limit, but its contribution to the
anomalous Ward identity does not.
It has been shown that the correct continuum result  (\ref{eq:anomaly})
is reproduced for any choice of lattice action, as long
as very general conditions are fulfilled.\cite{Rothe:1998ba}

Since the flavor-singlet axial-vector current is not conserved 
in the presence of the $U(1)_A$ anomaly the
$\eta'$ mass does not have to vanish in the limit $\hat m \to 0$.
However, in order to understand the experimentally observed mass
splitting between octet and singlet pseudoscalar mesons, Eq.~(\ref{eq:anomaly})
alone is not sufficient. In addition 
one needs a non-vanishing matrix element of the
topological charge density sandwiched between the $\eta'$ state and the
vacuum  
\beq
\langle 0 | \omega | \eta'\rangle &\neq& 0 \ .
\label{finite}
\eeq
Since $\omega$ is a total divergence, see Eq.~(\ref{eq:Kmu}),
the l.h.s.\ of Eq.~(\ref{finite})
vanishes to any finite order in perturbation
theory, i.e.\
the $U(1)_A$ problem cannot 
be solved by simply considering quark-antiquark annihilation into
(perturbative) gluons. The solution clearly lies in the
non-perturbative sector of QCD and is inevitably connected to
non-trivial topological features of the theory. 
't~Hooft\cite{tHooft} suggested
instantons as a possible solution to Eq.~(\ref{finite}).
Kogut/Susskind\cite{Kogut:1974ab} have argued that the
ninth Goldstone field is prevented from being realized in the
physical spectrum by the same mechanism that confines colored objects.
An alternative approach has been initiated by Witten\cite{Witten}
who proposed to consider QCD from the large-$N_C$ perspective,
where $N_C$ is the number of colors.
It turns out that in order to obtain a consistent picture
for the $\theta$-dependence of the pure Yang-Mills theory in
the formal limit $N_C \to \infty$,
the $\eta'$ mass squared should behave as ${\mathcal O}(1/N_C)$.
Veneziano\cite{Veneziano:1979ec} has found a
realization of Wittens general $1/N_C$ counting rules by
introducing a ghost state into the theory (the notion of ghost
states in this context has also been used by Kogut and Susskind;
it is also close to Weinberg's approach\cite{Weinberg:1975ui} that involves
negative metric Goldstone fields).
The ghost corresponds to an unphysical massless pole in the correlation
function $\langle K^\mu K^\nu \rangle$ which generates a
non-vanishing  topological susceptibility ({\em mean
square winding number per unit volume}\/)
\beq
  \tau_0 = \int d^4x \, 
  \langle 0 | T \left[ \omega(x) \, \omega(0)\right] | 0\rangle &\neq& 0
\nonumber \\[0.2em]
\Leftrightarrow \quad 
   q_\mu q_\nu \langle K^\mu K^\nu \rangle_{q \to 0} &\neq& 0   
\label{tau0}
\eeq
which is necessary to fulfill Eq.~(\ref{finite}), see 
also Eq.~(\ref{Leff0}) below. 
The ghost pole may be viewed as the result of an
infinite number of Feynman graphs contributing to
$\langle K_\mu K_\nu \rangle$, but does not correspond to an
observable glueball state, since the currents $K^\mu$ 
are gauge-variant. 
Diakonov and Eides\cite{Diakonov:1981nv} have discussed the
phenomenological consequences of the Witten/Veneziano ansatz by
considering the $\eta'$ meson as a mixture of the Veneziano
ghost and a flavor singlet would-be Goldstone boson and 
investigating the anomalous Ward identies.
They found that the gapless excitation given by the Veneziano
ghost is a consequence of the periodicity of the QCD potential
w.r.t.\ a certain generalized coordinate which in the gauge
$A_0=0$ can be chosen as
\beq
  X(t) &=& \int d^3 {\mathbf x} \, K_0(t,{\mathbf x}) \ .
\eeq
Under gauge transformations it transforms as $X \to X + n$,
where $n$ is the topological charge (\ref{Pontryagin})
of the transformation, while the potential remains unchanged.
The quasi-momentum related to $X$ is just the variable $\theta$,
see Eq.~(\ref{eq:QCD}), and
the connection of the Veneziano ghost to non-trivial topology,
as induced by e.g.\ instantons or other
finite-action field configurations, is obvious.
The $U(1)_A$ problem has also been studied
in lattice-QCD\cite{Fukugita:1995iw,Venkataraman:1997xi}, and both,
a non-vanishing $\eta'$ mass and a significant correlation with the topological
susceptibility and fermionic zero modes, have been observed.

Finally, 
the $U(1)_A$ problem has been investigated by considering
low-energy models of the strong interaction. In the framework
of the global color model Frank and Meissner\cite{Frank:1997ck}
follow the work of Kogut/Susskind\cite{Kogut:1974ab} and
require a certain non-trivial infrared behavior of an effective
(i.e.\ non-perturbative) gluon propagator which leads to
quark confinement as well as to a non-vanishing $\eta'$ mass. 
A low-energy  expansion of the effective action following from their ansatz
has been shown to reproduce the general results of Witten/Veneziano.
Dmitrasinovic\cite{Dmitrasinovic:1997te} investigated different
effective $U(1)_A$--breaking quark interactions as a low-energy
approximation of the t'Hooft or Veneziano/Witten mechanism,
which can be used to generate a non-vanishing $\eta'$ mass.

In the ideal world with three massless light quarks and three 
infinitely heavy quarks the $\eta'$ meson is a pure flavor singlet.
However, in the real world
the flavor symmetry is not perfect, and the neutral mesons 
mix among each other.
In the isospin limit ($m_u=m_d$) which is a very good
approximation to the real world, the $\pi^0$ is still a pure iso-triplet.
Without the $U(1)_A$ anomaly the
two iso-singlet mass eigenstates in the pseudoscalar sector would
consist of $u\bar u + d\bar d$ and $s\bar s$, respectively. 
The $U(1)_A$ anomaly mixes these {\em ideally mixed}\/ states towards
nearly flavor octet or singlet combinations which are to be identified
with the physical $\eta$ and $\eta'$ mesons, respectively.
{}From Eq.~(\ref{eq:path}) we see, that the anomalous term in 
Eq.~(\ref{eq:anomaly}) is independent of the quark masses.
Consequently, the $U(1)_A$ anomaly also induces mixing with heavier
pseudoscalar mesons ($\eta_c,\eta_b$) which is however 
less important since the non-anomalous terms in Eq.~(\ref{eq:anomaly})
dominate in case of heavy quark masses. Taking into account the mass difference
of up- and down-quarks as a source of
isospin-violation, also the $\pi_0$
receives a small iso-singlet admixture.
The quantification of the mixing parameters
is one of the main subjects of this review and will be discussed 
in Sections~3 and 4.

\section{Mixing in the $\eta$-$\eta'$ system}
In order to quantify the mixing in the $\eta$-$\eta'$ system,
one has to define appropriate mixing parameters which can be
related to physical observables.
One approach is based on chiral perturbation theory which 
traditionally leads to a description of $\eta$-$\eta'$ mixing
in terms of octet-singlet parameters.
Another useful concept to obtain well-defined quantities is 
to follow e.g.\ the work of Diakonov/Eides\cite{Diakonov:1981nv}
and to consider the operators appearing in  
the anomaly equation~(\ref{eq:anomaly})
sandwiched between the vacuum and physical meson states.
Making consequent use of the OZI-rule, one is led to the
quark-flavor basis where these matrix elements
are expressed in terms of a single mixing angle $\phi$
which can be determined from theory or phenomenology.\cite{Feldmann:1998vh}

For the sake of clarity, I keep -- if not otherwise stated --
isospin symmetry to be
exact ($m_u=m_d\ll m_s$) and neglect the contributions of heavy quarks
($m_Q \to \infty$). In this limit,
the $\pi^0$ meson is a pure iso-triplet,
and the $\eta_c$ meson a pure $c\bar c$ state.
I will discuss mixing phenomena with $\pi^0$ and $\eta_c$ in
sections 4.1 and 4.2, respectively.

\subsection{Decay constants in the octet-singlet basis and
$\chi$PT}
The low-energy physics of
light pseudoscalar mesons can be successfully described by
an effective Lagrangian which reflects a systematic expansion
in powers of small momenta and masses of the (almost) Goldstone
bosons $\pi,\eta,K$. This is the basis of
chiral perturbation theory\cite{GaLe85} ($\chi$PT).
Counter terms, arising from renormalization, can be absorbed into
higher order coefficients of the effective Lagrangian. 
In this sense
$\chi$PT is renormalizable order by order.
Since the $\eta'$ meson is not a Goldstone boson and its mass
is not small, it is usually not included as an explicit degree
of freedom in the Lagrangian.
Recently, Leutwyler and Kaiser\cite{Leutwyler97} have discussed
how to include the $\eta'$ into the framework of $\chi$PT in a
consistent way, and I shall briefly present their most important
results.

Starting point is the observation that in the formal limit
$N_C \to \infty$
the anomalous term in Eq.~(\ref{eq:anomaly}) vanishes, and
the $\eta'$ formally arises as a ninth Goldstone boson of 
$U(3)_L\times U(3)_R \to U(3)_V$.
One can therefore extend the counting rules for the
construction of the effective Lagrangian as follows:
$1/N_C = O(\delta)$, $p^2=O(\delta)$, $m_q=O(\delta)$,
where $\delta$ is the small expansion parameter.
In the standard framework the octet and singlet pseudoscalar mesons 
are 
parametrized in a non-linear way by a field
$U(x)=\exp[i\varphi(x)]$ where $\varphi(x)$ is a short-hand
for $\varphi^a(x) \lambda^a$ ($a=0..8$) and $\varphi^a$ are
the bare fields of the pseudoscalar nonet.
Considering only the leading order of the expansion in $\delta$,
one obtains 
the following three terms\cite{Leutwyler97,Moussallam:1995xp}
\beq
  {\mathcal L}^{(0)}&=& \frac{F^2}{4} \,
	{\rm tr}(\partial_\mu U^\dagger \, \partial^\mu U)
	+\frac{F^2}{4} \, 
        {\rm tr}(\chi^\dagger  U + U^\dagger  \chi)
        -6 \tau_0 \, \frac{1}{2} \, (\varphi^0)^2 \ .
\label{Leff0}
\eeq
To this order the parameter $F={\mathcal O}(\sqrt{N_C})$ 
is identified with the universal pseudoscalar
decay constant $F=F_a\simeq F_\pi=93$~MeV.
This follows immediately if one 
introduces source terms for the axial-vector currents~(\ref{eq:currents})
in a chirally-invariant way.
Finite meson masses 
are induced by the term proportional to
$\chi=2 B \hat m$. Here $\hat m$ is the matrix of (current)
quark masses and $B={\mathcal O}(1)$ a dimensional parameter which
reflects the non-vanishing quark condensate.
Finally, $6 \tau_0/F^2={\mathcal O}(N_C^{-1})$ is the contribution
of the $U(1)_A$ anomaly to the singlet mass.\footnote{For
convenience I
have changed the normalization of the singlet field
compared to Leutwyler et al.\cite{Leutwyler97}
$\psi \to \sqrt6 \, \varphi^0$} {}\
Here $\tau_0={\mathcal O}(1)$ is the topological susceptibility defined
in Eq.~(\ref{tau0}).
In order to account for the experimental fact that the 
decay constants of the light pseudoscalar mesons differ
substantially ($F_K =1.22 \, F_\pi$) one can include
the relevant terms of the next order in the
effective Lagrangian\cite{Leutwyler97}
\beq
  {\mathcal L}^{(1)} &=&
	L_5 \, {\rm tr} \left(
	\partial_\mu U^\dagger \, \partial^\mu U \,
	(\chi^\dagger  U + U^\dagger  \chi)\right)
      + L_8 \, {\rm tr} \left(
	(\chi^\dagger  U)^2 + h.c. \right) + \cr
&&
       \frac{F^2}{2} \, \Lambda_1 \,
	 \partial_\mu \varphi^0 \, \partial^\mu \varphi^0
      + \frac{F^2}{2 \sqrt6} \, \Lambda_2 \, i \varphi^0 
	\, {\rm tr} (\chi^\dagger  U - U^\dagger  \chi) 
      + {\mathcal L}_{WZW} + \ldots
\label{Leff1}
\eeq
Here $L_5={\mathcal O}(N_C)$ parametrizes corrections to the decay constants,
$L_8={\mathcal O}(N_C)$ the ones for the meson masses. $\Lambda_1$ and 
$\Lambda_2$ only influence the singlet sector and can
be attributed to OZI-rule violating contributions, i.e.\ they
are of order $1/N_C$. A possible 
source for these terms can be provided by e.g.\
glueball states which are not included in the effective Lagrangian,
see Section~4.3.
I emphasize that
the parameters $\Lambda_i$
are of different origin than 
the topological susceptibility $\tau_0$.
${\mathcal L}_{WZW}$, finally, denotes the
Wess-Zumino-Witten term which describes the
anomalous coupling to photons and will be discussed
in more detail below.
Taking e.g.\ the pion and kaon masses and decay constants
as input, one can -- to this order --
express the parameters $F,\ L_5$ and $B,\ L_8$ in terms
of physical quantities only.

Leutwyler and Kaiser have also discussed the next order,
${\mathcal L}^{(2)}$, see the  second paper in Ref.\cite{Leutwyler97}.
At this level, also loop-corrections calculated from the
Lagrangian ${\mathcal L}^{(0)}$ have to be taken into
account. They give
rise to the typical {\em chiral logs}\/,
$$
  \frac{M_P^2}{32 \, \pi^2 \, F^2} \, \ln \frac{M_P^2}{\mu_\chi^2}
\ ,
$$
which  contribute
at the order ${\mathcal O}(\delta^2)$.
For the understanding
 of the main features of $\eta$-$\eta'$ mixing the
higher order effects of chiral logs and ${\mathcal L}^{(2)}$
are not important.
In the following I will therefore concentrate
on the contributions from ${\mathcal L}^{(0)}$ and 
${\mathcal L}^{(1)}$ and discuss the important consequences
for the mixing parameters in the octet-singlet basis,
revealed by Leutwyler and Kaiser.

The decay constants in the $\eta$-$\eta'$ system are defined 
as matrix elements of axial-vector currents~(\ref{eq:currents}) 
\beq
\langle 0 | J_{\mu5}^a(0) | P(p)\rangle &=& i \, f_P^a \, p_\mu
\label{eq:fdef}
\eeq
where I have changed
the normalization convention to $f_\pi=
\sqrt2 F_\pi = 131$~MeV, and the currents are defined as in
Eq.~(\ref{eq:currents}).
Each of the two mesons $P=\eta,\eta'$ has both, octet and singlet
components, $a=8,0$. Consequently, Eq.~(\ref{eq:fdef}) defines
{\em four independent}\/ decay constants, $f_P^a$.
For a given current each pair of decay constants can be
used to define a separate mixing angle
\beq
&& \frac{f_\eta^8}{f_{\eta'}^8} = \cot\theta_8
\ , \qquad  \frac{f_\eta^0}{f_{\eta'}^0} = - \tan\theta_0 \ .
\label{eq:81angledef}
\eeq
Here I followed the convention of
Ref.\cite{Leutwyler97} 
and used a parametrization in terms of two basic
decay constants $f_8,f_0$ and two angles $\theta_8,\theta_0$
\beq
  \left\{ f_P^a\right\} &=&
\left(\matrix{ f_{\eta}^8 & f_{\eta}^0 \vspace{0.2em} \cr  
               f_{\eta'}^8 & f_{\eta'}^0}
\right)
= \left(\matrix{ f_8 \, \cos\theta_8 & - f_0 \, \sin\theta_0\vspace{0.2em} \cr
           f_8 \, \sin\theta_8 & \phantom{-} f_0 \, \cos\theta_0 }
\right) \ .
\label{eq:para81}
\eeq
The angles are chosen in such a way that $\theta_8=\theta_0=0$
corresponds to the $SU(3)_F$ symmetric world.

The matrix $\{f_P^a\}$ defined by Eq.~(\ref{eq:para81}) will play
a crucial role in the following discussion since it is exactly
the quantity that relates the physical fields $P=\eta,\eta'$
(which diagonalize the kinetic and mass terms in the effective
lagrangian) to
the bare octet or singlet fields 
$\varphi^a$ in the effective Lagrangian (\ref{Leff0})
and (\ref{Leff1})
\beq
  \varphi^a(x) &=& \sum_P (f^{-1})_P^a \, P(x)  \ .
\label{Pphiconnection}
\eeq
Note that Eq.~(\ref{Pphiconnection}) is
unique up to an unimportant overall
normalization which can be absorbed into the parameters of the
effective Lagrangian. Through the 
ansatz~(\ref{Pphiconnection}) it is guaranteed that the fields 
$\varphi^a$ have the proper behavior under
renormalization and $SU(3)_F$ transformations. 
Coupling the fields $\varphi^a$ to external octet and singlet
axial-vector currents
in an $SU(3)_F$-invariant way, one obtains the decay constants
from the matrix elements in Eq.~(\ref{eq:fdef}).
This leads to several important features
which shall be emphasized here: 
\begin{itemize}
\item One can define the matrix product 
\beq
 (f^T f)^{ab} = \sum_{P=\eta,\eta'} \, f_P^a \, f_P^b \ .  
\label{Fab}
\eeq
      The $88$- and $08$-elements 
      of this matrix are not effected by the parameters
      $\Lambda_i$ in the effective Lagrangian (\ref{Leff1})
      and can be expressed in terms of
      the parameters $F$ and $L_5$ only,
      i.e.\ to this order they are just
      fixed in terms of $f_\pi$ and $f_K$.
      This leads to
      the following relations among the decay constants
      and mixing parameters\cite{Leutwyler97}
\beq
 \sum_P \, f_P^8 \, f_P^8 
= f_8^2
       &=& \frac{4 f_K^2 - f_\pi^2}{3} \ ,
\label{eq:f8rel}
\\[0.1em]
\sum_P \, f_P^8 \, f_P^0   
= f_8 \, f_0 \, \sin(\theta_8-\theta_0) 
&=& - \frac{2 \sqrt2}{3} \, (f_K^2-f_\pi^2)
\ .
\label{eq:t8t1rel}
\eeq
Relation~(\ref{eq:f8rel}) follows from standard
$\chi$PT for the members of the pseudoscalar octet alone and
has been frequently used. 
The relation~(\ref{eq:t8t1rel}) stems from
the inclusion of the $\eta'$ meson into the chiral Lagrangian.
At first glance,
it is surprising
since it tells us that the matrix product of the decay constants
in the octet-singlet basis is not diagonal,
i.e.\ $\theta_8\neq \theta_0$. 
The decay constants of the charged pions and kaons, appearing
on the r.h.s.\ are well known from their leptonic decay\cite{PDG98}\
($f_\pi = 130.7$~MeV, $f_K = 1.22 \, f_\pi$). Thus the
strength of the flavor symmetry breaking
effects entering Eqs.~(\ref{eq:f8rel}) and 
(\ref{eq:t8t1rel}) is expected to be of the order of 20\%.
On the other hand, the mixing angles $\theta_8$ and $\theta_0$
themselves are small quantities, too, 
as long as the size of the anomalous contribution
to the singlet mass ($6\tau_0/F^2$) is large
compared to the effect of $SU(3)_F$ breaking ($M_K^2-M_\pi^2$).
One thus has
\beq
 \left|\frac{\theta_8-\theta_0}{\theta_8 + \theta_0}\right| 
   & \ll\hskip-1.05em/ & 1 \ ,
\eeq
and the difference between $\theta_8$ and $\theta_0$ 
should not be neglected.\footnote{It is amusing to note 
that already 25 years ago Langacker and
Pagels\cite{LaPa74}
have pointed out the inconsistencies resulting from using
only one mixing angle in Eqs.~(\ref{eq:para81},\ref{Pphiconnection}).
The strong interactions require a renormalization of the 
{\em bare}\/ octet and singlet fields that is more complicated than
a simple rotation.}

The singlet decay constants $f_0$ has an additional
contribution from the OZI-rule violating term in Eq.~(\ref{Leff1})
which is proportional to the parameter $\Lambda_1$
\beq
 \sum_P \, f_P^0 \, f_P^0 =  f_0^2 &=& 
  \frac{2 f_K^2 + f_\pi^2}{3} + f_\pi^2 \, \Lambda_1 \ .
\label{eq:f0rel}
\eeq		
The value of $\Lambda_1$ has to be determined from phenomenology.
Furthermore, the singlet decay constants $f_P^0$ are
renormalization-scale dependent,\cite{Leutwyler97} 
\beq
  \mu \, \frac{{\rm d}f_P^0}{{\rm d}\mu} &=& \gamma_A(\mu) \, f_P^0
\label{eq:f0scale} \ .
\eeq
The anomalous dimension $\gamma_A$ is of order $\alpha_s^2$
\beq
  \gamma_A(\mu) &=& - N_F \, \left( \frac{\alpha_s}{\pi}\right)^2 + 
  {\mathcal O}(\alpha_s^3) \ .
\eeq 
A comparison with Eq.~(\ref{eq:f0rel}) reveals that
the the behavior of $f_0$ under renormalization should be attributed to
the scale-dependence of the parameter $\Lambda_1 \to \Lambda_1(\mu)$. 
Numerically, the scaling of $f_P^0$, however, is only
a sub-leading effect.
Varying, for instance, the scale $\mu$ between $M_\eta$ and
$M_{\eta_c}$, the value of $f_P^0(\mu)$ changes by less than 10\%.
Note that the mixing angle $\theta_0$ is not scale-dependent since
it is defined as the ratio of two singlet decay constants, 
see Eq.~(\ref{eq:81angledef}).

\item In the past, in many 
      publications\cite{GaLe85,phen1,Gilman:1987ax,BaFrTy95,Burakovsky:1998vc}
      it has often been taken for granted 
      that the two angles $\theta_8$ and $\theta_0$ can be taken as equal,
      and can be identified with a universal
      mixing angle $\theta_P$ of the
      $\eta$-$\eta'$ system.
      From the above considerations it should be
      clear that neither of these assumptions is justified as soon
      as one includes flavor symmetry breaking effects in a
      systematic way (i.e.\ by taking into account 
	corrections to ${\mathcal L}^{(0)}$).

\item One can construct another matrix product
       \beq
        && (f\, f^T)_{P_1P_2}= \sum_{a=8,0} \, f_{P_1}^a \, f_{P_2}^a \cr
        &=&
        \left( \matrix{ f_8^2 \, \cos^2\theta_8 + f_0^2 \, \sin^2\theta_0
                       & f_8^2 \, \cos\theta_8 \sin\theta_8 -
                         f_0^2 \, \cos\theta_0 \sin\theta_0\vspace{0.2em} \cr
                          f_8^2 \, \cos\theta_8 \sin\theta_8 -
                         f_0^2 \, \cos\theta_0 \sin\theta_0 &
	f_8^2 \, \sin^2\theta_8 + f_0^2 \, \cos^2\theta_0 } \right)
      \nonumber\\[0.3em] &\neq& {\rm diag}[f_\eta^2, f_{\eta'}^2]
	\eeq
      which is defined in the basis of the {\em physical states}\/
      $P_1,P_2=\eta,\eta'$. Since it is non-diagonal,
      the $\eta$-$\eta'$ decay constants cannot be 
      adequately described by means of
      individual
      \lq{}decay constants $f_\eta$, $f_{\eta'}$\rq{} 
      if mixing is to be taken into account, 
      contrary to what is occasionally claimed in the literature.      
\end{itemize}


\subsection{Decay constants in the quark-flavor basis and OZI-rule}
The parametrization of
the decay constants can (and actually does) look much simpler in
another 
basis,
which is frequently
used,\cite{Isgur,Kazi:1976tu,Fritzsch,Diakonov:1981nv,Bramon97,Feldmann:1998vh,Feldmann:1998sh}
where the two independent axial-vector currents are taken as 
\beq
  J_{\mu5}^q &=& \phantom{-}\sqrt{\frac13} \, J_{\mu5}^8 +
                 \sqrt{\frac23} \, J_{\mu5}^0 
        = \frac{1}{\sqrt2} \left( \bar u \, \gamma_\mu \gamma_5 \, u
 + \bar d \, \gamma_\mu \gamma_5 \, d\right)\ , \nonumber \\[0.2em]
  J_{\mu5}^s &=& - \sqrt{\frac23} \, J_{\mu5}^8 +
                 \sqrt{\frac13} \, J_{\mu5}^0 
        = \bar s \, \gamma_\mu \gamma_5 \, s \ .
\eeq
In an analogous way
one defines new bare fields $\varphi^q$
and $\varphi^s$.
For obvious reasons, I call this basis the quark-flavor basis
for which I take indices $i,j=q,s$ to distinguish it from
the conventional octet-singlet basis with indices
$a,b=8,0$. In the quark-flavor basis the matrix $\chi$, which 
induces the explicit flavor symmetry breaking 
in the effective Lagrangian (\ref{Leff0}) and (\ref{Leff1}), is diagonal.
Without the $U(1)_A$ anomaly the physical states would thus be
close to the fields $\varphi^q$ and $\varphi^s$. Consider, for instance,
the analogous case in the vector meson sector where the $\omega$ and
$\phi$ meson are nearly pure $q\bar q$ and $s\bar s$ states, respectively.
The smallness of the $\phi$-$\omega$ mixing angle (about $3^\circ$)
is consistent with
the OZI-rule, i.e.\ amplitudes that involve quark-antiquark annihilation
into gluons are suppressed. The OZI-rule becomes rigorous in the
formal limit $N_C\to \infty$ and also for a vanishing strong
coupling constant at asymptotically large energies.
In the pseudoscalar sector, however, the $U(1)_A$
anomaly induces a significant mixing between the fields $\varphi^q$
and $\varphi^s$. As I have discussed in Section~2.1, the non-trivial
effect of the anomaly is connected with the topological properties
of the QCD vacuum and is not due to quark-antiquark annihilation.
We will see in the following how these obervations can be used
to obtain a powerful phenomenological scheme for the description
of $\eta$-$\eta'$ mixing. 

For this purpose let me first introduce an
 analogous parametrization as in Eq.~(\ref{eq:para81})
\beq
  \left\{ f_P^i\right\} &=&
\left(\matrix{ f_{\eta}^q & f_{\eta}^s \vspace{0.2em}\cr  f_{\eta'}^q & f_{\eta'}^s}
\right)
= \left(\matrix{ f_q \, \cos\phi_q & - f_s \, \sin\phi_s \vspace{0.2em}\cr
           f_q \, \sin\phi_q & \phantom{-} f_s \, \cos\phi_s }
\right) \ .
\label{eq:paraqsvar}
\eeq
From the effective Lagrangian~(\ref{Leff0}) and (\ref{Leff1}),
I obtain expressions for the basic parameters
$f_q$, $f_s$ and the difference between the two mixing angles
$\phi_q$ and $\phi_s$,
\beq
 \sum_P \, f_P^q \, f_P^q 
= f_q^2
       &=& f_\pi^2 + \frac{2}{3} \, f_\pi^2 \, \Lambda_1
\label{eq:fqrel} \ ,
\\[0.1em]
\sum_P \, f_P^q \, f_P^s   
= f_q \, f_s \, \sin(\phi_q-\phi_s) 
&=& \frac{\sqrt2}{3} \, f_\pi^2 \, \Lambda_1
\label{eq:phiqsrel} \ ,
\\[0.1em]
\sum_P \, f_P^s \, f_P^s =  f_s^2 &=& 
  2 f_K^2 - f_\pi^2 + \frac13 \, f_\pi^2 \, \Lambda_1 \ ,
\label{eq:fsrel} 
\eeq
which have to be compared with
relations~(\ref{eq:f8rel}),
(\ref{eq:t8t1rel}) and (\ref{eq:f0rel}).
The situation in the quark-flavor basis is different from
the one in the octet-singlet one: i) The difference between
the angles $\phi_q$ and $\phi_s$ is determined by an OZI-rule violating
contribution ($\Lambda_1\neq 0$) 
and not by $SU(3)_F$--breaking ($f_K\neq f_\pi$).
As already mentioned, the parameter $\Lambda_1$ has to
be estimated from phenomenology. Taking typical values for the
mixing parameters from the literature (see Tables~\ref{tab:fPa}
and \ref{tab:fPi} below),
one obtains the estimate  $|\phi_q-\phi_s| < 5^\circ$
 which 
translates into $|\Lambda_1| < 0.3$. 
ii) The values of $\phi_q$ and $\phi_s$ themselves are not
small quantities. In the $SU(3)_F$ symmetry limit they take
the {\em ideal}\/ value, $\arctan\sqrt2 \simeq 54.7^\circ$.
Phenomenological analyses give values around $40^\circ$
(see Table \ref{tab:fPi}).
We thus have
\beq
  \left|\frac{\phi_q-\phi_s}{\phi_q+\phi_s} \right| &\ll& 1 \ . 
\label{eq:phiest}
\eeq
Eq.~(\ref{eq:phiest}) can be taken as a justification to
treat the difference between the parameters $\phi_q$ and $\phi_s$
as a sub-leading correction. The advantage of this procedure
is obvious. One has to deal with only one mixing angle
$\phi\simeq \phi_q\simeq \phi_s$, 
now defined in the quark-flavor basis. In this basis the matrix
of decay constants is approximately diagonal, and we can write
\beq
  \left\{ f_P^i\right\} &=&
U(\phi) \, {\rm diag}[f_q,f_s] + {\mathcal O}(\Lambda_1) 
\label{eq:paraqs}
\eeq
where I have defined the rotation matrix 
\beq
\qquad U(\phi) = \left( \matrix{\cos\phi&-\sin\phi\cr \sin\phi & 
\phantom{-}\cos\phi}
	\right) \ .
\eeq
I stress that assuming an equation like~(\ref{eq:paraqs})
in one basis {\em necessarily}\/ leads to a more complicated 
parametrization like (\ref{eq:para81}) in another.

In the quark-flavor basis the assumption of a common mixing
angle is  directly related to the OZI-rule. 
The OZI-rule in this context means that
the $1/N_C$-suppressed parameters $\Lambda_i$ in the chiral
effective Lagrangian are dropped while e.g.\ the topological
susceptibility $\tau_0$ is kept. It is important to realize 
that the OZI-rule is a necessary ingredient for the
determination of {\em  process-independent}\/
mixing parameters.
Otherwise, we would encounter a problem, since
we had to introduce an extra unknown OZI-rule violating
parameter for each coupling or decay constant.
Turning the argument around this means that  {\em  process-independent}\/
mixing parameters can only be determined up to corrections of order
$1/N_C$.
 
The consequent application of the OZI-rule 
leads to the scheme which has been advocated for
in Ref.\cite{Feldmann:1998vh,Feldmann:1998sh}
I will refer to it as the FKS scheme in the following.
It is based on the following requirements
\begin{itemize}
\item All OZI-rule violating parameters $\Lambda_i$ of order $1/N_C$ in 
     the chiral effective Lagrangian are neglected. 
\item
     The parameters in the singlet channel ($f_0$ etc.)
     are not scale-dependent in the FKS scheme,
     since the additional renormalization
     effects, which are usually absorbed
     into the parameters $\Lambda_i$, also violate the OZI-rule.
\item 
     All other amplitudes that involve
     quark-antiquark annihilation but are not due to topological effects 
     are neglected (e.g.\ $\phi$-$\omega$ mixing, glueball admixtures
	in $\eta$ and $\eta'$, \ldots)
\end{itemize}      
For the phenomenological analysis of mixing parameters 
the FKS scheme provides the most useful concept. 
It can also be used as the starting point of a theoretical 
analysis\cite{Feldmann:1998vh}
of $\eta$-$\eta'$ mixing on the basis of the anomaly 
equation~(\ref{eq:anomaly}). In an earlier work, 
Diakonov/Eides\cite{Diakonov:1981nv} considered the anomalous
Ward identies by implicitely assuming the OZI-rule to apply in
the above sense which leads to completely analogous 
results as in the FKS scheme.

Of course, if precise data for e.g.\ the
decay constants $f_P^a$ or $f_P^i$ become available,
one will be able to
unambigously determine $\phi_q$ and $\phi_s$ (and therefore $\Lambda_1$)
from that one source.
This in turn could be taken as input in order to determine
the OZI-rule violating parameters in other processes.
In Ref.\cite{Leutwyler97} -- for instance --
the estimates $\Lambda_1-2 \Lambda_3 \simeq 0.25$
and $\Lambda_2 -\Lambda_3 \simeq 0.28$ at low renormalization
scales have been obtained from a 
phenomenological analysis. Varying $\Lambda_1$ in the interval
$(-0.3,0.3)$ this tranlates into $0 < \Lambda_2 < 0.3$ and 
$-0.28 < \Lambda_3 < 0.02$.

The connection between the mixing parameters 
in the FKS scheme and the mixing angles in the octet-singlet
basis reads\cite{Feldmann:1998vh} 
\beq
 \theta_8 &=&\phi - \arctan[\sqrt2 \, f_s/f_q] + {\mathcal O}(\Lambda_1) \ ,
\cr
 \theta_0 &=&\phi - \arctan[\sqrt2 \, f_q/f_s] + {\mathcal O}(\Lambda_1) \ . 
\label{eq:80rel}
\eeq
The deviation from the naive expectation $\theta_P=\phi+\theta_{\rm id}$
where $\theta_{\rm id}=-\arctan\sqrt2\simeq -54.7^\circ$ is the
ideal mixing angle, can be attributed to the deviation of the
ratio $f_q/f_s$ from unity which is induced by the
parameter $L_5$ in the effective Lagrangian (\ref{Leff1}).
Eq.~(\ref{eq:80rel}) is particularly useful since it allows to
translate results in the literature correctly from one basis into another.
As we will see below the usage of the correct relations between
$\theta_8$, $\theta_0$ and $\phi$ already resolves a big part of the
apparent discrepancies between the results of
different approaches mentioned in the introduction. It also reveals
that the notion of a single octet-singlet mixing angle $\theta_P$ is
more confusing than helpful in phenomenological analyses.

\subsection{Comparison of mixing parameters in the literature}

In Tables~\ref{tab:fPa} and \ref{tab:fPi} I list 
in chronological order some results for the
mixing parameters $f_a$ and $\theta_a$ in the octet-singlet basis,
as well as for the parameters $f_i$ and $\phi_i$ in the
quark-flavor basis, obtained in various 
analyses.\footnote{Of course, due to the long tradition of
the subject, the list in Tables~\ref{tab:fPa} and \ref{tab:fPi} is
far from being complete. The quoted references, therefore, should
be regarded as representative examples.} {}\
 A key ingredient for the comparison is the usage of
the general parametrization of decay constants, Eqs.~(\ref{eq:para81})
and (\ref{eq:paraqs}).
Note that some of the results of previous articles
have not been recognized,  simply because one only has concentrated
on finding a single mixing angle $\theta_P$. 
The importance of determining independently the four decay constants
in the $\eta$-$\eta'$ system (i.e.\
a set of four mixing parameters) 
has often not been realized.
Tables~\ref{tab:fPa} and \ref{tab:fPi}
reveal that in most of those cases that are not in conflict
with the general parametrization~(\ref{eq:para81}),
there is fair agreement among the results of different approaches.
The largest variation is found for the parameter $\theta_0$
which ranges from $-9^\circ$ to $0^\circ$.

\begin{table}[htbp]
\tcaption{Some results for mixing parameters of the
$\eta$-$\eta'$ system in the octet-singlet basis.
(The entries in parantheses $[..]$
have not been quoted in the original literature
but have been calculated from information given 
therein, assuming, for simplicity,
that all OZI-rule violating parameters except
for $\Lambda_1$ are zero.)}
\label{tab:fPa}
\centerline{\footnotesize\smalllineskip
\begin{tabular}{l c c c c  }\\
\hline
\\[-0.5em]
 source 
        & $f_8/f_\pi$ & $f_0/f_\pi$ 
        & $\theta_8$ & $\theta_0$  \\[-0.5em]
\\
\hline \\[-0.5em]
Mass matrix and radiative decays\cite{Kazi:1976tu,Fritzsch}
        & $[1.3]$  & $[1.2]$
        & $[-20^\circ]$ & $[-1^\circ]$ \\ 
$U(1)_A$ anomaly \&{} meson masses\cite{Diakonov:1981nv}
	& $[1.2]$  & $[1.1]$
        & $[-20^\circ]$ & $[-5^\circ]$ \\ 
Phenomenology\cite{GaLe85,phen1,Gilman:1987ax,BaFrTy95}
        & $1.2-1.3$ & $1.0-1.2$
        & \multicolumn{2}{c}{$-(23^\circ {\rm \ - \ } 17^\circ)$}  \\
NJL quark model \&{} phenom.\cite{Schechter:1993iz} 
        & $[1.24]$ & $[1.21]$
        & $[-19.5^\circ]$ & $[-5.5^\circ]$  \\
{\em Current mixing}\/ model \&{} phenom.\cite{KiPe93}
        & $[0.71]$ & $[0.94]$
        & $[-12.2^\circ]$ & $[-30.7^\circ]$ \\
Phenomenology\cite{Bramon97}
        & $[1.34]$ & $[1.21]$
        & $[-23.2^\circ]$ & $[-7.0^\circ]$ \\ 
GMO mass formula\cite{Burakovsky:1998vc}
        & 1.19 & 1.10
        &\multicolumn{2}{c}{ $-21.4^\circ$}  \\
$\chi$PT \&{} $1/N_C$ expansion \&{} phenom.\cite{Leutwyler97} 
        & 1.28 & 1.25
        & $-20.5^\circ$ & $-4^\circ$  \\
FKS scheme \&{} theory\cite{Feldmann:1998vh} 
        & 1.28 & 1.15
        & $-21.0^\circ$ & $-2.7^\circ$  \\
FKS scheme \&{} phenom.\cite{Feldmann:1998vh} 
        & 1.26 & 1.17
        & $-21.2^\circ$ & $-9.2^\circ$  \\
Vector meson dominance \&{} phenom.\cite{Benayoun:1999fv}
        & 1.36 & 1.32
        & $-20.4^\circ$ & $-0.1^\circ$ \\
Energy dependent scheme \&{} phenom.\cite{Escribano:1999nh} 
        & 1.37 & 1.21
        & $-21.4^\circ$ & $-7.0^\circ$
\end{tabular}}
\end{table}

\begin{table}[hbtp]
\tcaption{Same as Table~\ref{tab:fPa} but in the
quark-flavor basis.}
\label{tab:fPi}
\centerline{\footnotesize\smalllineskip
\begin{tabular}{l c c c c  }\\ 
\hline
\\[-0.5em]
 source 
        & $f_q/f_\pi$ & $f_s/f_\pi$ 
        & $\phi_q$ & $\phi_s$  \\[0.5em]
\hline \\[-0.5em]
Mass matrix and radiative decays\cite{Kazi:1976tu,Fritzsch}
        & $[1.0]$  & $[1.4]$
        & \multicolumn{2}{c}{$44^\circ$} \\ 
$U(1)_A$ anomaly \&{} meson masses\cite{Diakonov:1981nv}
	& 1.0  & 1.4
        &\multicolumn{2}{c}{$[42^\circ]$} \\ 
Phenomenology\cite{GaLe85,phen1,Gilman:1987ax,BaFrTy95}
        & $[1.1-1.2]$ & $[1.1-1.3]$
        & $[28^\circ$-$34^\circ]$ & $[35^\circ$-$41^\circ]$\\
NJL quark model \&{} phenom.\cite{Schechter:1993iz} 
        & $[1.07]$ & $[1.36]$
        & $[44.1^\circ]$ & $[40.6^\circ]$  \\
{\em Current mixing}\/ model \&{} phenom.\cite{KiPe93}
        & $[0.98]$ & $[0.66]$
        & $[35.9^\circ]$ & $[26.2^\circ]$ \\
Phenomenology\cite{Bramon97}
        & $[1.00]$ & $[1.45]$
        & \multicolumn{2}{c}{$39.2^\circ$} \\ 
GMO mass formula\cite{Burakovsky:1998vc}
        &$ [1.13]$ & $[1.16]$
        & $[31.2^\circ]$ & $[35.4^\circ]$  \\
$\chi$PT \&{} $1/N_C$ expansion \&{} phenom.\cite{Leutwyler97} 
        & $[1.08]$ & $[1.43]$
        & $[44.8^\circ]$ & $[40.5^\circ]$  \\
FKS scheme \&{} theory\cite{Feldmann:1998vh} 
        & 1.00 & 1.41
        & \multicolumn{2}{c}{$42.4^\circ$} \\ 
FKS scheme \&{} phenom.\cite{Feldmann:1998vh} 
        & 1.07 & 1.34
        & \multicolumn{2}{c}{$39.3^\circ$} \\ 
Vector meson dominance \&{} phenom.\cite{Benayoun:1999fv}
        & $[1.09]$ & $[1.55]$
        & $[47.5^\circ]$ & $[42.1^\circ]$ \\
Energy dependent scheme \&{} phenom.\cite{Escribano:1999nh} 
        & $[1.10]$ & $[1.46]$
        & $[38.9^\circ]$ & $[41.0^\circ]$
\end{tabular}}
\end{table}

For the following numerical discussion of phenomenological
observables I will often refer to the
set of mixing parameters obtained in
the phenomenologcial
analysis performed on the basis of the FKS scheme\cite{Feldmann:1998vh}
\beq
&& \, f_8 = (1.26 \pm 0.04) \, f_\pi \ , 
  	\qquad \theta_8 = -21.2^\circ \pm 1.6^\circ \ ,\cr 
&& \, f_0 = (1.17 \pm 0.03) \, f_\pi \ ,
        \qquad \theta_0 = -\phantom{1}9.2^\circ \pm 1.7^\circ 
\cr
& \Leftrightarrow& \cr
&& \matrix{ f_q = (1.07 \pm 0.02) \, f_\pi \ ,\cr
            f_s = (1.34 \pm 0.06) \, f_\pi \ ,}  \qquad
    \phi = 39.3^\circ \pm 1.0^\circ \ , \qquad \Lambda_1 \equiv 0 \ .
\label{eq:parset}
\eeq  
Note that the quoted errors refer to the experimental
uncertainty used in the determination of the mixing parameters,
only. A systematical error, arising from OZI-rule violations or
higher orders in the effective Lagrangian,
is not included. 
For given values, say, of the parameters $f_8$, $f_0$ and $\theta_8$
(which are rather well known) this error should be assigned to
the angle $\theta_0$ which may explain the rather large variation
of its value in different phenomenological approaches.

In some of the articles quoted in Tables~\ref{tab:fPa} and \ref{tab:fPi}
one can find different proposals for mixing parameters.
Escribano and Fr{\`e}re\cite{Escribano:1999nh}{} \ 
have suggested an alternative
parametrization
\beq
\left(\matrix{ f_{\eta}^8 & f_{\eta}^0 \vspace{0.2em}\cr  f_{\eta'}^8 & f_{\eta'}^0}
\right)
&=& \left(\matrix{ \hat f_8 \, \cos\theta_\eta & 
			- \hat f_0 \, \sin\theta_\eta \vspace{0.2em}\cr
                   \hat f_8 \, \sin\theta_{\eta'} & 
		\phantom{-} \hat f_0 \, \cos\theta_{\eta'} }
\right) 
\label{eq:energypar}
\eeq
where $\theta_\eta$ and $\theta_{\eta'}$ are interpreted as
(energy-dependent)
mixing angles of $\eta$ and $\eta'$, respectively.
It can be mapped onto the one in Eq.~(\ref{eq:para81}) via
\beq
&&  \tan\theta_8 = \frac{\sin\theta_{\eta'}}{\cos\theta_\eta} \ , \qquad
     f_8 = \hat f_8 \, \sqrt{\cos^2\theta_\eta + \sin^2\theta_{\eta'}} 
     \ , \cr 
&& \tan\theta_0 = \frac{\sin\theta_\eta}{\cos\theta_{\eta'}} \ , \qquad
    f_0 = \hat f_0 \, \sqrt{\sin^2\theta_\eta + \cos^2\theta_{\eta'}} 
    \ .
\label{eq:frere}
\eeq
The authors claim that the phenomenological success of the
ansatz Eq.~(\ref{eq:energypar})
with substantially different values of $\theta_\eta$ and
$\theta_{\eta'}$ gives
evidence for the energy-dependence of the mixing angles.
Of course, a similar concept could be introduced 
for the quark-flavor basis, defining in an analogous
way energy-dependent
mixing angles $\phi_\eta$ and $\phi_{\eta'}$.
The actual values for $\theta_{\eta}$ and $\theta_{\eta'}$
found in the phenomenological
analysis\cite{Escribano:1999nh} translate
into almost equal values for  $\phi_\eta$ and $\phi_{\eta'}$.
In this basis the apparent energy-dependence cannot be
observed. This is in line with the above results: 
The necessity to use
two mixing angles is primordially a consequence of $SU(3)_F$--breaking, 
not of energy-dependence.

Kisselev and Petrov\cite{KiPe93} have introduced still another
parametrization in terms of an averaged mixing angle $\bar \theta$ and
an explicit symmetry breaking parameter $\varepsilon$. It can be
related to the parametrization~(\ref{eq:para81}) by
$$ \varepsilon = \sqrt{\frac{\tan\theta_8}{\tan\theta_0}}
\ , \qquad
   \tan\bar\theta = - \sqrt{\tan\theta_8 \, \tan\theta_0}{} \ .
$$ However, the authors fix the
value of $\varepsilon$  by model assumptions
to values smaller than one, 
while the analyses based on
Eq.~(\ref{eq:t8t1rel}) find $\varepsilon>1$.

Bramon et al.\cite{Bramon97} used constituent quark masses 
($\tilde m$)
instead of the decay constants
for the parameterization of $SU(3)_F$--breaking.
Since constituent quarks obey a Goldberger-Treiman
relation for the effective quark-meson coupling constant
$g_i \, f_i = \tilde m_i$ with $g_q \simeq g_s$,
one has $f_q/\tilde m_q \simeq f_s/ \tilde m_s$
which has been used in Tables~\ref{tab:fPa} and \ref{tab:fPi}.

\subsection{Two-photon decays and ${\mathcal L}_{WZW}$ Lagrangian}

One important source of information about the decay constants
are the two-photon decays $P\to\gamma\gamma$.
They 
are driven by the chiral 
anomaly\cite{ABJ} in QED
and have been used in most of the phenomenological analyses.
In particular, for
 the determination of the parameter set (\ref{eq:parset}),
the parameters $f_q$ and $f_s$ have been adjusted to the
$\eta(\eta')\to \gamma\gamma$ decay widths for a given
value of $\phi$ (see Table~\ref{fig:phi} below).
In the effective Lagrangian the chiral anomaly enters via
the Wess-Zumino-Witten term in the following way
\beq
  {\mathcal L}_{WZW} &=& - \frac{N_C \, \alpha_{\rm em}}{4\pi} 
 	\, F_{\mu\nu} \, \tilde F^{\mu\nu} \,
	{\rm tr} [ {\mathcal Q}^2  \varphi] \ ,
\eeq
where ${\mathcal Q}={\rm diag}[2/3,-1/3,-1/3]$ denotes
the matrix of quark charges. 
As has been discussed in detail by Leutwyler
and Kaiser,\cite{Leutwyler97} in
the flavor singlet channel one has again to allow for an OZI-rule
violating correction, which essentially corresponds to
replacing $f_0 \to f_0 / (1+\Lambda_3)$ after inserting
Eq.~(\ref{Pphiconnection}) into the WZW~Lagrangian.
(In the FKS scheme, $\Lambda_3$ is assumed to be small and neglected.)
Using the general parametrization of decay constants~(\ref{eq:para81})
and the connection between basis and physical fields~(\ref{Pphiconnection}),
the chiral anomaly prediction reads 
\beq
\Gamma[\eta\phantom{'}\to\gamma\gamma] &=&
\frac{9\alpha_{\rm em}^2}{16 \pi^3} \, M_{\eta}^3 \,
\left[\frac{C_8\, \cos\theta_0 }{f_8 \, \cos(\theta_8-\theta_0)} 
    - \frac{(1+\Lambda_3) \, C_0\, \sin\theta_8 }
           {f_0 \, \cos(\theta_8-\theta_0)} 
\right]^2 \nonumber \ , \\[0.3em]
\Gamma[{\eta'}\to\gamma\gamma] &=&
\frac{9\alpha_{\rm em}^2}{16 \pi^3} \, M_{{\eta'}}^3 \,
\left[\frac{C_8 \, \sin\theta_0 }{f_8 \, \cos(\theta_8-\theta_0)}
    + \frac{(1+\Lambda_3) \, C_0 \, \cos\theta_8 }{
          f_0 \, \cos(\theta_8-\theta_0)} 
\right]^2 \, .
\label{eq:gammapred}
\eeq
Here $C_8=(e_u^2+e_d^2 - 2 e_s^2)/\sqrt6$ 
and $C_0=(e_u^2+e_d^2+e_s^2)/\sqrt3$
are charge factors which are multiplied
by the elements of the {\em inverse matrix}\/ $\{f^{-1}\}{}_P^a$
obtained from Eq.~(\ref{eq:para81}).
In leading order the scale-dependence of $\Lambda_3$
cancels the one of $f_0$.

There is an alternative approach to obtain
a renormalization group
invariant prediction for $\eta(\eta')\to\gamma\gamma$.
In the two-component scheme of Veneziano/Shore\cite{ShVe91}
the singlet part of the decay amplitude is written as a sum of
a contribution from the would-be Goldstone boson, 
$g_{\tilde \eta_0 \gamma\gamma}$,
and a gluonic part which includes the Veneziano ghost,
$g_{\tilde G \gamma\gamma}$.
The residual effect of the gluonic contribution, after the
ghost field has been combined with $\tilde \eta_0$ and $\eta_8$
to yield the physical $\eta$ and $\eta'$ states should
be related to the parameter $\Lambda_3$ in $\chi$PT.

It is to be stressed, that Eq.~(\ref{eq:gammapred}) is
only rigorously valid in the chiral limit $\hat m \to 0$.
For the $\pi^0\to\gamma\gamma$ decay corrections to the chiral limit
are indeed small ($m_\pi^2 \ll 1~$GeV$^2$), 
and the value of the pion decay constant obtained
in this way is compatible\cite{PDG98} with the one measured in 
$\pi^\pm \to \mu^\pm \nu_\mu$. In the $\eta$ and $\eta'$ case,
however, Eq.~(\ref{eq:gammapred}) may receive 
additional corrections.
The phenomenological success of  Eq.~(\ref{eq:gammapred}),
on the other hand, indicates that these corrections cannot be
too large.

\subsection{Radiative transitions between light pseudoscalar and vector mesons}

The transitions $P\to V\gamma$ or $V \to P\gamma$ with
$P=\eta,\eta',\ldots$ and $V=\rho,\omega,\phi \ldots$
provide another possibility to  investigate
the mixing scenario in the pseudoscalar 
meson sector. 
A comparison of theory and experimental data
may also yield interesting information about the properties of
light vector mesons.
The relevant coupling constants 
are defined by matrix elements of the
electromagnetic current\cite{Dumbrajs:1983jd}
\beq 
\langle P(p_P)| J_\mu^{\rm em} | V(p_V,\lambda)\rangle |_{q^2=0}
&=&  
- g_{VP\gamma} \, \epsilon_{\mu\nu\rho\sigma} \, p_P^\nu \,
  p_V^\rho \varepsilon^{\sigma}{(\lambda)} \ .  
\eeq
The decay widths in terms of
these coupling constants read
\begin{eqnarray}
&& \Gamma[ P \to V \gamma ] = \alpha_{\rm em} \, g_{PV\gamma}^2 \, 
      k_{V}^3 \ , \qquad
\Gamma[ V \to P \gamma ] = \frac{\alpha_{\rm em}}{3} \, 
g_{PV\gamma}^2 \, 
              k_{P}^3 ~.
\label{PVgamma} 
\end{eqnarray}
An early discussion of these reactions in connection with
$\eta$-$\eta'$ mixing can be found in Refs.\cite{Kazi:1976tu,Fritzsch}
The $g_{PV\gamma}$ coupling constants have also been used in
the analysis of the $\eta$-$\eta'$ mixing angle
in Ref.\cite{Gilman:1987ax}
The subject has been
reconsidered by
Ball/Fr\`ere/Tytgat\cite{BaFrTy95} who investigated the coupling
constants $g_{PV\gamma}$ as a function of the mixing angle
in the naive octet-singlet scheme,
and by Bramon et al.\cite{Bramon97} who used the experimental information
on $g_{PV\gamma}$ coupling constants for a fit of the mixing angle
$\phi$ in the quark-flavor basis. 
In both analyses a (small) $\omega$-$\phi$ mixing angle
has been taken into account, too.

The theoretical estimates for the $g_{PV\gamma}$ coupling constants
are obtained by combining the chiral anomaly prediction 
for the decays $P\to \gamma\gamma$ (\ref{eq:gammapred}) with vector meson 
dominance.
The results look particularly simple in the FKS scheme.
Since in many analysis also OZI-rule violating contributions
have been (partly) taken into account, I quote here the
expressions that include both, the OZI-rule violating 
contribution to the Wess-Zumino-Witten Lagrangian ($\Lambda_3$)
and the $\phi$-$\omega$ mixing angle ($\phi_V$).
The following expressions represent
a generalization of the formulas quoted, for instance,
in Refs.\cite{BaFrTy95,Feldmann:1998sh} 
\beq
  g_{\eta\rho\gamma} &=& \frac{3 m_\rho}{2 \pi^2 f_\rho}
\left( 
\frac12 \, \frac{\cos\phi_s}{f_q   }
- \frac{\Lambda_3}{\sqrt6}\,\frac{\sin\theta_8}{f_0 } 
\right) \nonumber \ , \\[0.2em]
  g_{\eta'\rho\gamma} &=& \frac{3 m_\rho}{2 \pi^2 f_\rho}
\left( 
\frac12 \, \frac{\sin\phi_s}{f_q   }
+ \frac{\Lambda_3}{\sqrt6}\,\frac{\cos\theta_8}{f_0 } 
\right) \nonumber \ , \\[0.2em]
  g_{\eta\omega\gamma} &=& \frac{3 m_\omega}{2 \pi^2 f_\omega}
\left( 
\frac16 \, \frac{\cos\phi_s 
}{f_q   }
-\frac13 \, \frac{\sin\phi_q \sin\phi_V}{f_s   }
- \frac{\Lambda_3}{3\sqrt6}\,\frac{\sin\theta_8}{f_0 } 
\right) \nonumber \ , \\[0.2em]
 g_{\eta'\omega\gamma} &=& \frac{3 m_\omega}{2 \pi^2 f_\omega}
\left( 
\frac16 \, \frac{\sin\phi_s 
}{f_q   }
+\frac13 \, \frac{\cos\phi_q \sin\phi_V}{f_s   }
+\frac{\Lambda_3}{3\sqrt6}\,\frac{\cos\theta_8}{f_0 } 
\right) \nonumber \ , \\[0.2em]
  g_{\eta\phi\gamma} &=& \frac{3 m_\phi}{2 \pi^2 f_\phi}
\left( 
\frac16 \, \frac{\cos\phi_s \sin\phi_V}{f_q   }
+\frac13 \, \frac{\sin\phi_q 
}{f_s   }
+ \frac{\Lambda_3}{3\sqrt3}\,\frac{\sin\theta_8}{f_0 } 
\right) \nonumber \ , \\[0.2em]
 g_{\eta'\phi\gamma} &=& \frac{3 m_\phi}{2 \pi^2 f_\phi}
\left( 
\frac16 \, \frac{\sin\phi_s \sin\phi_V}{f_q   }
-\frac13 \, \frac{\cos\phi_q 
}{f_s   }
-\frac{\Lambda_3}{3\sqrt3}\,\frac{\cos\theta_8}{f_0 } 
\right) \ .
\label{eq:gpvformula}
\eeq
Here $\phi_V$ is 
defined in the same manner as $\phi$, see Eq.~(\ref{eq:stateqs}). 
For simplicity, I have neglected
terms of the order $(\Lambda_3 \, \sin\phi_V)$ and have
set
$\cos(\phi_q-\phi_s)$, $\cos(\theta_8-\theta_0)$ and
$\cos\phi_V$ to unity in
Eq.~(\ref{eq:gpvformula}).

Three recent numerical analyses
are summarized in Table~\ref{tab:gPV},
referring to a determination in the FKS scheme,\cite{Feldmann:1998sh} 
and two investigations\cite{Escribano:1999nh,Benayoun:1999fv}
where different $\eta$-$\eta'$ 
mixing schemes have been used and a non-vanishing $\phi$-$\omega$
mixing angle have been taken into account.
The numerical values of the mixing parameters 
in the three analyses are, however, not too different
from each other, see Table~\ref{tab:fPa},
and the obtained estimates
for $g_{PV\gamma}$ are very similar and turn out to be in 
fair agreement
with the experimental findings.

\begin{table}[htbp]
\tcaption{Some recent estimates of the coupling constants $|g_{PV\gamma}|$
compared to experiment.}
\label{tab:gPV}
\centerline{\footnotesize\smalllineskip
\begin{tabular}{l c c c c c  }\\
\hline\\[-0.5em]
$PV$ & 
          FKS\cite{Feldmann:1998sh} &
          Escribano/Fr\`ere\cite{Escribano:1999nh} &
          Benayoun et al.\cite{Benayoun:1999fv} &
           Experiment\cite{PDG98} \\[0.5em]
\hline
\\[-0.5em]
$\eta\rho$ & 1.52 & 1.43 & 1.69 & $1.47^{+0.25}_{-0.28}$  \\ 
$\eta'\rho$ & 1.24 & 1.23 & 1.38 & $1.31 \pm 0.06$  \\ 
$\eta\omega$ & 0.56 & 0.54 & 0.58 & $0.53 \pm 0.04$  \\ 
$\eta'\omega$ & 0.46 & 0.55 & 0.44 & $0.45 \pm 0.03$  \\ 
$\eta\phi$ & 0.78 & 0.73 & 0.70 & $0.69 \pm 0.02$  \\ 
$\eta'\phi$ & 0.95 & 0.83 & 0.70 & $1.00^{+0.29}_{-0.21}$   
\end{tabular}}	  
\end{table}

\subsection{Light-Cone Wave Functions}

The decay constants defined in Eq.~(\ref{eq:fdef}) play an important role
in exclusive reactions at large momentum transfer.
In this case it is useful to consider an expansion of the physical
meson states in terms of Fock states 
with increasing number of partons. 
Schematically, for the $\eta$ and $\eta'$ mesons,
one may write\cite{Feldmann:1998vh,Feldmann:1998sh}
\beq
 |\eta\rangle &=& \sum_{a=8,0} \, \Psi_\eta^a(x,{\mathbf k_\perp})
     \, |a\rangle + \ldots \cr
 |\eta'\rangle &=& \sum_{a=8,0} \, \Psi_{\eta'}^a(x,{\mathbf k_\perp}) 
    \, |a \rangle + \ldots
\label{eq:Fock}
\eeq
where $|a\rangle= |\bar q \,\frac{\lambda^a}{\sqrt2} \, q \rangle$ 
is a partonic quark-antiquark Fock state. 
Each Fock state has
an individual light-cone wave functions $\Psi_P^a$.
Here $x$ denotes the ratio of the
quark and meson momenta in the light-cone plus-direction, and
${\mathbf k_\perp}$ is the quark momentum transverse to the meson one.
The dots in Eq.~(\ref{eq:Fock}) stand for the higher
Fock states which may include  additional gluons or
quark-antiquark pairs and, in principle, also a two-gluon component
$|gg\rangle$. Since the matrix elements
in  Eq.~(\ref{eq:fdef}) correspond to the annihilation of two
quarks at one space-time point, the decay constants
$f_P^a$ are related\footnote{I follow the normalization convention
of Ref.\cite{FeKr97b}} \ to the values
of the light-cone wave functions $\Psi_P^a$ \lq{}at the origin\rq{}
\beq
  f_P^a &=& 2 \sqrt 6 \, \int \frac{ dx \, d^2{\mathbf k_\perp}}{16 \pi^3} \, \Psi_P^a
\label{eq:lcrel} \ .
\eeq
An analogous relation is valid for the decay constants $f_P^i$ and
light-cone wave functions $\Psi_P^i$ in the quark-flavor 
basis. The relation~(\ref{eq:lcrel}) underlines the importance
of the decay constants for the description of the $\eta$-$\eta'$
system: They enter both, the effective Lagrangian relevant for
low-energy physics and the light-cone wave functions utilized
in high-energy reactions.

I remark at this point, that in the
flavor singlet channel, quark-antiquark and two-gluon parton states
can mix {\em perturbatively}\/. The evolution equations
to first order in $\alpha_s$ have been derived by
Baier and Grozin.\cite{BaGr81} This mixing is a true
OZI-rule violating process and should be neglected in
the FKS scheme.

At large energies one often integrates out
the intrinsic transverse momenta to obtain the
(scale-dependent) distribution amplitudes $\Phi_P^a(x;\mu)$
which are defined by non-local matrix elements 
\beq
i \, f_P^a \, \Phi_P^a(x;\mu) &=& \int \frac{dz^-}{2\pi} \, e^{ixp^+z^-}
     \, \langle 0 | \bar q(0) \, \gamma^+ \gamma_5 \,
	\frac{\lambda^a}{\sqrt2} \, q(z^-) | P(p)\rangle 
\, \Big|_{\mu} \ .
\label{Phidef}
\eeq
They can be expanded about Gegenbauer polynomials, which are the
eigenfunctions of the QCD evolution equation for mesons
\beq
 \Phi(x;\mu) &=& 6 \, x \, (1-x) \, \left(
	1 + \sum_{n=2,4,\ldots} \, B_n(\mu) \, C_n^{(3/2)}(2x-1) \right)
\label{gegenbauer} \ .
\eeq
In the limit $\mu\to\infty$, the coefficients
$B_n$ evolve to zero with anomalous dimensions
increasing with $n$, and one is left
with the asymptotic distribution amplitude 
$\mbox{$\phi_{\rm AS}(x)=6x(1-x)$}$.
Usually one keeps only a finite number of non-zero
Gegenbauer coefficients in Eq.~(\ref{gegenbauer})
which are then determined from phenomenology, QCD sum rules, 
low-energy models etc.
Typical QCD sum rule estimates\cite{Braun:1989qv} 
lead to distribution amplitudes for the pion which are somewhat
broader than the asymptotic one ($ B_2 \simeq 0.44$
and $B_4\simeq 0.25$ at $\mu = 1$~GeV).
Ball\cite{Ball:1998je} also considered the
distribution amplitudes for the $\eta$ meson and
finds a smaller value of the first Gegenbauer coefficient,
$B_2 \simeq 0.2$, which
follows the general trend that heavier mesons have narrower
distribution amplitudes.
The pion distribution amplitude has also been calculated in
the instanton model,\cite{Petrov:1998kg} and values 
$B_2\simeq 0.06$ and $B_4 \simeq 0.01$ have been found.

The Fock state expansion is also related to 
 the parton distributions
which are extracted from the
structure functions
measured in deep inelastic scattering.
Formally, the parton distributions arise from an infinite
sum over all Fock state wave functions\cite{BrLe80} squared and
integrated over transverse momenta and all but
the momentum fraction $x_j$ of the struck quark with a certain
flavor, e.g.\
\beq
  f_{u/\pi}(x) &=& \sum_{\beta N} \int [d^2k]_N \, [dx]_N \,
	\left|\Psi_{\beta N}\right|^2 \, \delta(x-x_j)
\label{parton} \ .
\eeq
The sum runs over all Fock states with parton number $N$ being
in a color/spin/flavor combination labeled by $\beta$.
A detailed analysis in the nucleon case\cite{Diehl:1998kh}
 has revealed that Fock states  higher than
the leading $q\bar q$ one lead to higher powers of $(1-x)$
in Eq.~(\ref{parton}), under the 
reasonable assumption that the distribution amplitudes
of higher Fock states can be described by their asymptotic form
multiplied by polynomials in
the light-cone momentum fractions $x_i$.
Restricting oneself to a few Fock states 
is thus sufficient to
predict the parton distributions at large $x$. 


\begin{figure}[htbp]
\vspace*{13pt}
\begin{center}
\epsfclipon
\psfig{file=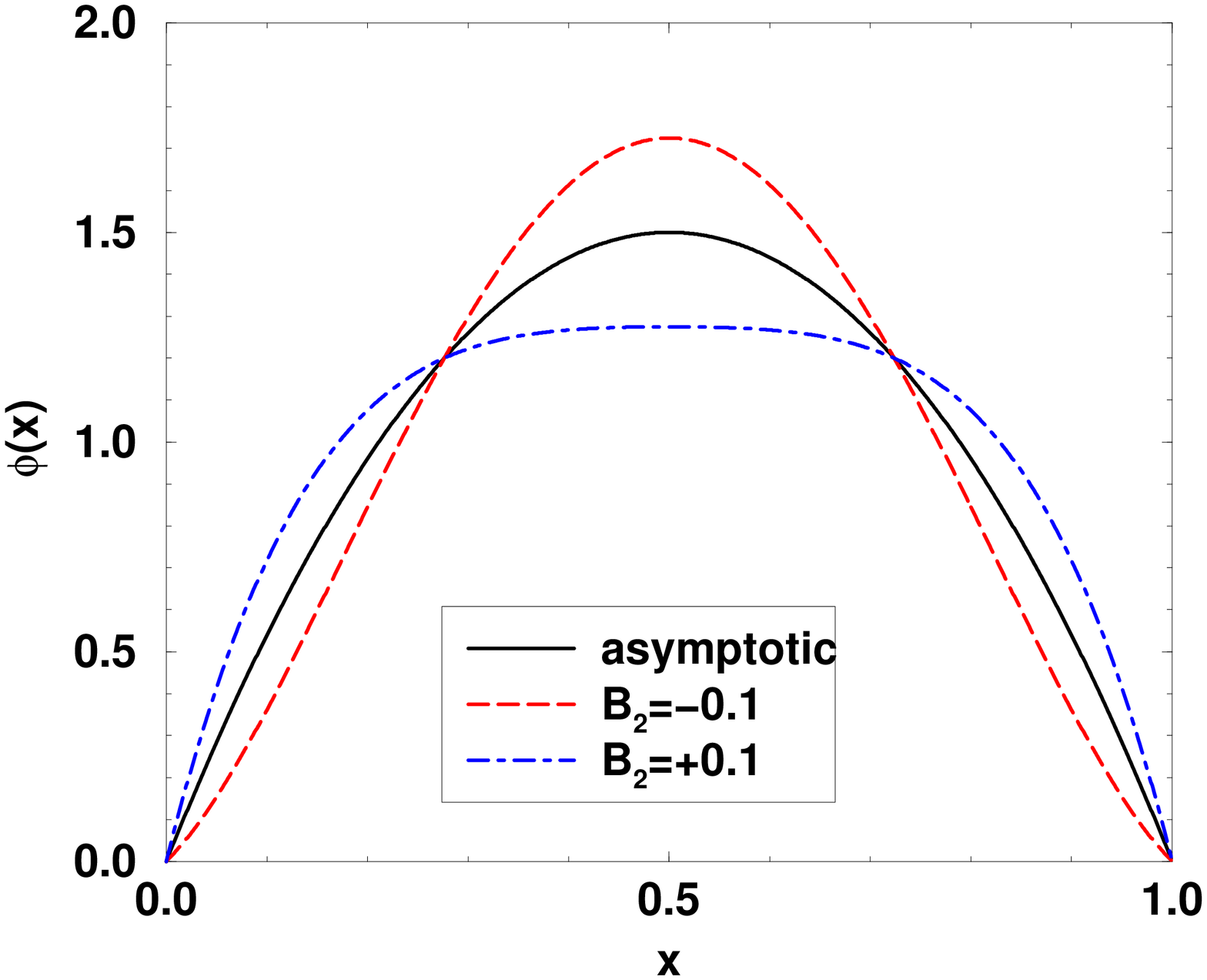,bb=90 35 590 645 ,width=5cm,angle=-90}
\psfig{file=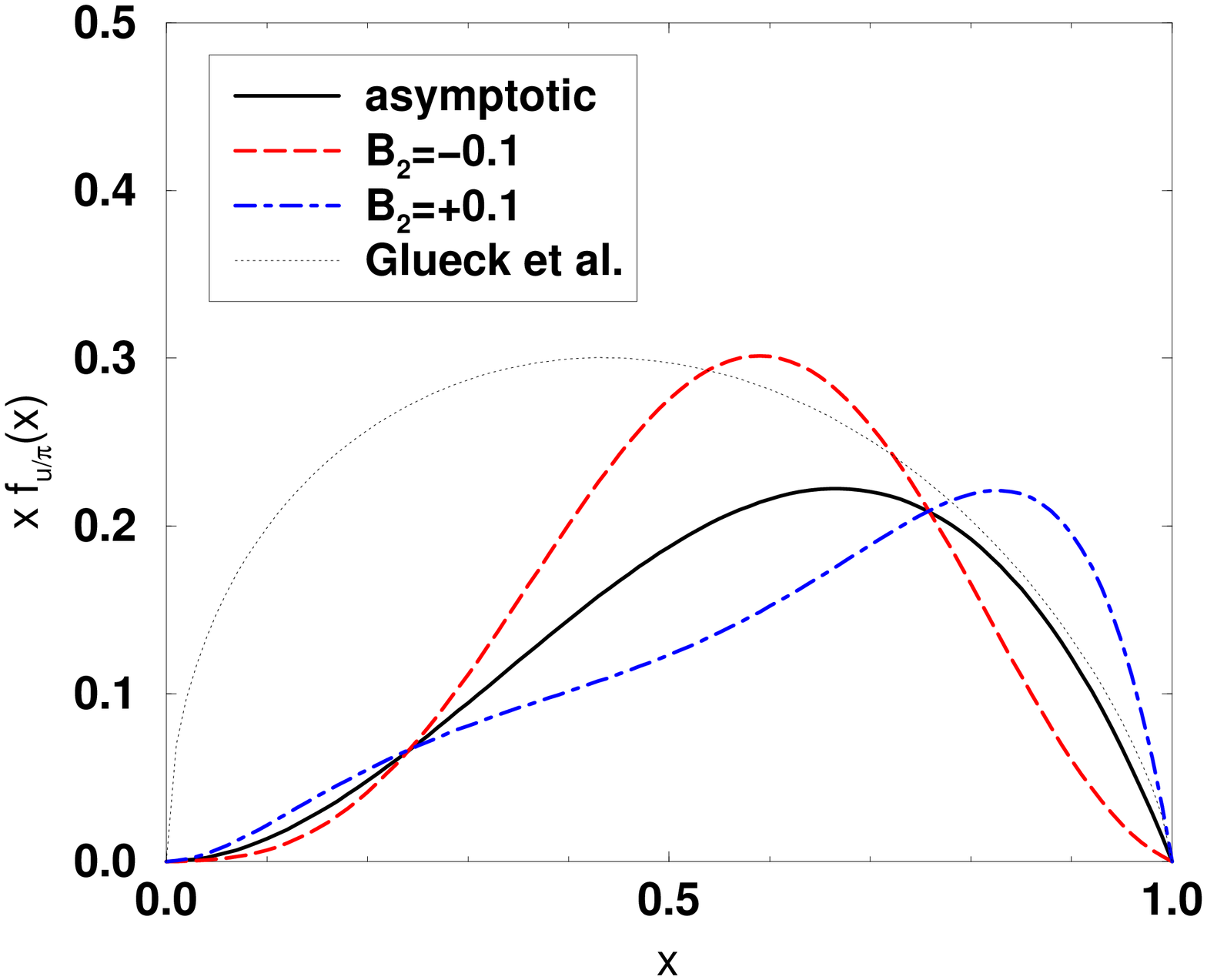,bb=90 35 590 645 ,width=5cm,angle=-90}
\end{center}
\vspace*{13pt}
\fcaption{The distribution amplitude $\Phi(x)$ of the
$q\bar q$ Fock state in a pseudoscalar meson (l.h.s.)
and its contribution
to the valence parton distribution $x \, f_{u/\pi}(x)$
(r.h.s.) for different
values of the Gegenbauer coefficient $B_2$ at $\mu=1$~GeV
(solid line: $B_2=0$; 
 dashed line: $B_2=-0.1$;
 dot-dashed line: $B_2=+0.1$).
The parametrization for $x f_{u/\pi}(x)$ (dotted line) has been
taken from Gl\"uck et al.\cite{GRS}
}
\label{fig:da}
\end{figure}

Only for the pion experimental data are available but suffers from
rather large errors. 
The recent analysis by Gl\"uck et al.\cite{GRS}
leads to the following (LO) parametrization of the
valence quark distribution inside the
pion at 1~GeV$^2$
\beq
  x \, f_{u/\pi}(x) &=& 0.745 \, (1-x)^{0.727} 
	\left(1-0.356 \, \sqrt{x} + 0.379 \, x\right) x^{0.506}
\label{eq:GRS} \ .
\eeq
Despite of the uncertainties, it
can be used as a cross-check of
the $x$ distribution in the $q\bar q$ Fock
state.\cite{JaKrRa96,CVogt} 
In Fig.~\ref{fig:da} I plotted the 
asymptotic distribution amplitude and slight deviations from it
(using $B_2 = \pm 0.1$) as a function of $x$. 
In the same figure I have shown the contribution of
the $q\bar q$ Fock state\footnote{For the transverse part of
the wave function I assumed a Gaussian. Its width is fixed
by the $\pi^0\to\gamma\gamma$ decay\cite{BHL}.} {}\ to the valence quark
distribution function of the pion, confronted with the 
phenomenological parametrization by Gl\"uck et al.\cite{GRS}
As one observes, a distribution amplitude $\Phi_\pi(x)$
which is  close to the asymptotic form already at low renormalization
scales, $\mu \simeq 1$~GeV, is preferred.

Once, the pion distribution amplitude has been
determined,
it can be applied to other hard exclusive reactions
(for instance, soft and hard 
contributions to the pion electromagnetic form 
factor\cite{Braun:1994ij,JaKr93,Stefanis:1998dg},
charmonium decays into light pseudo\-scalars,\cite{Bolz:1997ez}
$B$ meson decays into light 
pseudoscalar mesons\cite{Ball:1998tj,Khodjamirian:1998ji,Feldmann:1999sm}
).
The most important experimental
information on the distribution amplitude of the pion comes
from the $\pi\gamma$ transition form factor at large momentum
transfer.
In the following paragraph I will compare the $\pi\gamma$
form factor with the $\eta\gamma$ and $\eta'\gamma$
form factors.

\subsection{$P\gamma$ transition form factors}

\begin{figure}[htbp]
\vspace*{13pt}
\begin{center}
\epsfclipon
\unitlength0.95cm
a)
\begin{picture}(6,5)
\put(0,5){
\psfig{file=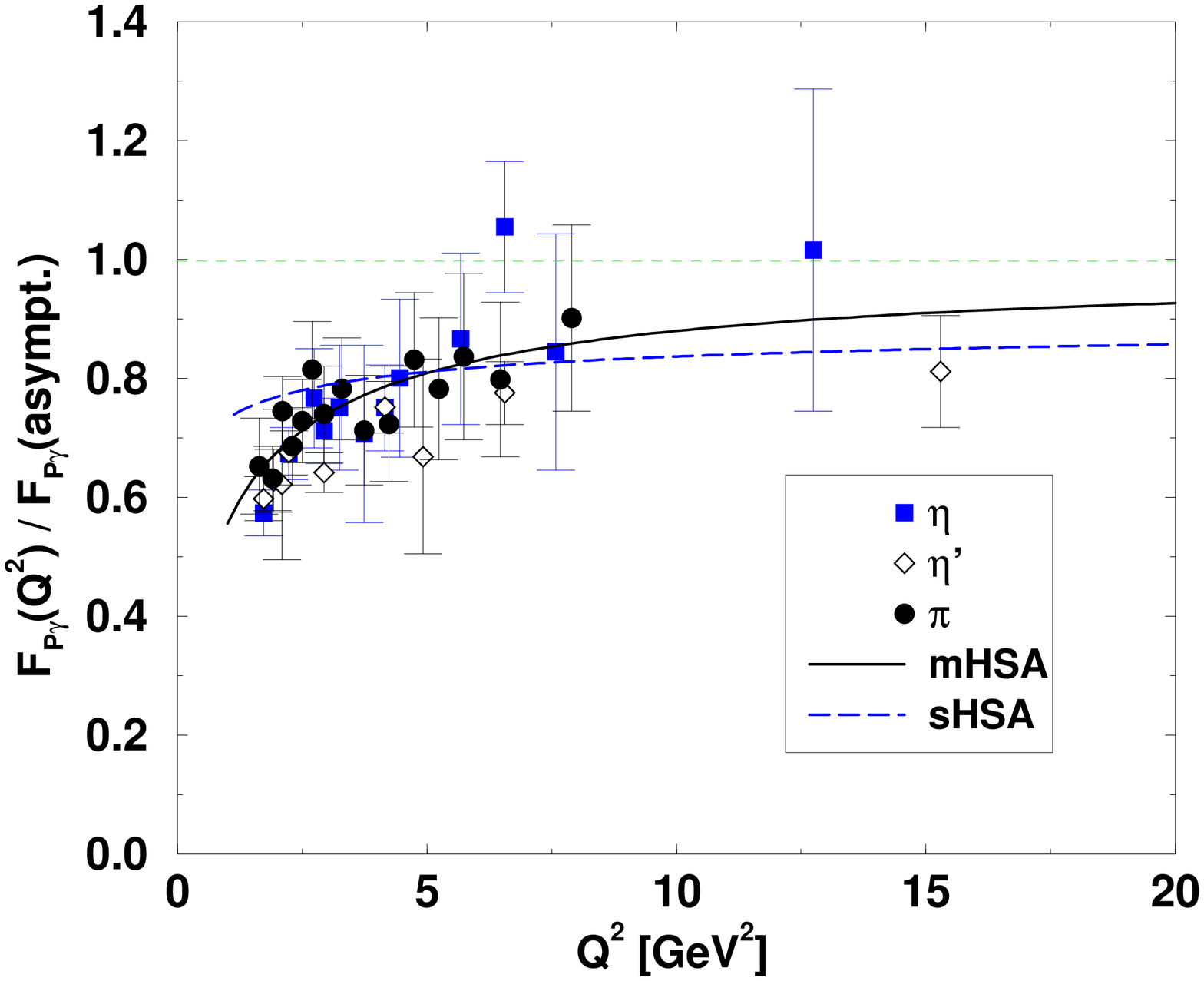,bb=90 35 590 645 ,width=4.6cm,angle=-90}}
\end{picture}
b)
\begin{picture}(6,5)
\put(0,5){
\psfig{file=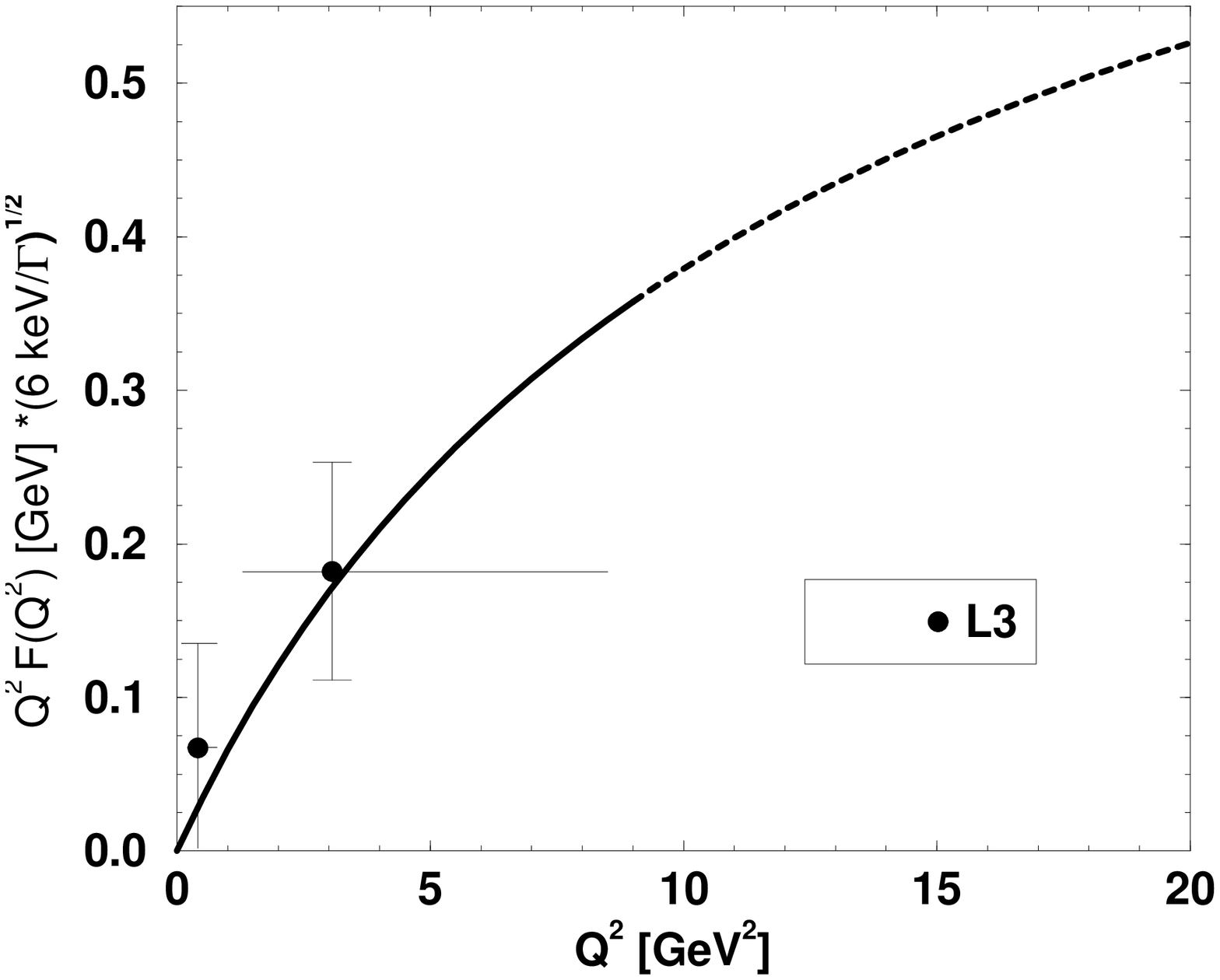,bb=120 115 547 650 ,width=4.5cm,angle=-90}}
\end{picture}
\end{center}
\vspace*{13pt}
\fcaption{a)
         Results for the photon-transition form factors
         of light pseudoscalar mesons $\pi^0,\eta,\eta'$,
         divided by their asymptotic behavior~(\protect\ref{eq:FPgamma}).
         The sHSA result has been calculated according to
         Brodsky\cite{Brodsky:1997ia}; the mHSA prediction
         is based on a calculation
         taken from Ref.\cite{FeKr97b} The asymptotic distribution amplitude
         has been used, and meson masses have been neglected. 
	 Experimental data are taken from CLEO\cite{CLEO97}.
         b) 
         Result for the $\eta_c\gamma$ transition form factor.
         The theoretical prediction 
         is taken from Ref.\cite{FeKr97a}
         Experimental data are taken from a recent
	analysis by the L3 collaboration.\cite{L3etac}}
\label{fig:Pgamma}
\end{figure}

For neutral pseudoscalar mesons the process $\gamma\gamma^*\to P$
with (at least) one highly virtual photon offers a possibility
to test the wave functions $\Psi_P^a$ and the
decay constants $f_P^a$. The meson-photon transition form factor
measured in this process can be expressed as a convolution of 
the light-cone wave functions
with a perturbatively calculable hard-scattering 
amplitude.\cite{BrLe80,Chernyak:1984ej,JaKrRa96,FeKr97b,Ong:1995gs,KrRa96,Cao:1996di,MuRa97}
For asymptotically large momentum transfer the form factor is 
solely determined by the decay constants
\beq
Q^2 \, F_{P\gamma}(Q^2)&=& 6 \sum_{a} \, C_a \, f_P^a 
\qquad (Q^2 \to \infty) 
\label{eq:FPgamma}
\eeq
where $C_{8,0}$ are defined after Eq.~(\ref{eq:gammapred}) and
$C_3=(e_u^2-e_d^2)/\sqrt2$.
If we had experimental data in the asymptotic region of momentum
transfer Eq.~(\ref{eq:FPgamma}) could be taken as a rigorous
way to determine two of four mixing parameters in 
Eq.~(\ref{eq:para81}).
However, as the example of the pion (where the decay constant
$f_\pi$ is known) reveals, the recent experimental data from
CLEO\cite{CLEO97} at $Q^2$ values of a few GeV$^2$ are
still 15-20\% below the asymptotic value.
One therefore has to take into account corrections to
Eq.~(\ref{eq:FPgamma}).
A calculation of the $\eta(\eta')\gamma$
transition form factor within
the modified hard-scattering approach (which takes into account
transverse momenta and Sudakov suppressions)
has been performed in Ref.\cite{FeKr97b}
The usage of the two-angle parametrization~(\ref{eq:para81})
turns out to be crucial to obtain a simultaneous description of both,
the two-photon decays (depending on the {\em inverse}\/ of
the decay constant matrix) and
the transition form factor at large momentum transfer (depending 
linearly on the $f_P^a$).
A very good description of 
the CLEO data is obtained
if the values of mixing parameters in Eq.~(\ref{eq:parset})
and the asymptotic distribution amplitudes
are used.
The result of that analysis is plotted in
Fig.~\ref{fig:Pgamma}a) where I have divided the transition
form factors for $\pi^0,\eta,\eta'$
by their asymptotic behavior~(\ref{eq:FPgamma}), using 
the mixing parameters in Eq.~(\ref{eq:parset}).
As one observes, within 
the errors the results for the three different mesons nearly fall
on top of each other. This indicates, that the octet and singlet
pseudoscalar mesons behave very similarly at large energies,
or, in other words, the $x$ distributions in the light-cone
wave functions of the 
$q\bar q$ Fock state for  $\pi$, $\eta$ and $\eta'$ mesons
are not very different from each other.
This is to be confronted with the $\eta_c\gamma$ transition
form factor which behaves differently,\cite{L3etac,FeKr97a}
see Fig.~\ref{fig:Pgamma}b),
due to the suppression
with the heavy quark mass at intermediate energies and a
different distribution amplitude which can be approximated
by a Gaussian around $x_0=1/2$,
\beq
  \Phi_{\eta_c}(x) &=& {\mathcal N} \, x \, (1-x) \, 
	\exp\left[ - a_c^2 \, M_{\eta_c}^2 \, (x - x_0)^2\right]
\ .
\eeq
For comparison I have also plotted in Fig.~\ref{fig:Pgamma}a)
the result of the
standard hard-scattering approach (sHSA), 
following the work of Brodsky\cite{Brodsky:1997ia}
and using the asymptotic distribution amplitude
\beq
  Q^2 \, F_{\pi\gamma}(Q^2) &=& \sqrt2 f_\pi \,
\left(1 - \frac 53 \, \frac{\alpha_V(e^{-3/2} Q)}{\pi} \right)	\label{shsa}
\label{eq:shsa} \ .
\eeq
Here the deviation from the asymptotic limit~(\ref{eq:FPgamma}) is due
to the first order QCD correction to the
hard-scattering amplitude.  Note that the argument
of  $\alpha_V$ in Eq.~(\ref{eq:shsa}) 
reflects a rather low renormalization scale.
One is thus sensitive to the infrared behavior of the strong
coupling constant.
Brodsky uses a particular choice  $\alpha_V(\mu)$ 
that {\em freezes}\/ for $\mu \to 0$. 
In this special form 
the sHSA also yields a good description of the data above, say, 3~GeV$^2$.


\begin{figure}[htbp]
\vspace*{13pt}
\begin{center}
\epsfclipon
\psfig{file=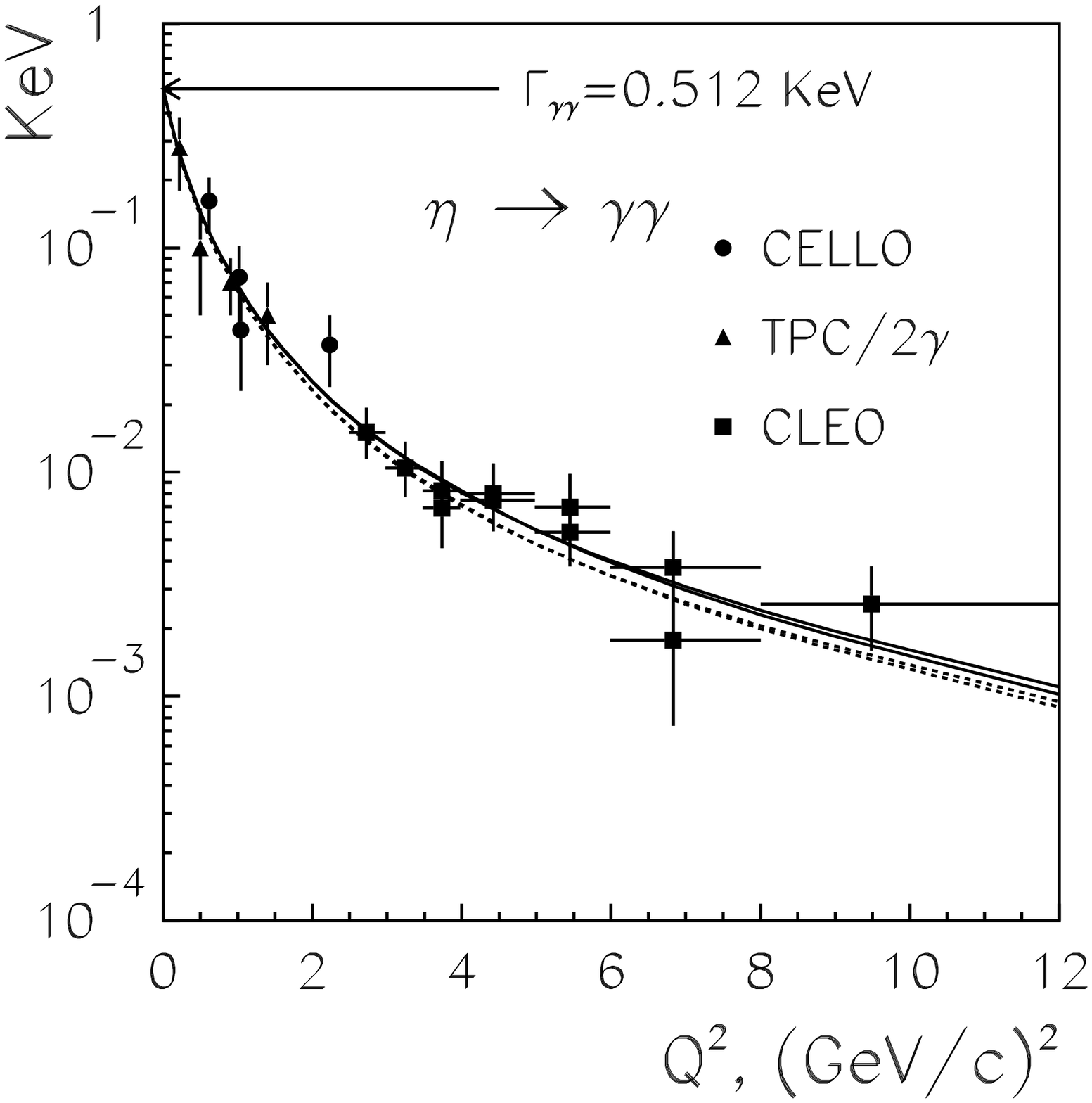 ,width=5.2cm}
\psfig{file=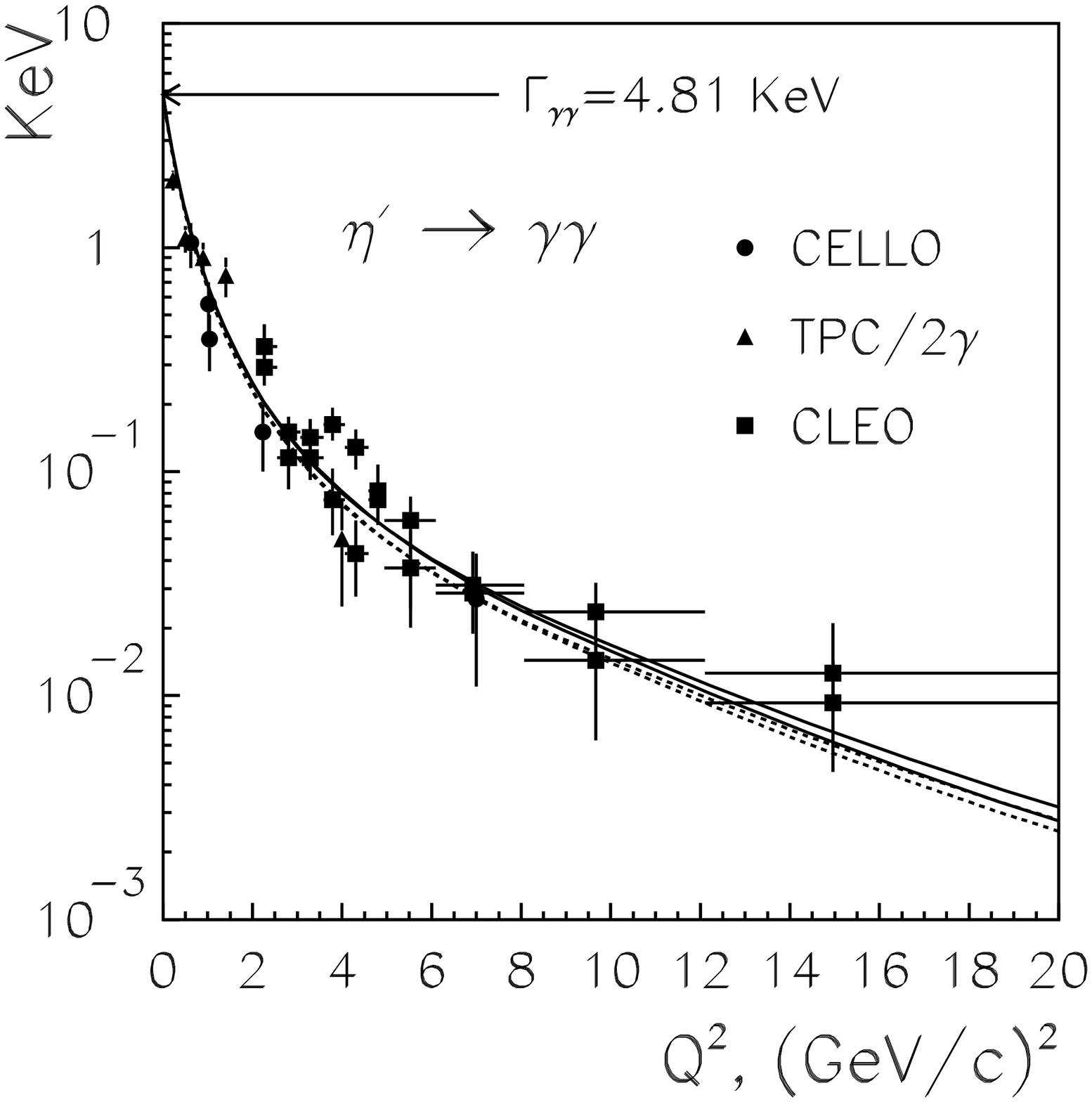 ,width=5.2cm}
\end{center}
\vspace*{13pt}
\fcaption{
         Results for the photon-transition form factors
         of light pseudoscalar mesons $\eta,\eta'$
         (Figures taken from Ref.\cite{Anisovich97}).
	 The theoretical result 
         is based on a wave function model 
	 taken from Anisovich et al.\cite{Anisovich97}
        (Different curves correspond to different parameter
          sets of the model.)
 	 Experimental data are taken from CLEO\cite{CLEO95},
         CELLO\cite{CELLO91} and TPC/2$\gamma$.\cite{TPC90}
        }
\label{fig:anisovich}
\end{figure}

The photon-transition form factors with light pseudoscalar mesons
have also been investigated by
Anisovich et al.\cite{Anisovich97}
Also in that analysis the universality of the $q\bar q$ wave
functions of $\pi^0$, $\eta$ and $\eta'$ with a distribution
amplitude close to the asymptotic form has been confirmed.
The $\eta$-$\eta'$ mixing has been treated in the FKS scheme
with a mixing angle $\phi=37.5^\circ$ which is not too different from the
value quoted
in Eq.~(\ref{eq:parset}). 
In addition to the HSA analyses
a soft hadronic part of the photon has been modeled, which
leads to a prescription of the data at low $Q^2$ with a
behavior close to vector meson dominance (VDM). The result of that analysis is
plotted in Fig.~\ref{fig:anisovich}. 


The $\eta(\eta')\gamma$ transition form factors have also been
treated in the conventional mixing scheme,
using  one common mixing angle in the octet-singlet 
basis.\cite{JaKrRa96,Cao:1998jm}
In this case a decent description of the data can only be achieved
by choosing different parameter values for the decay constants
and wave functions than the ones favored by other processes
and $\chi$PT.

It has 
often been tried to infer information on the decay constants
from fitting a pole formula for the $P\gamma$ transition form
factor to experimental data. 
In the case of the pion this is motivated by an interpolation
formula which has been proposed by Brodsky/Lepage,\cite{BrLe80}
\beq
F_{\pi\gamma}^{BL}(Q^2) & = &
\frac{6 \, C_3 \, f_\pi}{Q^2 + 4\pi^2 f^2_\pi}
 \ . 
\label{BLint}
\eeq
Obviously, it has the correct asymptotic limit~(\ref{eq:FPgamma}).
Furthermore for $Q^2 \to 0$ it coincides with the prediction
from the chiral anomaly. It {\em happens}\/ to have a similar form as
the vector dominance model (VDM) if one identifies $M_V=2 \pi f_\pi$.
Astonishingly, one has the approximate
equality $M_{\rho,\omega}  \simeq 2 \pi f_\pi$,
but there is no theoretical justification to assume that this
relation has to be exact.
Nevertheless, most of the experiments quote pole mass values
extracted from a fit to the data.
For the $\pi\gamma$ form factor this mass comes out to be not too different
from $M_\rho$, but of course it can not be used as a measurement
of $f_\pi$ by requiring Eq.~(\ref{BLint}) to hold exactly.
For the  $\eta\gamma$ and $\eta'\gamma$
transition form factors the situation is even more complicated
due to mixing, and even on the approximate level one does not 
find a simple relation of the experimental pole-mass fits with
the $\eta$-$\eta'$ mixing parameters.
Consequently, these fits should be viewed rather
as an effective parametrization of experimental data (which even
depends on the measured range of momentum transfer)
than a determination of process-independent quantities,
see the second paper of Ref.\cite{FeKr97b}
and references therein.

There are some additional processes which are similar to
the $P\gamma$ transition form factor and allow
for an independent determination of the mixing angles,
in particular they may be helpful to fix the value
of $\theta_0$.
One may, for instance, think of central
$\eta$ or $\eta'$ production in $pp$ collisions,\cite{Frere:1998ez,Close}
where the transition form factors for $g^* g^* \to \eta(\eta')$
are assumed to be relevant. 
The ratio of these form factors at large momentum transfer exactly
gives\cite{Feldmann:1998sh} $-\tan\theta_0$.
A complementary decay mechanism is provided by 
$\gamma$-Odderon-$\eta(\eta')$ processes in diffractive
$ep$ scattering as discussed by Kilian/Nachtmann.\cite{Kilian:1997ew} 
The ratio of form factors
in this case turns out to be given by\cite{Feldmann:1998sh} $\cot\theta_8$.
The experimental determination of the form factors from such
processes may, however, be very difficult.

A very academic process is the decay $Z \to \eta(\eta')\gamma$
which is similar to the $P\gamma$ transition form factors.
Taking into account the electroweak charges of the involved quarks
one obtains\cite{Feldmann:1998sh} 
$\Gamma[Z\to\eta\gamma]/\Gamma[Z\to\eta'\gamma]\simeq \tan^2\theta_0$.
Due to the smallness of the individual branching ratios\cite{Manohar:1990hu}
experimental data should not be expected in the near future.
The same vertex
is involved in the decays
$\eta(\eta')\to \gamma\mu^+\mu^-$, but at small momentum
transfers where the ratio gives a
measure for the angle $\theta_8$. As discussed by Bernabeu 
et al.\cite{Bernabeu:1998hy} the $P\gamma Z$ vertex can be measured by
extracting the $\gamma$-$Z$ interference  term from suitably
chosen asymmetries.

\subsection{Matrix elements with pseudoscalar quark currents}

Let me proceed with considering the matrix elements of the
pseudoscalar currents, entering
the anomaly equation (\ref{eq:anomaly}). They determine the
quark mass contribution to the meson masses.
It is natural to take 
the matrix elements in the quark-flavor basis where
the quark mass matrix is diagonal.
Let me define the four parameters 
$h_P^i$ for $i=q,s$ and $P=\eta,\eta'$ as
\beq
2 \, m_i \,  \langle 0 |  j_5^i(0) |P\rangle
  &=& h_P^i \ .
\label{eq:hPi}
\eeq
Here the pseudoscalar currents
in the quark-flavor basis are given as
$j_5^q = (\bar u i \gamma_5 u + \bar d i \gamma_5 d)/\sqrt2$ and
$j_5^s = \bar s i \gamma _5 s$, respectively.
Following the chiral effective Lagrangian 
(\ref{Leff0}) and (\ref{Leff1}), we can express
the matrix elements
in Eq.~(\ref{eq:hPi}) in terms of the decay constants and
the parameters
$B$, $L_8$ and $\Lambda_2$. The values of $B$ and $L_8$
are fixed in terms of  the pion and kaon masses and decay constants.
The parameter $\Lambda_2$ is needed to cancel the scale-dependence
of $\Lambda_1$ (the singlet pseudoscalar current is not
renormalized.) In the FKS scheme
both, $\Lambda_1$ and $\Lambda_2$, are neglected.
In this case one obtains
the simple representation
\beq
  \left\{ h_P^i\right\} &=&
\left(\matrix{ h_{\eta}^q & h_{\eta}^s\vspace{0.2em} \cr  h_{\eta'}^q & h_{\eta'}^s}
\right)
= U(\phi) \, {\rm diag}[ f_q \, M_\pi^2, \ f_s \, (2 M_K^2 - M_\pi^2) ] \ .
\label{eq:hpara}
\eeq
The validity of this simplification has again
to be tested against phenomenology.
With the same assumptions that lead to the FKS scheme
it makes sense to introduce  basis 
states\footnote{One 
sometimes finds the notation $|ns\rangle$
and $|s\rangle$ for non-strange and strange $q\bar q$ combinations,
respectively.} {}\ $|\eta_q\rangle$ and $|\eta_s\rangle$.
These are connected to the physical states by
the same mixing angle $\phi$ as the decay constants~(\ref{eq:paraqs})
and the quark mass contributions~(\ref{eq:hpara})
\beq
\left( 
 \matrix{ 
 |\eta\phantom{{}'}\rangle \cr
 |\eta'\rangle
 }
\right) &=& 
U(\phi) \, \left(
 \matrix{ 
 |\eta_q\rangle \cr 
 |\eta_s\rangle
 }
\right) \ .
\label{eq:stateqs}
\eeq
This connection has to be confronted with Eq.~(\ref{Pphiconnection}).
The basis states in Eq.~(\ref{eq:stateqs}) have a definite decompositon in
terms of quark-antiquark Fock states
\beq
|\eta_q \rangle &=&
\Psi_q^{}(x,{\mathbf k_\perp}) \, |u\bar u + d\bar d\rangle/\sqrt2 + \ldots 
\cr 
|\eta_s \rangle &=& 
\Psi_s^{}(x,{\mathbf k_\perp}) \, |s\bar s \rangle + \ldots
\label{eq:Fockqs}
\eeq
This is to be compared with Eq.~(\ref{eq:Fock}).
In this form~(\ref{eq:stateqs}) the FKS scheme 
has been utilized in a number of 
analyses.\cite{Isgur,Kazi:1976tu,Fritzsch,Bramon97,Feldmann:1998vh,Feldmann:1998sh,Munz:1994si} 
The separation into strange and non-strange quarks is natural
in those reactions where either the $s\bar s$ or $u\bar u + d\bar d$
component is probed, e.g.\ by light vector mesons, which 
show nearly ideal mixing, or by Cabbibo-favored weak transitions, $c \to s$.
Considering appropriate ratios of observables with $\eta$ and $\eta'$ in
such decay modes, one obtains 
an almost model-independent determination
of the mixing angle $\phi$. Such an investigation
on the basis of the FKS scheme  has been performed 
in Ref.\cite{Feldmann:1998vh}
and led to the result quoted in Table~\ref{fig:phi} (for
the decays $J/\psi \to \eta(\eta')\gamma$ included in
that table see Eqs.~(\ref{eq:anglerel}) and (\ref{rjpsi}) below).
Bramon et al.\cite{Bramon97} analyzed even more decay modes,
like those of tensor mesons or higher spin-states into pairs
of pseudoscalar and the whole class of radiative transitions between
vector and pseudoscalar mesons (see Section~3.5). 
This requires at some stage some
additional (but plausible) 
model-assumptions about $SU(3)_F$--breaking  and mixing angles of
vector and tensor mesons. Nevertheless, 
almost the same value, $\phi=39.2^\circ\pm1.3^\circ$, 
as in the FKS analysis has been found.

\begin{table}[htbp]
\tcaption{Determination of the mixing angle $\phi$ from
different decay channels, according to 
Ref.\cite{Feldmann:1998vh} and references therein.
The quoted error refers to the experimental uncertainties, only.
}
\label{fig:phi}
\begin{center}
\epsfclipon
\psfig{file=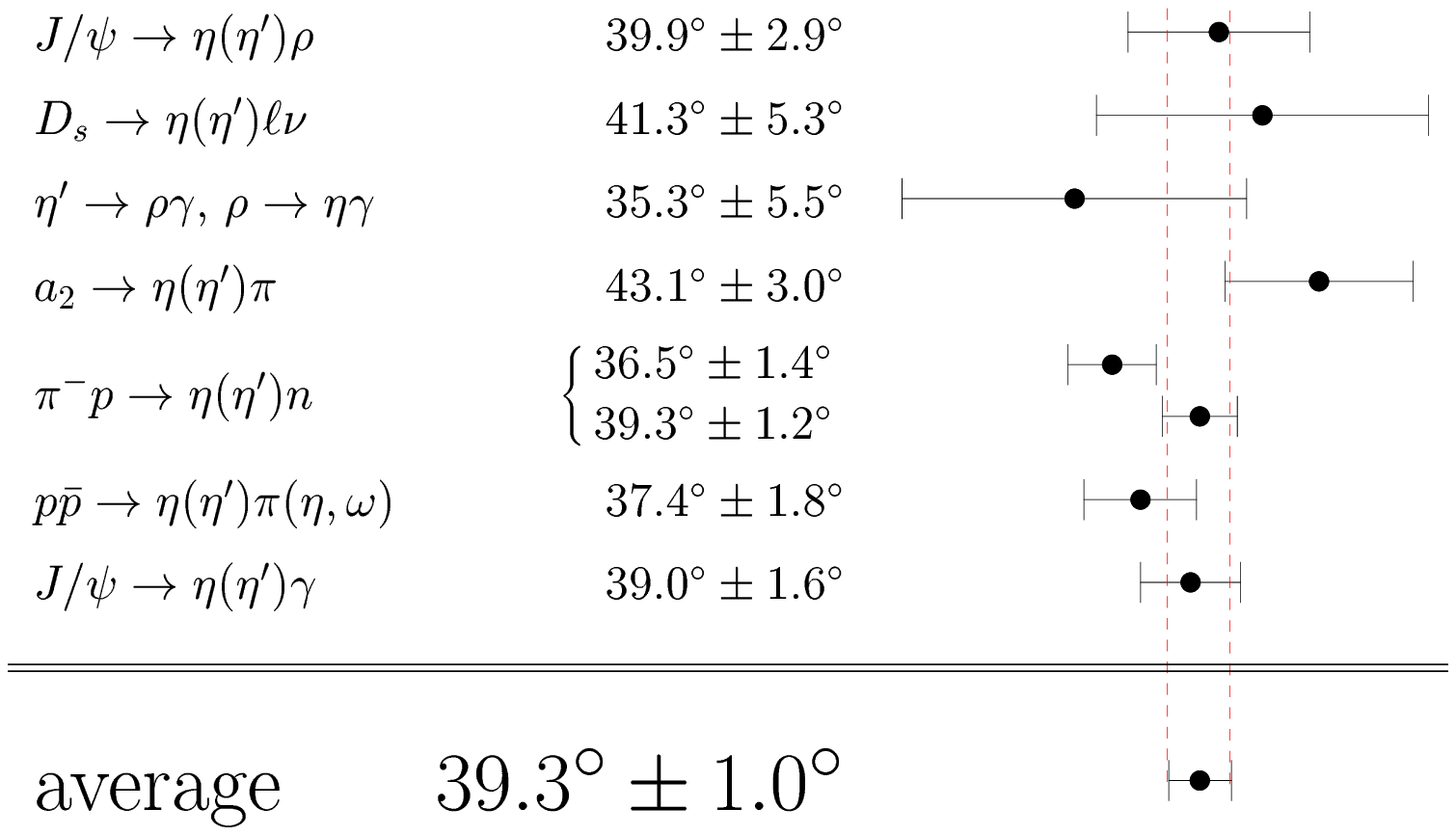,bb=95 460 545 730, width=8.3cm}
\end{center}
\end{table}

Values for the 
parameters $h_P^i$ can be
estimated using the pion and kaon masses and the
phenomenological results for the mixing angle $\phi$
and the decay parameters $f_q$ and $f_s$, see
Eq.~(\ref{eq:parset})
\beq
\left(\matrix{ h_{\eta}^q & h_{\eta}^s \vspace{0.2em}\cr  h_{\eta'}^q & h_{\eta'}^s}
\right) &=&
\left(\matrix{   0.0020~{\rm GeV}^3 & -0.053~{\rm GeV}^3\vspace{0.2em} \cr
                 0.0016~{\rm GeV}^3 &\phantom{-} 0.065~{\rm GeV}^3 }
\right) \ .
\label{eq:hPi_est}
\eeq

The matrix elements $h_P^i$ are, for example,
an important ingredient in the calculation
of $B$~meson decays into light mesons
in the factorization approach,
see e.g.\ Refs.\cite{Ali97,Chen:1999nx}
Occassionally,
it has been popular in this context to use a simplified
treatment of $\eta$-$\eta'$ mixing.
For instance, in the analysis of $B$ decays into $\eta$ or $\eta'$
in Ref.\cite{Dighe:1996gq} the $\eta$ meson is approximated as
$\sim |u\bar u + d \bar d -  s \bar s\rangle/\sqrt3$ and the
$\eta'$ meson as $\sim |u\bar u + d \bar d + 2 s \bar s\rangle/\sqrt6$.
This would correspond to taking
a mixing angle $\phi=\arctan\sqrt2/2\simeq 35.2^\circ$ and
ignoring $SU(3)_F$--breaking effects ($f_s=f_q=f_\pi$)
which would lead to significantly different values than in
Eq.~(\ref{eq:hPi_est}).

\subsection{Matrix Elements with the topological charge density}

Besides the decay constants, the matrix elements of 
the topological charge density $\omega$ in Eq.~(\ref{eq:anomaly}) 
\beq
A_P \equiv
\langle 0 | 2 \, \omega |P\rangle
\label{eq:AP}
\eeq
play an important role in the understanding of $\eta$-$\eta'$ mixing.
They can be used to define yet another mixing angle
\beq
  \frac{A_\eta}{A_\eta'} &=& - \tan\theta_y \ .
\eeq
Through the anomaly equation~(\ref{eq:anomaly}) the
quantities $A_\eta$ and $A_{\eta'}$
are directly related to the decay constants
and the mass parameters $h_P^i$.
If one sets the up- and down-quark masses in
Eqs.~(\ref{eq:anomaly}) and (\ref{eq:hPi}) 
to zero which is equivalent to
neglecting $M_\pi^2$ compared to $M_K^2$,
one obtains\cite{BaFrTy95,Feldmann:1998vh,KiPe93}
\beq
A_P 
&\simeq&
\frac{1}{\sqrt2} \,
\langle 0 | \partial^\mu  J_{\mu5}^q |P\rangle
= M_P^2 \, \frac{ f_P^q}{\sqrt2} 
= M_P^2 \, \frac{ f_P^8 + \sqrt2 \, f_P^0}{\sqrt 6}
\label{eq:gluerel} \ .
\eeq
The remaining pair of equations,
\beq
\langle 0 | 2 \, \omega |P\rangle
&=&
M_P^2 \, f_P^s - h_P^s \ ,
\eeq
when supplied with the ans\"atze~(\ref{eq:paraqs}) and
(\ref{eq:hpara}) for the decay constants and the parameters
$h_P^i$ in the FKS scheme,
can be transformed into an additional relation between the
angles $\theta_y$, $\theta_8$ and $\phi$, namely\cite{Feldmann:1998vh}
\begin{equation}
\matrix{
\tan\theta_8 &=& \tan\theta_y &=& -\frac{M_\eta^2}{M_{\eta'}^2} \, \cot\phi \cr
\uparrow && \uparrow && \uparrow \cr
{\rm octet\ (quarks)} && 
{\rm singlet\ (gluons)}  && {\rm quark\ flavor}}
\label{eq:anglerel}
\end{equation}
Here I have indicated the noteworthy fact that Eq.~(\ref{eq:anglerel})
connects the three different angles $\theta_8$, $\theta_y$ and $\phi$, 
and thus three different aspects of $\eta$-$\eta'$ mixing.
A prominent example where the angle $\theta_y$ enters is
the radiative $J/\psi$ decay into $\eta$ or $\eta'$.
Novikov et al.\cite{Novikov:1980uy} have argued
that the annihilation of the $J/\psi$ into light quarks
is dominated by the anomaly, i.e.\ the
matrix elements in Eq.~(\ref{eq:gluerel}). One then obtains\cite{Feldmann:1998vh}\
for the ratio of decay widths
\beq
R(J/\psi)= \frac{\Gamma[J/\psi \to \eta'\gamma]}
                   {\Gamma[J/\psi \to \eta\gamma]}
&=&
\left| \frac{\langle 0|\omega|\eta'\rangle}{
             \langle 0|\omega|\eta \rangle}\right|^2 \,
\left(\frac{k_{\eta'}}{k_\eta}\right)^3 
= \cot^2\theta_y\,
\left(\frac{k_{\eta'}}{k_\eta}\right)^3
\label{rjpsi}  
\eeq
where $k_P$ is the three-momentum of the final state meson in
the rest-frame of the $J/\psi$. The radiative $J/\psi$ decays
together with Eq.~(\ref{eq:anglerel}) provide an essential
cross-check of the self-consistency of the whole mixing approach.
From the experimental measurement\cite{PDG98} of the ratio~(\ref{rjpsi}),
$5.0\pm0.6$, one actually finds 
the following values of the mixing parameters,
$\theta_y=\theta_8=-22.0^\circ \pm 1.2^\circ$ and 
$\phi=39.0^\circ\pm 1.6^\circ$ 
where the result for the angle $\phi$ has already been 
used in Table~\ref{fig:phi} and turns out to be consistent
with the values obtained from other processes.
This is to be confronted with
the  naive but incorrect
expectation $\theta_8=\theta_0=\theta_P=\phi+\theta_{\rm id}$.
This formula is 
one of the sources of the discrepancies between different
determinations of mixing angles which have been mentioned in the
introduction.

Taking values for the decay constants $f_P^a$ from
Eq.~(\ref{eq:parset}) and using Eq.~(\ref{eq:gluerel}), 
one can give absolute numbers for
the matrix elements of the topological charge density
\beq
   A_{\eta\phantom{{'}}} =
   \langle 0 | 2 \, \omega|\eta\phantom{{}'}\rangle
     &=& 0.023~{\rm GeV}^3 \ , \cr
   A_{\eta'} = \langle 0 | 2 \, \omega |\eta'\rangle
     &=& 0.058~{\rm GeV}^3 \ .
\label{eq:APvalues}
\eeq
The ratio $A_{\eta'}/A_\eta$ can also be used to 
determine\cite{Feldmann:1998sh}
the ratios $R(\psi')=5.8$, analogously to
Eq.~(\ref{rjpsi}).  A recent measurement of the BES
collaboration\cite{Bai:1998ny}
yields $R(\psi')=2.9^{+5.4}_{-1.8}$.

\subsection{Mass formulas}

The anomaly equation~(\ref{eq:anomaly}) connects the masses
and decay constants of pseudoscalar mesons. Supplied with 
the ans\"atze for the decay constants~(\ref{eq:paraqs})
and the quark mass contributions~(\ref{eq:hpara}) 
one can  obtain several relations
that connect the masses of the physical states $\eta$ and $\eta'$
with the parameters in a given basis.

\subsubsection{$U(1)_A$ mass shift}

First, it is convenient to consider the trace of the physical
meson mass matrix
\beq
  M_\eta^2 + M_{\eta'}^2 &=& 2 \, M_K^2 + \sqrt3\,
 \frac{ \cos\theta_8 \, A_{\eta'} - \sin\theta_8 \, A_{\eta}}
       {f_0 \, \cos[\theta_8-\theta_0]} 
 \equiv 2 \, M_K^2 + M_{U(1)_A}^2 \ .
\label{eq:trace}
\eeq
Here I have defined the mass shift $M_{U(1)_A}$, which parametrizes
the deviation from the $U(1)_A$ symmetric world.
Note the similarity with the formula for
the two-photon decay widths~(\ref{eq:gammapred}): The matrix
elements $A_P$ are weighted with the elements of the {\em inverse
matrix}\/ $\{f^{-1}\}_P^0$. 
Using Eq.~(\ref{eq:trace}) and physical kaon and
$\eta,\eta'$ masses or using
the phenomenological values
for decay constants~(\ref{eq:parset}) and
gluonic matrix elements~(\ref{eq:APvalues}), respectively,
one obtains the value 850~MeV for $M_{U(1)_A}$.

The $U(1)_A$ mass shift has also been calculated on the lattice,
using quenched QCD with Wilson fermions for
a single massive flavor.\cite{Kuramashi:1994aj}
The values for $M_{U(1)_A}$ extracted from the $\eta'$ propagator
on the lattice show to a good approximation
a linear rise with decreasing quark mass $m$. 
Qualitatively,
this behavior is expected from the structure of the 
effective chiral Lagrangian (\ref{Leff0}) and (\ref{Leff1}),
namely $M_{U(1)_A}^2 \propto \tau_0/F_0^2$
with $F_0^2 = F_\pi^2 (1 + 2 (F_K^2/F_\pi^2-1) \, m/m_s )$,
where I have expressed the parameter $L_5$ in terms of
the kaon and pion decay constants and set $m_u=m_d=0$.
Extrapolating the lattice data
to the case of three massless flavors (using $M_{U(1)_A}^2 \propto N_F$) 
a value of $M_{U(1)_A}=751 \pm 39 $~MeV is found which is
somewhat smaller than the phenomenological value. 
The dependence of the singlet meson mass on the quark masses
is stronger than expected from $\chi$PT.
At an effective quark mass of $m_s/3$, which should be taken for
comparison with the real world,
the lattice simulation yields only $M_{U(1)_A} \simeq 650$~MeV.
Calculating on the other hand the mass shift from the
topological susceptibility,
see Eq.~(\ref{eq:tau0det}) below, a rather large value
$M_{U(1)_A}=1146 \pm 67 $~MeV is found on the lattice.
Another lattice calculation has used quenched and unquenched
staggered fermions.\cite{Venkataraman:1997xi}
In that analysis it has been shown that the $\eta'$ mass is
particularly sensitive to fermionic zero-modes, indicating the
strong connection with the topological properties of the theory.
The result for $M_{U(1)_A}$ in the limit of massless quarks
is found to be $876 \pm 16$~MeV.

\subsubsection{Dashen's theorem and GMO formula}

One is often interested in features of the
meson mass matrix in
the octet-singlet basis. However, 
the connection between the physical fields $P=\eta,\eta'$ and
the bare fields $\varphi^8$ and $\varphi^0$ in the effective
chiral Lagrangian is not simple.
One has to use Eq.~(\ref{Pphiconnection}) which
involves both mixing angles, $\theta_8$ and $\theta_0$.
Thus, instead of considering an octet-singlet mass matrix
one is led to considering the following product of 
decay constants and masses\cite{Leutwyler97,Moussallam:1995xp}
\beq
(f^T M^2  f)^{ab} \equiv \sum_P \,  f^a_P \, M_P^2 \, f_P^b 
\label{eq:fmf} \ .
\eeq
This result is equivalent to the ideas proposed by several
other groups\cite{Dmitrasinovic:1997te,Burakovsky:1998vc,Donoghue}
on the basis of a current algebra theorem
proposed by Dashen.\cite{Dashen:1971et}
With the correct treatment of the decay constants,
which is essential to obtaining an object 
with well-defined octet-singlet quantum numbers, Dashen's theorem
reads 
\beq
\sum_{P}   f_P^a \, M_P^2 \, f_P^b =
- \langle 0 | \left[ {\mathcal Q}_a^5, \left[ {\mathcal Q}_b^5, {\mathcal H}_{\chi SB}(0)
	\right]\right] | 0 \rangle \qquad (a,b = 1\ldots 8)
\eeq
where ${\mathcal Q}_a^5$ are the pseudoscalar charges and
${\mathcal H}_{\chi SB}$ the  chiral symmetry breaking part 
of the QCD Hamiltonian. 
I repeat that single \lq{}decay constants\rq{}
$f_\eta,f_{\eta'}$, which are sometimes used in this context,
have no process-independent interpretation.

In particular, one may consider
the diagonal elements of the
matrix in Eq.~(\ref{eq:fmf}).
The octet element does not receive contributions 
from the anomalous or OZI-rule violating terms
in the effective chiral Lagrangian and can be
expressed in terms of masses and decay constants of 
the pion and kaon only.
In the FKS scheme,
using Eqs.~(\ref{eq:paraqs}) and (\ref{eq:hpara}),
the result can be written as
\beq
\sum_P \, f_P^8 \, M_P^2 \, f_P^8 &=&
f_8^2 \, (\cos^2\theta_8 \, M_\eta^2 +\sin^2\theta_8 \, M_{\eta'}^2) 
\nonumber \\[0.3em]
&=&
\frac{f_q^2 \, M_\pi^2 +2 \, f_s^2 \, (2 M_K^2 - M_\pi^2)}{3} 
\label{gmotilde} \ .
\eeq 
This formula allows
to derive an improved Gell-Mann--Okubo (GMO) formula\cite{GMO}
with the angle $\theta_8$
playing a distinguished role.\cite{Feldmann:1998sh} 
In the usual form it reads
\beq
\cos^2\theta_8 \, M_\eta^2 +\sin^2\theta_8 \, M_{\eta'}^2
&\simeq& 
\frac{4 M_K^2 - M_\pi^2}{3} - 
 \Delta_{\rm GMO} \, \frac{M_K^2-M_\pi^2}{3}
\label{gmoexp} \ .
\eeq 
With the ansatz~(\ref{eq:hpara}) 
one obtains a contribution
$\Delta_{\rm GMO} = 4 \, (f_q^2-f_s^2)/(3\,f_8^2)$. 
It is to be stressed that the $SU(3)_F$
corrections in Eq.~(\ref{gmoexp}) enter in second
order, and thus additional corrections can be important,
e.g.\ from higher order contributions in $\chi$PT.\cite{GaLe85,LaPa74}
In any case, investigations of the GMO formula and its
corrections provide an estimate of the mixing parameter $\theta_8$.
Most of the analyses lead to values of
about $-20^\circ$, which is consistent with the number 
for $\theta_8$ in Eq.~(\ref{eq:parset}), 
 see also Table~\ref{tab:fPa}.

\subsubsection{Topological susceptibility}

In the FKS scheme
the singlet-singlet matrix element, as defined in Eq.~(\ref{eq:fmf}), reads
\beq
\sum_P f^0_P \, M_P^2 \, f_P^0
&=& f_0^2 \, (\sin^2\theta_0 \, M_\eta^2 + \cos^2\theta_0 \, M_{\eta'}^2)
\nonumber \\[0.3em]
&=&  \frac{2 f_q^2 \, M_\pi^2+f_s^2 \, (2 M_K^2-M_\pi^2)}{3}
    + \sqrt3 \, f_0 \left(\cos\theta_0 \, A_{\eta'} - \sin\theta_0 \, A_{\eta}
      \right) \ . \cr &&
\eeq
It has a contribution from finite quark masses
(i.e.\ $M_\pi,M_K \neq 0$), similar
as the octet-octet matrix element in
Eq.~(\ref{gmotilde}). Additional contributions appear due to the
$U(1)_A$ anomaly; comparison with Eq.~(\ref{Leff0}) 
yields an expression for
the topological susceptibility $\tau_0$
\beq 
\tau_0 &=& \frac{\sqrt3 \, f_0}{12}
     \left(\cos\theta_0 \, A_{\eta'} - \sin\theta_0 \, A_{\eta}
      \right) 
\label{eq:tau0det} \ .
\eeq
The value of $\tau_0$ following from the phenomenological values
for $f_0$, $\theta_0$ and $A_P$ comes out as ($192$~MeV)$^4$. 
Note that for $\theta_8\neq\theta_0$ the connection 
between the topological
susceptibility $\tau_0$ and the mass shift $M_{U(1)_A}$,
as defined in Eq.~(\ref{eq:trace}), is no longer
simple, $f_0^2 \, M_{U(1)_A}^2 \neq 12 \, \tau_0$.
The original Witten-Veneziano formula is recovered for
$\theta_0=\theta_8$.

\subsubsection{Mass matrix and Schwinger's formula}

One can consider a similar construction as Eq.~(\ref{gmotilde})
in the quark-flavor basis. 
Here, the situation is simplified if one adopts the
ansatz~(\ref{eq:paraqs}) for the
decay constants in the FKS scheme. In this case (and only then)
one can unambiguously define a mass matrix 
${\mathcal M}^2$ in the quark-flavor basis via
\beq
(f^T M^2 f)^{ij} &=& \sum_P \, f^i_P \, M_P^2 \, f_P^{j} 
\cr
&=&
\sum_P \, f_i  \, \big(U^\dagger(\phi)\big){}_P^i \, M_P^2 \, 
		\big(U(\phi)\big){}_P^j  \, f_j
\ \equiv \  f_i  \, ({\mathcal M}^2)^{ij}  \, f_j
\label{mqsdef} \ .
\eeq
The structure of this matrix follows solely from the
anomaly equation~(\ref{eq:anomaly}). Following the notation
of Ref.\cite{Feldmann:1998vh} one has
\beq
 {\mathcal M}^2
 &=& 
\left( \matrix{ m_{qq}^2 + 2 \, a^2 & \sqrt 2 \, y \, a^2\vspace{0.2em} \cr
                \sqrt 2 \, y \, a^2 & m_{ss}^2 + y^2 \, a^2 } \right)
\label{eq:mass}
\eeq
with $m_{qq}^2 \simeq M_\pi^2$, $m_{ss}^2\simeq 2 M_K^2 - M_\pi^2$,
$y=f_q/f_s$ and $a^2 = M_{U(1)_A}^2/(2+y^2)$.
The anomaly contribution $\propto a^2$ is
manifestly \lq{}non-democratic\rq{} due to the appearance
of the flavor symmetry breaking term $y$.
The consideration of the mass matrix in Eq.~(\ref{eq:mass})
turns out to be completely analogous to the 
analysis of the anomalous Ward identities performed earlier by
Diakonov/Eides.\cite{Diakonov:1981nv} 
This comes as no surprise since both methods rely on the
same assumption about OZI-rule violation and $SU(3)_F$ breaking.
(The correspondence between
notations used here and in Ref.\cite{Diakonov:1981nv} reads
$a^2 = \mu_1^2/2$ and $y=f_1/f_2$. Diakonov and Eides used
the estimates $f_q \simeq f_\pi$ and $f_s \simeq 2 f_K - f_\pi$
in order to obtain predictions for the various matrix elements
of $\eta$ and $\eta'$.) 
The diagonalization of the mass matrix~(\ref{eq:mass})
gives a theoretical estimate of the mixing angle $\phi$, which
comes out as about $42^\circ$, see Refs.\cite{Diakonov:1981nv,Feldmann:1998vh}
and Table~\ref{tab:fPi}.
Within the uncertainties this value 
is compatible with the phenomenological value quoted in Eq.~(\ref{eq:parset}).

The matrix~(\ref{eq:mass}) can also be used to obtain an improved
version of Schwinger's mass formula\cite{Schwinger:1964}.
It is most easily derived by considering the trace and the determinant of
the mass matrix in the quark-flavor basis and in the physical one
and solving for $M_\eta'$. A crucial point is that 
the two matrices have to be connected by a simple rotation which
is approximately true for the quark-flavor basis (\ref{mqsdef}) but
not for the octet-singlet one. This yields\footnote{A 
variant of Schwinger's formula has recently  been discussed
by Burakovsky and Goldman.\cite{Burakovsky:1998vc} It differs
from the one presented in Eq.~(\ref{Schwinger}) due to the usage of a 
simplified treatment of the decay constants.}
\beq
  M_{\eta'}^2 &=& M_\pi^2 + \frac{(M_K^2-M_\pi^2) \, 
			(2 M_K^2 - M_\eta^2-M_\pi^2)}{
			M_K^2 - (2+y^2) \, M_\eta^2/4 -
				(2-y^2) \, M_\pi^2/4} \ .
\label{Schwinger}
\eeq
Schwinger's original formula is recovered in the limit $y\to 1$.
It has also been re-derived in another context 
by Veneziano.\cite{Veneziano:1979ec}.
It is noticeable that,
in the latter analysis, the anomalous mass contribution can be
expressed in terms of the Veneziano
ghost coupling constant, $a^2=\lambda_\eta^2/N_C$.  
Schwinger used his formula to predict the mass of the $\eta'$ meson.
For $y=1$ he obtains a too high value of about 1600~MeV.
As is obvious from Eq.~(\ref{Schwinger}) the formula is
very sensitive to deviations of the flavor symmetry breaking parameter
$y$ from unity since $y^2$ enters in the
difference of two terms in the denominator. Indeed, for values of $y$ about
0.8, following from the phenomenological determination of the
decay constants~(\ref{eq:parset}), 
the correct value of the $\eta'$ mass is found.

\subsection{Pseudoscalar coupling constants of the nucleon}

The coupling constants of the nucleon with the light
pseudoscalar mesons $\pi^0$, $\eta$ and $\eta'$ are basic ingredients
for the low-energy description of hadronic physics, especially
for the description of nucleon-nucleon scattering data.
A phenomenological determination of the coupling constants can
be achieved via
dispersion relations (see e.g.\ Grein/Kroll\cite{Grein:1980nw}),
potential models (see e.g.\ Nagels et al.\cite{Nagels:1978ze}),
or effective low-energy Lagrangians 
of the nucleon (see e.g.\ Stoks/Rijken\cite{Stoks:1996yj}).
While the value for $g_{\pi NN}\simeq 13$ is well-established,
the values for $g_{\eta NN}$ and $g_{\eta' NN}$ are not so
well-known, see e.g.\ Dumbrajs et al.\cite{Dumbrajs:1983jd}
and references therein. 

An alternative way to estimate the couplings $g_{PNN}$ is to
relate them to the axial-vector coupling constants of the nucleon
which are defined as
\beq
  \langle  N(p,s)| J_{\mu5}^a | N(p,s)\rangle &=& G_A^a \, s_\mu 
\eeq
where $s_\mu$ is the spin-vector of the nucleon.
A different normalization convention, often found in the literature,
is given by $a_3=\sqrt2 \, G_A^3$, $a_8=\sqrt 6 \, G_A^8$,
$a_0=\sqrt3 \, G_A^0$.
The parameter $a_3$ can be determined from neutron $\beta$\/-decay
by using isospin symmetry. This yields\cite{PDG98} 
$a_3=1.267 \pm 0.0035$.
Conventionally, one rewrites $a_3$ and $a_8$ in terms of the
coupling constants $D$ and $F$, namely $a_3=F+D$ and
$a_8=3F-D$. The ratio $F/D=0.575 \pm 0.016$ is determined
from the hyperon $\beta$\/-decays and
$SU(3)_F$ flavor symmetry relations.\cite{Close:1993mv}
Together with the phenomenological
value of $a_3$ one obtains $a_8 = 0.58\pm 0.03$.
Finally, the singlet axial charge $a_0$ can be extracted from a
measurement of the first moment of the structure function
$g_1^p(x,Q^2)$ measured in polarized lepton-nucleon scattering,
\beq
\Gamma_1^p(Q^2)= \int dx \, g_1^p(x,Q^2) &=& 
\frac{C_1^{NS}}{12} \left[a_3 + \frac{a_8}{3}\right]
+\frac{C_1^{S}}{9} a_0(Q^2) \ ,
\eeq
where $C_1^{(N)S}$ are known coefficients, calculable in perturbation
theory.
The latest analysis of the Spin-Muon Collaboration\cite{SMC97} (SMC) yields
$a_0(5~{\rm GeV}^2) = 0.28 \pm 0.17$. 
The axial charges have  recently been determined from
a lattice simulation with dynamical Wilson fermions\cite{Guesken97}
which leads to slightly smaller values for $a_3$, $a_8$ and $a_0$.

In the parton model
the axial charges can be expressed in terms of
the integrated polarized parton distributions
\beq
a_3 &=&\Delta u - \Delta d \ , \nonumber\\[0.1em]
a_8 &=&\Delta u + \Delta d - 2\Delta s \ , \nonumber\\[0.1em]
a_0(Q^2) &=&\Delta u + \Delta d +\Delta s - 
  3 \, \frac{\alpha_s}{2\pi} \, \Delta g(Q^2) \ .
\label{eq:polpd}
\eeq
The appearance of the polarized gluon distribution 
in the expression for the 
nucleon's singlet axial charge  is a consequence of the
$U(1)_A$ anomaly.\cite{anomalyspin}
In the Adler-Bardeen scheme\cite{Adler:1969er}
the integrated polarized quark distributions $\Delta q$
are scale-indepen\-dent.
The smallness of the singlet axial charge $a_0$ compared to $a_8$,
which is the origin of the nucleon {\em spin puzzle}, can be
explained by a substantial (negative) contribution of the
polarized gluon distribution, $\Delta g$. 

Let me return to the pseudoscalar coupling constants
of the nucleon.
For the pion one has the well-known Goldberger-Treiman (GT) relation
\beq
  2 \, m_N \, G_a^3 &=& f_\pi \, g_{\pi N N}
\eeq
which has been successfully tested against phenomenology.
Its derivation is based on the usual PCAC assumptions.
The obvious generalization for $G_A^8$ and $G_A^0$ reads
\beq
   2 \, m_N \, G_A^a &=& \sum_{P=\eta,\eta'}
\,  f_P^a \, g_{PNN} \ . \qquad (a=8,0)
\label{eq:GTansatz}
\eeq
Here I have only taken into account the contributions from the
{\em physical}\/ $\eta$ and $\eta'$ states.
Note that
both, $G_A^0$ and $f_P^0$ have 
the same anomalous dimension $\gamma_A$ determined by
the renormalization of the singlet axial-vector current (\ref{eq:f0scale}). 
Consequently, the quantities $g_{PNN}$ in Eq.~(\ref{eq:GTansatz})
are scale-independent as
they should be. The introduction of an additional OZI-rule violating
parameter besides $\Lambda_1$ (present in $f_P^0$) is not mandatory
for a consistent behavior of the singlet couplings under
renormalization. However, higher excited pseudoscalar states 
or glueballs can be included, in principle, but
-- following the FKS scheme -- are assumed to be negligible
in Eq.~(\ref{eq:GTansatz}).

In the literature
the GT relation in the singlet channel is often formulated
in a different way. Shore and Veneziano\cite{Shore:1990zu}
(see also Ref.\cite{Narison:1998aq})
established a two-component
description of the singlet axial-charge which has been investigated in
a number of phenomenological 
applications (e.g.\ Refs.\cite{Schechter:1993iz,Cheng:1996ri};
see, however, also Refs.\cite{Efremov:1991nj,Bass:1999is})
In this picture, the singlet GT relation is modified by an additional
direct coupling of the Veneziano-ghost (more precisely the
operator $G \tilde G$) to the nucleon ($g_{\tilde GNN}$).
Neglecting $\eta$-$\eta'$ mixing for the moment, this yields
\beq
  2 \, m_N \, G_A^0 &=& \tilde f \, g_{\eta' NN} 
		+ \frac{ \tilde f^2 \, M_{\eta'}^2 
		\, g_{\tilde G N N}}{\sqrt 3}
\label{eq:GTansatz2} \ .
\eeq
It is to be stressed that
the quantity $\tilde f$ does not coincide with the
decay constant $f_0$. It is defined via the 
pseudoscalar singlet current in such a way that
it would coincide with $f_\pi$ if the $U(1)_A$ anomaly were
turned off, and it is scale-independent.
The correct behavior of $G_A^0$ 
under renormalization requires a non-trivial
scale-dependence of the coupling $g_{\tilde GNN}$.  
It has also been shown in the analyses of Shore and Veneziano
that the singlet axial charge $a_0$
can be written as the product of the
first moment of the topological susceptibility
and the vertex function of the would-be singlet Goldstone boson
$\tilde \eta_0$. Let me investigate the
phenomenological consequences of the two alternatives
(\ref{eq:GTansatz}) and (\ref{eq:GTansatz2}), respectively.

Making use of the GT relation~(\ref{eq:GTansatz}) and inserting the 
phenomenological values for the meson decay constants~(\ref{eq:parset}),
one finds the following connection
between axial charges and pseudoscalar coupling constants of the nucleon
\beq
\begin{array}{rcl}
     a_3 = \sqrt 2 \, G_A^3 &=& 1.267 \pm 0.004 \ \
\end{array}
 &\Rightarrow  &
	\ \ \ \, g_{\pi NN\phantom{{}'}}= 12.86 \pm 0.06 \ ,\cr
\left.
\begin{array}{rcl}          &\downarrow&  F/D \cr
     a_8 = \sqrt6 \, G_A^8  &=& 0.58 \pm 0.03 \cr
                            &\downarrow& {\rm  SMC} \cr
     a_0 = \sqrt3 \, G_A^0  &=& 0.28 \pm 0.17 
\end{array}
\ \ \right\}
 &  \Rightarrow  &
  \Bigg\{
  \begin{array}{l}
    g_{\eta NN\phantom{{}'}} = 3.4 \pm 0.5  \ ,
\cr
    g_{\eta'NN} = 1.4 \pm 1.1  \ .
  \end{array} \ 
\label{gPNN}
\eeq 
The errors take into account the uncertainties w.r.t.\
the axial charges and the meson decay constants.
This may be compared to the
$SU(3)_F$ symmetry 
prediction,\cite{Dumbrajs:1983jd}
$g_{\eta_8 NN}= g_{\pi NN} \, (3-4 \alpha)/\sqrt3 \simeq 3.4$ 
with $\alpha=D/(D+F)=0.635$. 
The value of $g_{\eta NN}$ in Eq.~(\ref{gPNN}) 
already saturates the bound obtained
from the analysis of $NN$ forward-scattering data
by Kroll/Grein,\cite{Grein:1980nw}
$g_{\eta(\eta') NN} \leq 3.5$. 
In this case, 
one would expect $g_{\eta' NN}$  to be
close to zero which is in accord 
with the value quoted in Eq.~(\ref{gPNN}).
Recently, from a measurement of $\eta'$ production in proton-proton
collisions close to threshold at COSY\cite{Moskal:1998pc} the
bound $g_{\eta'NN} \leq 2.5$ has been deduced.
On the other hand, fits of $NN$ data using potential models
occasionally lead to significantly larger
values for $g_{\eta NN}$ and $g_{\eta' NN}$
(see Dumbrajs et al.\cite{Dumbrajs:1983jd}
and references therein).

It is  illustrative to rewrite the solution for
the pseudoscalar
couplings of the nucleon from Eqs.~(\ref{eq:GTansatz})
and (\ref{eq:polpd}) as follows,
\beq
  \frac{g_{\pi NN}}{2m_N} &=& \, \frac{1}{f_\pi} \,  
		\, \frac{\Delta u - \Delta d}{\sqrt2}\ , \nonumber\\[0.2em]
  \frac{g_{\eta NN}}{2 m_N} &=&
  \frac{\cos\phi}{f_q} \, \frac{\Delta u + \Delta d}{\sqrt2}
  - \frac{\sin\phi}{f_s} \, \Delta s + 
     \frac{\alpha_s}{2 \pi} \, \Delta g \,
	\frac{\sqrt3 \, \sin\theta_8}{ f_0 \, \cos[\theta_8-\theta_0]}
\ , \nonumber \\[0.2em]
 \frac{g_{\eta' NN}}{2 m_N} &=&
  \frac{\sin\phi}{f_q} \, \frac{\Delta u + \Delta d}{\sqrt2}
  + \frac{\cos\phi}{f_s} \, \Delta s   
   - \frac{\alpha_s }{2 \pi} \, \Delta g \, 
	\frac{ \sqrt3 \, \cos\theta_8}{ f_0 \, \cos[\theta_8-\theta_0]}
\ ,
\label{eq:getapNN}
\eeq
where I have used the FKS scheme to express the decay constants
in terms of the mixing parameters.
The first terms on the r.h.s.\ in Eq.~(\ref{eq:getapNN})
are the quark contributions which
have the form of the standard GT relation.
The ratio of the additional gluon contribution
to $g_{\eta'NN}$ and $g_{\eta NN}$ is given by $-\cot\theta_8$, the same
ratio as e.g.\ found in the ratio of $J/\psi \to P\gamma$
decay amplitudes.
In the picture based on the GT relation~(\ref{eq:GTansatz})
the effect of $\Delta g$ is thus to reduce both, the singlet axial charge
$a_0$ and the pseudoscalar coupling $g_{\eta' NN}$ compared to their
nonet symmetry values.

In case of the GT relation~(\ref{eq:GTansatz2}) one 
usually assumes that the coupling $g_{\eta' NN}$ is related
to the polarized quark distributions and $g_{\tilde G NN}$ to the
polarized gluon distribution, respectively.\footnote{This
is to be confronted with Eq.~(\ref{eq:GTansatz}) 
where the contribution from the Veneziano ghost is present in
$\Delta q$ and $\Delta g$, separately, but is assumed to cancel
in the sum for the scheme-independent quantity $G_A^0$.} \ 
This results in a similar
representation as in Eq.~(\ref{gPNN}) but without
the $\Delta g$ contributions and $f_q,f_s$ replaced by $\tilde f$.
If one assumes $\tilde f \simeq f_\pi$, 
the extracted value for $g_{\eta'NN}$ in
this picture comes out larger and close to its nonet symmetry
value. For example, Cheng\cite{Cheng:1996ri} obtained 
$g_{\eta' NN}=4.7$ which happens to be closer to the
value obtained in potential models but violates the bounds
of Ref.\cite{Grein:1980nw}
The answer to the {\em spin puzzle}\/ on the basis of the
GT relation~(\ref{eq:GTansatz2}) remains unchanged, namely  that $a_0$ is small
due to the additional nucleon-ghost coupling which is related
to $\Delta g$.

In summary, both alternatives (\ref{eq:GTansatz}) and
(\ref{eq:GTansatz2}) give a consistent description
of the axial charges of the nucleon, but with significantly
different values of the coupling constant $g_{\eta' NN}$.
In any case, the GT relations rely on
the  PCAC hypothesis. The corresponding
uncertainties are
on similar footing as the ones for the chiral anomaly prediction of
the $P \to \gamma\gamma$ decay widths, and the same attention
concerning corrections is to be paid, see the comments after
Eq.~(\ref{eq:gammapred}).

\subsection{Summary of $\eta$-$\eta'$ mixing parameters}

$\eta$-$\eta'$ mixing can be described
by different mixing angles. Let me briefly summarize  
their definition, their determination from experiment
and their relations among each other:
\begin{description}
\item[\rm $\theta_8$ and $\theta_0$: ]
  These mixing parameters
  are defined as the ratio of the octet and singlet
        {\em decay constants}\/ of
	$\eta$ and $\eta'$  mesons through
        axial-vector currents (\ref{eq:81angledef}).
  The decay constants 
  enter the decays 
  $P\to \gamma\gamma$, see Eq.~(\ref{eq:gammapred}),
  as well as 
    the $P\gamma$ transition form factors
  at large momentum transfer,
  see Eq.~(\ref{eq:FPgamma}).
  They also connect the bare fields in the
	chiral effective Lagrangian with the physical 
        ones~(\ref{Pphiconnection}).
  Furthermore, the angle $\theta_8$ enters the (improved)
  Gell-Mann--Okubo formula (\ref{gmotilde}). 

\item[$\theta_y$: ] This angle measures the ratio of 
	matrix elements of the topological charge density $\omega$
        with $\eta$ and $\eta'$ mesons.
        It can easily be extracted from the radiative
        $J/\psi$ decays (\ref{rjpsi}).

\item[$\phi$: ] Up to OZI-rule violating corrections
        of order $1/N_c$, this is the universal mixing
	angle that parametrizes ratios of matrix elements 
        of quark currents with light ($u,d$) or strange quarks
        in the FKS scheme,  see Eqs.~(\ref{eq:paraqs}) and
        (\ref{eq:hpara}).
        There are several reactions
	where this angle can be probed, see Table~\ref{fig:phi}.
        It can also be estimated from the diagonalization of
        the mass matrix in
	the quark-flavor basis~(\ref{eq:mass}).
\end{description}
From a combined expansion in small momenta and masses and
in powers of $1/N_C$ in the framework of $\chi$PT one obtains the
relation~(\ref{eq:t8t1rel}) which predicts the difference
between the angles $\theta_8$ and $\theta_0$.
In the FKS scheme an analysis of the anomaly equation 
provides another important relation~(\ref{eq:anglerel})
which connects the three angles $\theta_y$, $\theta_8$ and
$\phi$ in the limit $m_u,m_d \to 0$.
The independent determination of the mixing parameters from
phenomenology and
theory\cite{Diakonov:1981nv,Leutwyler97,Feldmann:1998vh} 
supports the validity of Eqs.~(\ref{eq:t8t1rel})
and (\ref{eq:anglerel}) and the internal
consistency of the mixing approach.

The reader may wonder why I have not discussed the value
of {\em the}\/ mixing angle $\theta_P$ between octet
and singlet states. I stress again that the correct 
representation of the physical fields $\eta$ and $\eta'$ 
in terms of bare octet and singlet fields is given
by Eq.~(\ref{Pphiconnection}) which cannot be written as
a simple rotation, once the next-to-leading order in
the chiral effective Lagrangian~(\ref{Leff1}) is taken
into account. Thus the usage of $\theta_P$ is restricted
to the leading order Lagrangian~(\ref{Leff0}) which gives
only a poor approximation to the real world. 
Nevertheless, one can define {\em approximate}\/ octet
and singlet fields by {\em requiring}\/ a simple connection 
with the physical fields via a rotation matrix $U(\theta_P)$. 
Let me for illustration make the conventional choice
\beq
\theta_P &:=& \phi - \arctan\sqrt2 \ .
\label{thetaP}
\eeq 
The mixing
angles $\theta_8$ and $\theta_0$ can then be expanded
as
\beq
\theta_{8,0} &=& \theta_P \mp 
\frac{\sqrt2}{3} \, \left(1-y\right)
+ {\mathcal O}(1-y)^2 \ .
\eeq
Thus, in principle, all the results can alternatively be
written in terms of a single (approximate)
octet-singlet mixing angle $\theta_P
\simeq (\theta_8+\theta_0)/2$
and explicit $SU(3)_F$ symmetry corrections proportional to $(1-y)$.
Such a procedure has been suggested by Benayoun et al.\cite{Benayoun:1999fv}
In that approach the transition from pure (bare) octet and singlet
fields to the physical ones is accomplished by a non-diagonal
matrix which results from a renormalization of the meson fields
in an effective Lagrangian and yields an analogous relation to
Eq.~(\ref{Pphiconnection}). 

The  {\em effective}\/
octet and singlet states defined implicitly via
Eq.~(\ref{thetaP}) have the following Fock state 
expansion\cite{Feldmann:1998sh}
\beq 
|\eta_8 \rangle  &=&
 \frac{\Psi_q + 2 \, \Psi_s}{3} \, 
 \frac{|u\bar u + d\bar d - 2 s\bar s\rangle}{\sqrt6} 
+ \frac{\sqrt2 \, (\Psi_q - \Psi_s)}{3} \,
 \frac{|u\bar u + d\bar d + s\bar s\rangle }{\sqrt3} 
+ \ldots
\nonumber \\[0.2em] 
|\eta_0 \rangle &=& 
 \frac{\sqrt2 \, (\Psi_q - \Psi_s)}{3} \, 
 \frac{|u\bar u + d\bar d - 2 s\bar s\rangle}{\sqrt6} 
+ \frac{2 \,\Psi_q + \Psi_s}{3} \, 
 \frac{|u\bar u + d\bar d + s\bar s\rangle}{\sqrt3} 
+ \ldots \cr &&
\label{81def}
\eeq
which is to be confronted with Eq.~(\ref{eq:Fockqs}).

In any case, the value of $\theta_P$ alone is not sufficient
to describe $\eta$-$\eta'$ mixing.
Since the angles $\theta_8$, $\theta_0$, $\theta_y$ and $\phi$ can be 
obtained by simple ratios of $\eta$ and $\eta'$ observables
and obey the useful relation~(\ref{eq:anglerel}) I
prefer to present the results in terms of these angles.

\section{Mixing with other states}

\subsection{Mixing in the $\pi^0$-$\eta$-$\eta'$ system}

In the real world isospin is not an exact symmetry, and thus
$\pi^0$ is to be viewed as a mixture of  
pure isospin-triplet and -singlet components.
Of course, the magnitude of isospin-symmetry violation which
is due to the differences in the
masses of up- and down-quarks
is small, and we should not expect the same order of accuracy
for phenomenological predictions as in the $\eta$-$\eta'$ case.
Furthermore, the comparison with experimental data is more
difficult due to the interference with
isospin-violating effects from 
the electromagnetic charges which can be important, too.

Nevertheless, it is a straightforward task to generalize the
results of Section~3 and consider mixing
in the $\pi^0$-$\eta$-$\eta'$ system. 
As already mentioned, the deviations
of $\pi^0$ from a pure isospin-triplet (which I denote
as $\varphi_3$) are small and the related mixing angles, $\epsilon$
and $\epsilon'$,
can be treated as an expansion parameter. 
Moreover, a possible difference in basic 
decay constants for up- and down-quark states should be
negligible. Therefore I consider
\beq
  |\pi^0\rangle &=& |\varphi_3\rangle +
	\epsilon \, |\eta\rangle +
        \epsilon' \, |\eta'\rangle
\label{eq:pietaetap}
\eeq
where $\epsilon$ and $\epsilon'$ parametrize the $\eta$ and
$\eta'$ admixtures in the pion.
The mixing parameters $\epsilon$ and $\epsilon'$ receive contributions
from the $U(1)_A$ anomaly very similarly as the mixing angles
in  the $\eta$-$\eta'$ sub-system. An analysis within the 
FKS scheme  leads to the following estimate,
\beq
 &&
\epsilon  = \cos\phi \, \frac{m_{dd}^2 - m_{uu}^2}
                             {2 \, (M_\eta^2 - M_\pi^2)} 
\ , \qquad 
\epsilon'  = \sin\phi \, \frac{m_{dd}^2 - m_{uu}^2}
                             {2 \, (M_{\eta'}^2 - M_\pi^2)}
\label{epsepsp} \ , 
\eeq
 where the difference $m_{dd}^2 - m_{uu}^2$ is defined
in an analogous way as in Eq.~(\ref{eq:mass}) and
can be estimated from
$2 (M_{K^0}^2 - M_{K^\pm}^2 + M_{\pi^\pm}^2 - M_{\pi^0}^2)$ to amount
to $0.0104$~GeV$^2$. In the latter formula, the specific
combination of meson masses guarantees, that the leading order
contributions from electromagnetic self-energy corrections drop out.
Inserting the phenomenological number for the mixing angle
$\phi$ (\ref{eq:parset}),
one obtains $\epsilon=0.014$ and $\epsilon'=0.0037$.
Similar values have been obtained by Chao et al.\cite{Chao:1989yp}
The values of $\epsilon$ and $\epsilon'$ are needed, for instance,
for the description of the decay $\eta \to 3 \pi$ (see e.g.\ Ref.\cite{Leutwyler96}). Gardner\cite{Gardner:1998gz} has recently emphasized the
importance of $\pi^0$-$\eta$-$\eta'$ mixing in connection with
the determination of the CKM-matrix elements via 
unitarity triangles from $B$ decays into
light pseudoscalar mesons.

It is instructive to consider the ratio $\epsilon'/\epsilon$
and to express it in the following way, 
using Eq.~(\ref{eq:anglerel}), 
\beq
\frac{\epsilon'}{\epsilon} 
&=& 
- \tan\theta_8 \, \tan^2\phi
\label{epspeps} \ .
\eeq
This formula would coincide with the result obtained earlier
in the conventional octet-singlet scheme\cite{Leutwyler96}
if one identified $\theta_8  \to \theta_P = \phi - \arctan\sqrt2$.
However, as we learned, this relation is significantly 
spoiled by the $SU(3)_F$ corrections arising from $f_q/f_s\neq 1$,
which are also relevant in Eq.~(\ref{epspeps}).

It is also possible to estimate the matrix elements of the
pion with the topological charge density in the anomaly equation~(\ref{eq:anomaly}),
\beq
\langle 0 | \omega|\pi^0\rangle
&=&
\epsilon \ \langle 0 | \omega|\eta\rangle
+
\epsilon' \ \langle 0 | \omega | \eta'\rangle
 \nonumber \\[0.2em]
&=& \frac{1}{\cos\phi} \, \frac{m_{dd}^2 - m_{uu}^2}
                             {2 \, M_\eta^2} \,
   \langle 0 | \omega|\eta\rangle
\label{GGpi} \ ,
\eeq
where I have used Eq.~(\ref{eq:anglerel}), and $M_\pi^2$
has been neglected compared to $M_\eta^2$ for simplicity.
The ratio $\langle 0|\omega|\pi^0\rangle/
\langle 0|\omega|\eta\rangle$ has also been derived
in a leading order approach
by Leutwyler\cite{Leutwyler96}.
In that work $\eta$-$\eta'$ 
mixing has been neglected completely.
This corresponds to
taking 
$\cos\phi\simeq\cos\arctan\sqrt2=1/\sqrt3$ and 
$M_\eta^2\simeq 4/3 M_K^2 \simeq 4/3 m_s B$ in
Eq.~(\ref{GGpi}), which results in
\beq
r = \frac{\langle 0 | \omega|\pi^0\rangle}
{\langle 0 | \omega|\eta\rangle}
&\simeq&
\frac{ 3 \sqrt3}{4} \, \frac{m_d-m_u}{m_s}
\label{GGpiestimate} \ .
\eeq
This ratio is frequently 
discussed\cite{Leutwyler:1996qg,Donoghue:1992ac} in
connection with a determination of light quark mass ratios 
from the decays $\psi'\to J/\psi\,\pi^0$ and $\psi'\to J/\psi\,\eta$,
which are assumed to be dominated by the gluonic matrix elements
in a similar way as the radiative $J/\psi$ decays discussed 
around Eq.~(\ref{rjpsi})
\beq
  \frac{\Gamma[\psi'\to J/\psi\,\pi^0]}
       {\Gamma[\psi'\to J/\psi\,\eta ]} &\simeq &
	|r|^2 \, \frac{k_\pi^3}{k_\eta^3} \ .
\eeq
Taking the experimental result for the decay width\cite{PDG98},
one obtains $r=0.043 \pm 0.006$ from which
Donoghue and Wyler\cite{Donoghue:1992ac} obtain an estimate for the
quark mass ratio $(m_d-m_u)/m_s = 0.036 \pm 0.009$.
Leutwyler,\cite{Leutwyler:1996qg}
on the other hand, estimated the quark mass ratio
from other processes, $(m_d-m_u)/m_s = 0.025$, and predicted
$r = 0.032$.  Formula~(\ref{GGpi}) leads to an
even smaller result $r=0.022$. 
One has to conclude\cite{Leutwyler:1996qg}
that some (higher-order or electro-magnetic)
corrections to either Eq.~(\ref{GGpi}) or the decay mechanism
for $\psi'\to J/\psi\,\pi^0$ are still not under control.\footnote{
Note that $e^4$ and $(m_d-m_u)/m_s$ are both of the order
of a few percent only; thus a contribution to $\psi'\to J/\psi\,\pi^0$
via two photons may be of similar size as 
the isospin-suppressed gluonic mechanism.}

\subsection{Heavy quark admixtures in light pseudoscalar mesons}

Initiated by the CLEO measurement\cite{Behrens:1998dn}\
of an unexpectedly large
branching fraction of $B\to K \eta'$, the subject of heavy
quark components in light pseudoscalar mesons has recently regained 
interest.
Early investigations of this subject already date back to
1976/77, when Kramer et al.\cite{Kazi:1976tu} and
independently Fritzsch and Jackson\cite{Fritzsch}
discussed mixing of light
and heavy pseudoscalar mesons in the context of radiative $J/\psi$
decays. A rather exhaustive phenomenological analysis of the mixing
parameters in the $\eta$-$\eta'$-$\eta_c$ system
has also been performed by Chao.\cite{Chao:1989yp}
In all cases, a rather small admixture of heavy quarks in
light pseudoscalar mesons has been inferred.
On the other hand, more recently,
Halperin/Zhitnitsky\cite{HaZh97} and Cheng/Tseng\cite{ChTs97}\
proposed a prominent intrinsic charm component in $\eta'$ to explain
the $B\to K \eta'$ decay width. 
Their results have been criticized for both
theoretical
and phenomeno\-logical
reasons.\cite{FeKr97b,Feldmann:1998vh,Ali97,Franz:1998hw}
Also, it has been realized that a modification of the
effective parameters in the factorization approach
combined with a variation of the $B \to \eta'$ form factor
and the usage of the improved
$\eta$-$\eta'$ mixing parameters (\ref{eq:parset})
can easily increase the theoretical prediction for
BR$[B\to K\eta']$ by a factor 2 to 3, see the recent 
report of Cheng et al.\cite{Chen:1999nx} and references therein.
This means that a large intrinsic charm component in $\eta'$
is not really needed to explain the data.

Let me present the theoretical arguments in favor of a
small charm admixture in $\eta$ and $\eta'$.
In the FKS scheme
the analysis of the anomaly equation~(\ref{eq:anomaly})
can easily be extended to the charmed axial-vector current
\beq
  \partial^\mu \bar c \, \gamma_\mu\gamma_5 \, c &=& 
	2 m_c \,\bar c \, i\gamma_5 \, c
 + 2 \, \omega \ .
\label{eq:charmanomaly}
\eeq
In the following it is convenient to treat $1/m_c$ as a small
parameter and expand all quantities to first non-trivial order
in this parameter.
Taking matrix element with light pseudoscalar mesons,
$P=\eta,\eta'$,  one obtains
\begin{equation}\matrix{
M_P^2 \, f_P^c &=& h_P^c &+& A_P \cr
\uparrow && \uparrow && \uparrow \cr
{\mathcal O}(1/m_c^2) && 
{\mathcal O}(1) &&{\mathcal O}(1)}
\label{anomal1}
\end{equation}
where the notation of Section~3, see Eqs.~(\ref{eq:hPi})
and (\ref{eq:AP}), has been generalized to include
three independent flavor combinations $i=q,s,c$.
From Eq.~(\ref{anomal1}) 
one immediately obtains an estimate for the parameters 
$h_P^c$ which provides a measure for the $c\bar c$ admixture
in $\eta$ or $\eta'$
\begin{equation}
h_P^c = - A_P (1 + {\mathcal O}(1/m_c^2))  \ .
\label{thetactilde}
\end{equation}
In particular, Eq.~(\ref{eq:anglerel}) fixes the ratio
$h_{\eta'}^c/h_{\eta}^c = A_{\eta'}/A_{\eta'} =
- \cot\theta_8$. It is the same quantity that enters
the ratio of radiative $J/\psi$ decays~(\ref{rjpsi}). Thus the anomaly
picture of Novikov et al.\cite{Novikov:1980uy}, 
see Eq.~(\ref{rjpsi}) and the intrinsic charm
picture
give an equivalent description of these decays.
It is convenient to parametrize the quantities $h_P^c$ as follows
\beq
 h_{\eta\phantom{{}'}}^c &=& - \theta_c \, \sin\theta_8 \, f_{\eta_c} 
				\, M_{\eta_c}^2 \ , \cr
 h_{\eta'}^c            &=& \phantom{-} \theta_c \, \cos\theta_8 \, f_{\eta_c} 
				\, M_{\eta_c}^2
\label{eq:thetac} \ .
\eeq
Taking the values of $A_P$ quoted in Eq.~(\ref{eq:APvalues})
and $f_{\eta_c} \simeq f_{J/\Psi} = 410$~MeV, one 
estimates $\theta_c\simeq -1.0^\circ$.
This number is reasonably small, and it can be tested by 
comparing the radiative $J/\psi$ decays to, say, $\eta'$ or
$\eta_c$ mesons 
\begin{eqnarray}
\frac{\Gamma[J/\psi \to \eta'\gamma]}{\Gamma[J/\psi\to \eta_c\gamma]}
&=& \theta_c^2 \, \cos^2\theta_8 \,
\left(\frac{k_{\eta'}}{k_{\eta_c}}\right)^3
\label{etacetap} \ .
\end{eqnarray}
The experimental number for this ratio, $0.33 \pm 0.10$,\cite{PDG98}
corresponds to $|\theta_c\cos\theta_8|=0.014 \pm 0.002$, which is
in good agreement with the theoretical estimate
(FKS: $\theta_c\cos\theta_8 =- 0.016$). 
Chao et al.\cite{Chao:1989yp} argue that also higher 
excitations of the $\eta_c$ meson (e.g.\ the $\eta'_c$)
should be taken into account. In order to quantify the effect
a certain hierarchy of the related mixing angles has to
be assumed. In total it leads to a slightly smaller value for the
mixing angle $\theta_c$ which is still compatible with the radiative
$J/\psi$ decays, (Chao: $\theta_c\cos\theta_8=-0.012$).

There is another parameter which can be used to quantify the charm
component of a light pseudoscalar, namely the decay constant
$f_P^c$. However, we observe that 
it enters Eq.~(\ref{anomal1}) only at sub-leading order
in the $1/m_c$ expansion, which
makes its determination non-trivial.
A simple way out is to assume the same mixing behavior as it has
been proven successful in the light meson sector.\cite{Feldmann:1998vh}
Following Eq.~(\ref{eq:hpara}), 
one defines an extended mixing matrix via
\beq
\left( \matrix{ 
 |\eta\phantom{{}'}\rangle \cr
 |\eta'\rangle \cr
 |\eta_c\rangle
 }\right)
&=& 
\left( \begin{array}{ccc}
\cos \phi & - \sin\phi & - \theta_{c} \sin\theta_y \\
\sin \phi & \phantom{-{}}\cos\phi 
& \phantom{-{}} \theta_{c}  \cos\theta_y \\
- \theta_{c} \sin(\phi-\theta_y) &
- \theta_{c}  \cos(\phi-\theta_y) & 1 
\end{array} \right)
\left( \matrix{ 
 |\eta_q\rangle \cr 
 |\eta_s\rangle \cr
 |\eta_c^0\rangle
 }\right)
\nonumber \\[0.3em]
&=& 
\ U(\phi,\theta_y,\theta_c) \
\left( \matrix{ 
 |\eta_q\rangle \cr 
 |\eta_s\rangle \cr
 |\eta_c^0\rangle
 }
\right)
\label{eq:stateqsc}
\eeq
with $|\eta_c^0\rangle = \Psi_c(x,{\mathbf k_\perp}) \, |c \bar c\rangle +
\ldots$ having no light quark valence components.
The ansatz $\{f_P^i\} = U(\phi,\theta_y,\theta_c) \, {\rm diag}
[f_q,f_s,f_{\eta_c}]$ and $\{h_P^i\} = U(\phi,\theta_y,\theta_c) \, {\rm diag}
[h_q,h_s,h_c]$ 
with $h_c = f_{\eta_c} \, M_{\eta_c}^2$ yields
\beq
&& f_{\eta\phantom{{}'}}^c = -\theta_c \, \sin\theta_8 \, f_{\eta_c} 
	\simeq -2.4~{\rm MeV} \ , \cr
&& f_{\eta'}^c = \phantom{-} \theta_c \, \cos\theta_8 \, f_{\eta_c}
\simeq -6.3~{\rm MeV}  \ .
\label{eq:fPc}
\eeq
From Eq.~(\ref{eq:stateqsc}) one can also read off the
light quark admixtures in the $\eta_c$ meson: 1.5\% of $\eta_q$
and 0.8\% of $\eta_s$.
A similar description of $\eta$-$\eta'$-$\eta_c$ mixing
which leads to comparable results has been given
by Petrov.\cite{Petrov:1997yf}

Another estimate of the charm decay constant
of $\eta'$ has been suggested by Franz et al.\cite{Franz:1998hw}
who integrate out the heavy quarks from the QCD Lagrangian to obtain
an approximative {\em operator}\/ relation between the heavy quark
axial-vector current and the topological charge density which is valid
at low scales to leading order in $1/m_c^2$.
Their final result for the decay constants $f_P^c$ reads
\beq
  f_P^c &=& - \frac{1}{12 \, m_c^2} \, A_P \ .
\eeq
This result is
equivalent to a perturbative computation of the $c\bar c$
triangle graph performed by Ali et al.\cite{Ali97}
The so-obtained values for $f_P^c$ are
exactly a factor of three smaller than the ones presented
in Eq.~(\ref{eq:fPc}). The explanation for this discrepancy is
not yet clear. Either the ansatz~(\ref{eq:stateqsc})
is too naive or the {\em perturbative}\/ calculation of $f_P^c$
receives additional non-perturbative contribution
which may emerge from $c\bar c$ modes in the $\eta'$ {\em state}\/ itself.

In any case, the charm component inside the $\eta$ or $\eta'$ turns
out to be rather small, and obviously it is not the dominant
contribution to the $B\to\eta'K$ decay via the elementary
weak process $b \to s c \bar c$. 
The values in Eq.~(\ref{eq:fPc}) also lie
comfortably within the phenomenological bounds, obtained from
the consideration of the $\eta\gamma$ and $\eta'\gamma$ 
transition form factors (see Ref.\cite{FeKr97b} for
details), $ -65\,{\rm MeV} \leq f_{\eta'}^c \leq 15\, {\rm MeV}$.

The mixing with the even heavier bottomonium states is obtained
by scaling Eq.~(\ref{eq:thetac}) 
with the quarkonium masses and decay constants, 
and turns out to be tiny
\beq
  \theta_b &=& \theta_c \, \frac{M_{\eta_c}^2 \, f_{\eta_c}}
			        {M_{\eta_b}^2 \, f_{\eta_b}}
  \simeq - 0.06^\circ \ .
\eeq
Nevertheless, the small admixture of $b\bar b$ in
$\eta$ and $\eta'$ can provide the leading contribution
to the radiative $\Upsilon$ decays in full analogy 
to Eqs.~(\ref{rjpsi}) and (\ref{etacetap}), and
the ratio for the decay widths of $R(\Upsilon)=
\Gamma[\Upsilon\to\eta'\gamma]/\Gamma[\Upsilon\to\eta\gamma]$ 
is again given in terms of
$\theta_y$ only, and its value can be predicted\cite{Feldmann:1998sh} 
to amount to
$6.5$. 

One can also combine the results from Section~4.1 about
$\pi$-$\eta$-$\eta'$ mixing with $\eta$-$\eta'$-$\eta_c$ mixing.
Let me define (cf.~Eq.~(\ref{eq:pietaetap}))
\beq
  |\pi^0\rangle &=& |\varphi_3\rangle +
	\epsilon_q \, |\eta_q\rangle +
        \epsilon_s \, |\eta_s\rangle+
        \epsilon_c \, |\eta_c\rangle \ .
\eeq
Since both, $m_d-m_u$ and $1/m_c^2$ are small expansion
parameters, one can combine Eqs.~(\ref{eq:pietaetap}) 
and (\ref{eq:stateqsc}) and
simply obtains
\beq
  \epsilon_c&=& \epsilon \, (-\theta_c \sin\theta_y) +
              \epsilon' \, (\theta_c \cos\theta_y) \simeq - 1.5 \cdot 10^{-4}
\eeq
which is compatible with the value found by Chao et al.\cite{Chao:1989yp}
but probably too small to be of phenomenological relevance.

\subsection{Mixing with pseudoscalar glueballs and/or higher excitations}

In QCD color singlet combinations of only gluon fields
may appear as hadronic bound states (glueballs).
Lattice calculations, see e.g.\ Ref.\cite{Morningstar:1997ff}, indeed find
some indications for glueballs with masses of a few GeV,
and also some experimental candidates are
heavily discussed in the literature, see e.g.\ Refs.\cite{PDG98,Frere:1998ez,Close}
and references therein.
In particular, pseudoscalar glueballs with masses
of 2~GeV or higher are expected
from QCD sum rules and lattice 
calculations, see e.g.\ Refs.\cite{Narison:1996fm,Gabadadze:1998zc}
On the other hand the pseudoscalar state $\eta(1440)$ (the former $\iota$)
is often considered as a glueball candidate, too.

So far, the chiral effective Lagrangian presented in Section~3.1
has been the basis of my discussion of $\eta$-$\eta'$ mixing.
It does contain neither higher excitations
of light pseudoscalar quarkonium states nor glueballs as explicit 
degrees of freedom. 
However, implicitly the possible effect of additional states is
encoded in the parameters of the effective Lagrangian.
Consider, for example, 
the phenomenological fact that the OZI-rule violating 
coefficients $\Lambda_{1,2,3}$ are small and compatible with zero.
From this one may deduce 
that the glueball admixture in the $\eta$ and $\eta'$
meson is small, say of the order of a few percent, comparable with
the mixing in the $\phi$-$\omega$ system. 
This conclusion has been drawn, for instance,
by Anisovich et al.\cite{Anisovich97} from the analysis of the
$\eta(\eta')\gamma$ transition form factors.
It is important to understand that such a statement does not 
necessarily mean
that there are no pronounced
gluonic effects in the $\eta$ or $\eta'$ meson.
Only, as I discussed in Section~2.1, these effects are not
to be interpreted as admixtures of conventional glueball fields
but rather as topological effects, connected to e.g.\ instantons.
Remember, that the ghost states of the Veneziano-
or Kogut/Susskind-type or the negative metric states
\`a la Weinberg --   which give rise
to a non-vanishing topological susceptibility,
the singlet mass shift, $\eta$-$\eta'$ mixing
itself, an enhanced $J/\psi\to \eta'\gamma$ decay width etc.\
 -- are not physical.

Nevertheless, in order
to obtain a complete picture of the pseudoscalar sector 
one may, of course, include glueballs and higher states
in the mixing scenario, 
e.g.\ the $\eta(1295)$ which is approximately degenerated
with the $\pi(1300)$ and thus likely to be a radial
excitation of an $|\eta_q\rangle$ state.
In principle, a determination of mixing parameters from 
phenomenological considerations similar to the ones performed
in the analysis of $\eta$-$\eta'$ mixing should be possible.
This requires the same attention concerning the
definition of mixing angles, inclusion of $SU(3)_F$--breaking
effects etc.
In particular, the OZI-rule violating contributions have to
be under control.
Without going into further detail,
I refer to the literature (see e.g.\ Refs.\cite{Narison:1996fm,gluemix}
and references therein) where
one can find some analyses in simplified schemes of mixing
between $\eta$ and $\eta'$ mesons, glueballs etc. 
Due to the reasons mentioned
above some of the conclusions therein have to be taken with care,
in particular concerning the glueball nature of mesons
in the pseudoscalar spectrum.

\section{Summary}

In the last few years we have 
achieved a very consistent picture of (strong) mixing phenomena
in the pseudoscalar meson sector, in particular for $\eta$ 
and $\eta'$ mesons. 
The main progress, which has been discussed in detail in this
review, is based on the definition of process-independent
mixing parameters that can be used in different phenomenological
situations and at different energy scales.
The phenomenological determination of
these parameters allows us to understand the properties
of pseudoscalar mesons in terms of their underlying 
quark (and eventually gluon) structure.
An important role is played by the pseudoscalar meson decay constants.
On the one hand, they connect the bare fields in the chiral low-energy
Lagrangian to the physical ones~(\ref{Pphiconnection}). On the other
hand, they enter the light-cone wave functions of
quark-antiquark Fock states in the parton picture.

For the description of the decay constants in the $\eta$-$\eta'$
system a universal octet-singlet mixing angle
$\theta_P$ has been shown to be not sufficient. 
The ratios of $\eta$ and $\eta'$
decay constants with octet or singlet axial vector currents 
define two independent mixing angles $\theta_8$ and
$\theta_0$.  In
chiral perturbation theory the 
difference between the two angles can be expressed 
in terms of the parameter $L_5$ which
in turn is determined by the difference of pion and kaon decay
constants~(\ref{eq:t8t1rel}).
This implies that the connection between bare octet and singlet
fields and physical $\eta$ and $\eta'$ states is not
a simple rotation.

In phenomenological analyses one often does not measure
$\theta_8$ or $\theta_0$ directly, but rather extract
particular ratios of $\eta$ and $\eta'$ matrix elements:
First, there are processes which are induced by the 
topological charge density $\omega = \frac{\alpha_s}{8\pi} \, G \tilde G$,
like, for instance, the radiative $J/\psi$ decays into $\eta$ or 
$\eta'$.\cite{Novikov:1980uy}
Matrix elements of $\omega$ between the vacuum and $\eta$ or $\eta'$
fields can be used to define another mixing angle ($\theta_y$).
Secondly, one considers matrix elements of quark currents with only light
($u$, $d$) or only strange quarks, respectively, for example in the decays
$J/\psi\to \rho\eta(\eta')$. 
A priori, every ratio of independent $\eta$ and $\eta'$ observables
defines an independent observable.
In order to keep the predictive power of the whole mixing approach one
has to apply the usual OZI-rule, i.e.\
contributions from quark-antiquark annihilation are
neglected while the effect of topologically non-trivial field
configurations (e.g.\ instantons) is kept.
These assumptions lead to the FKS scheme where 
mixing in the 
quark-flavor basis is described by a single mixing angle $\phi$.
Moreover,
by exploiting the Ward identities, including the $U(1)_A$ anomaly,
one obtains the important relation~(\ref{eq:anglerel})
which connects the angles  
$\theta_8$, $\theta_y$ and $\phi$.
Using these relations instead of the naive one, $\theta_P=\phi-\arctan\sqrt2$,
resolves a big part of the inconsistencies which have
been present in the literature for many years.
With the set of mixing parameters~(\ref{eq:parset})
one reproduces 
an abundance of phenomenological data 
and fulfills important theoretical constraints.
The uncertainties in the determination of the
mixing parameters, arising from the experimental errors,
is rather small. The systematical error 
due to the neglect of OZI-rule violating contributions
is of order $1/N_C$. Empirically,
the comparison of different phenomenological
approaches indicates that the value of the mixing angle
$\theta_0$ may be rather sensitive to the treatment
of $1/N_C$ corrections.

An important test of the mixing approach is provided
by the simultaneous consideration of 
the decays $\eta(\eta')\to\gamma\gamma$ on the
basis of the Wess-Zumino-Witten term
      and the $\eta(\eta')\gamma$ transition form factors
      at large momentum transfer
in the hard-scattering approach.
The two-angle
      parameterization of decay constants turned out to be crucial.
In this context we have also seen that the light-cone wave functions
of the quark-antiquark Fock states of $\pi$, $\eta$ and $\eta'$ mesons
can not be very different\cite{Anisovich97,FeKr97b} 
which implies that -- up to charge factors --
the members of the light
pseudoscalar nonet behave similarly in hard exclusive reactions.

Moreover, I have considered various mass formulas:
The $U(1)_A$ mass shift and its connection
to the topological susceptibility has been determined and
compared to lattice QCD analyses. An $SU(3)_F$ improved Gell-Mann--Okubo
formula on the basis of Dashen's theorem has been shown to
give an estimate of the angle $\theta_8$.
Also the influence of flavor symmetry breaking on Schwinger's mass
formula has been discussed.
The diagonalization of a mass matrix
in the FKS scheme\cite{Feldmann:1998vh}
turns out to lead to equivalent results
as the analysis of anomalous Ward identities performed in 
Ref.\cite{Diakonov:1981nv}

The $\eta$-$\eta'$ mixing approach has been
generalized to include also other states, for
instance $\pi^0$ and/or $\eta_c$ mesons. 
In particular, I find a reasonably small intrinsic
charm component in the $\eta'$ meson which cannot
provide a dominant contribution to the decay $B\to \eta'K$.
Concerning the role of pseudoscalar glueballs and/or higher
excitations, I have stressed the importance 
of distinguishing  physical glueball fields from topological
effects related to e.g.\ instantons.
In $\chi$PT their effect
is encoded in the parameters
of the effective Lagrangian. With the present amount of 
data there is no signal for a sizeable pseudoscalar glueball admixture
in $\eta$ or $\eta'$ mesons.

The question of how to treat gluonic contributions 
has been shown to be also of relevance for
the description of the
singlet axial charge of the nucleon by means of
generalized Goldberger-Treiman relations with
the pseudoscalar coupling constants. 
Saturating the axial charges by the physical $\eta$ and $\eta'$
fields only, I obtained values for $g_{\eta NN}$ and $g_{\eta' NN}$
which are consistent with the bounds from an analysis
of $N$-$N$ scattering data on the basis of dispersion 
relations.\cite{Grein:1980nw} On the
other hand, introducing an additional 
contribution of the Veneziano-ghost 
to the singlet axial charge of the nucleon
 and relating it to $\Delta g$ leads to a larger value of
$g_{\eta'NN}$ which exceeds the bounds in Ref.\cite{Grein:1980nw}

The phenomenological values of mixing parameters 
have been used to fix several matrix elements involving 
pseudoscalar mesons which may be useful in future applications.
Of present interest are, for instance, the decays of $B$ mesons
with $\eta$ or $\eta'$ in the final state.\cite{Ali97,Chen:1999nx}
Another topic, which has been discussed recently,
is the electro-production of 
$\pi^0$, $\eta$ and $\eta'$ mesons off nucleons or 
deuterons.\cite{Eides:1998ch}
Also, for the description of
 charmonium decays into light pseudoscalar mesons\cite{Bolz:1997ez} 
a decent knowledge
of the mixing parameters is necessary.\cite{Feldmann:1998sh}
As I discussed, there are more exotic cases, namely 
couplings of $\eta$ or $\eta'$ to Pomerons, Odderons or
$Z$-bosons, which in principle may provide a direct 
determination of the mixing angles $\theta_8$ and,
in particular, $\theta_0$.
The theoretical lessons we learned from
the pseudoscalar sector 
may also be helpful for our understanding
of mixing phenomena 
in other channels (scalar mesons, vector mesons, \ldots).

\nonumsection{Acknowledgements}
\noindent
It is a pleasure to thank Peter Kroll
and Berthold Stech for a most enjoyable collaboration
on the subject of $\eta$-$\eta'$ mixing.
Furthermore, I want to express my gratitude to Heiri Leutwyler and
Roland Kaiser for interesting and inspiring discussions.
I thank Vladimir Savinov whose analysis of the
CLEO data on the $\eta(\eta')\gamma$ form factors motivated
us to reconsider the $\eta$ and $\eta'$ properties.
I appreciate helpful comments
from  Maurice Benayoun, Rafel Escribano, Stephan G\"usken, Maxim Polyakov,
Thorsten Struckmann, and Christian Weiss.
I am grateful for financial support from
{\it Deutsche Forschungsgemeinschaft}\/.

\nonumsection{References}

\end{document}